\documentclass[journal]{IEEEtran}

\usepackage{cite}
\usepackage{amsmath,amssymb,amsfonts,amsthm}
\usepackage{algorithmic}
\usepackage{graphicx}
\usepackage{textcomp}
\usepackage{url}
\usepackage[justification=centering]{caption}

\usepackage{textcomp} 

\usepackage{enumitem}
\setlist{nosep}
\usepackage{enumitem}

\usepackage{subfloat} 

\usepackage{multirow}
\usepackage{pifont}
\newcommand{\cmark}{\ding{52}}  
\newcommand{\dmark}{\ding{108}} 
\newcommand{\amark}{\ding{212}} 

\usepackage{hyperref}
\hypersetup{colorlinks=true,linkcolor=blue,}

\usepackage[table, svgnames, dvipsnames]{xcolor}

\begin{document}

\title{A Survey of Rate-optimal Power Domain NOMA with Enabling Technologies of Future Wireless Networks}

\author{Omar~Maraqa, Aditya S. Rajasekaran~\IEEEmembership{(Member,~IEEE)}, Saad Al-Ahmadi, Halim Yanikomeroglu~\IEEEmembership{(Fellow,~IEEE)}, Sadiq M. Sait~\IEEEmembership{(Senior Member,~IEEE)}%

\thanks{O.~Maraqa and S. Al-Ahmadi are with the Department of Electrical Engineering, King Fahd University of Petroleum \& Minerals, Dhahran-31261, Saudi Arabia (e-mail: g201307310@kfupm.edu.sa; saadbd@kfupm.edu.sa).}%
\thanks{A.~S.~Rajasekaran is with the Department of Systems and Computer Engineering, Carleton University, Ottawa, ON K1S 5B6, Canada and with Ericsson Canada Inc, Ottawa, ON K2K 2V6, Canada (email: aditya.sriram.rajasekaran@ericsson.com).}%
\thanks{H. Yanikomeroglu is with the Department of Systems and Computer Engineering, Carleton University, Ottawa, ON K1S 5B6, Canada (email: halim@sce.carleton.ca).}%
\thanks{S.~M.~Sait is with the Department of Computer Engineering, and Center of Communications and IT Research, Research Institute, King Fahd University of Petroleum \& Minerals, Dhahran-31261, Saudi Arabia (e-mail: sadiq@kfupm.edu.sa).}%
\thanks{$\copyright$ 2020 IEEE. Personal use of this material is permitted. Permission from IEEE must be obtained for all other uses, in any current or future media, including reprinting/republishing this material for advertising or promotional purposes, creating new collective works, for resale or redistribution to servers or lists, or reuse of any copyrighted component of this work in other works. Accepted by IEEE Communications Surveys \& Tutorials, July 2020.}%
}

\markboth{Accepted by IEEE Communications Surveys \& Tutorials, VOL. xx, NO. xx, July 2020}%
{Maraqa~\MakeLowercase{\textit{et al.}}: A Survey of Rate-optimal Power Domain NOMA with Enabling Technologies of Future Wireless Networks}

\maketitle

\begin{abstract}
The ambitious high data-rate applications in the envisioned future beyond fifth-generation (B5G) wireless networks require new solutions, including the advent of more advanced architectures than the ones already used in 5G networks, and the coalition of different communications schemes and technologies to enable these applications requirements. Among the candidate communications schemes for future wireless networks are non-orthogonal multiple access (NOMA) schemes that allow serving more than one user in the same resource block by multiplexing users in other domains than frequency or time. In this way, NOMA schemes tend to offer several advantages over orthogonal multiple access (OMA) schemes such as improved user fairness and spectral efficiency, higher cell-edge throughput, massive connectivity support, and low transmission latency. With these merits, NOMA-enabled transmission schemes are being increasingly looked at as promising multiple access schemes for future wireless networks. When the power domain is used to multiplex the users, it is referred to as the power domain NOMA (PD-NOMA). In this paper, we survey the integration of PD-NOMA with the enabling communications schemes and technologies that are expected to meet the various requirements of B5G networks. In particular, this paper surveys the different rate optimization scenarios studied in the literature when PD-NOMA is combined with one or more of the candidate schemes and technologies for B5G networks including multiple-input-single-output (MISO), multiple-input-multiple-output (MIMO), massive-MIMO (mMIMO), advanced antenna architectures, higher frequency millimeter-wave (mmWave) and terahertz (THz) communications, advanced coordinated multi-point (CoMP) transmission and reception schemes, cooperative communications, cognitive radio (CR), visible light communications (VLC), unmanned aerial vehicle (UAV) assisted communications and others. The considered system models, the optimization methods utilized to maximize the achievable rates, and the main lessons learnt on the optimization and the performance of these NOMA-enabled schemes and technologies are discussed in detail along with the future research directions for these combined schemes. Moreover, the role of machine learning in optimizing these NOMA-enabled technologies is addressed.
\end{abstract}


\section{Introduction}\label{sec:intro}

\IEEEPARstart{F}{}uture wireless networks such as beyond fifth-generation (B5G) or 6G  networks are expected to support extremely high data rates (up to 1 Tbps) and a very large number of users or nodes (up to $10^7$ nodes per $km^2$) with a variety of applications and services~\cite{8808168}. However, multiple access schemes used in the past generations of cellular networks will not scale to meet these unprecedented demands for user density and network traffic. The multiple access schemes used to date in cellular networks include frequency division multiple access (FDMA) in 1G systems, time division multiple access (TDMA) in 2G, code division multiple access (CDMA) in 3G, and orthogonal frequency division multiple access (OFDMA) in 4G networks. The common theme in all these multiple access schemes is \lq\lq orthogonality\rq\rq~where, theoretically, the different users are allocated distinct frequency channels or time slots or signature codes or resource blocks so that they do not interfere with one another when they access the network resources. However, this insistence on orthogonality poses a limit on the number of users that can access the network resources and thereby reduces the overall spectral efficiency (SE). Non-orthogonal multiple access (NOMA) schemes, on the other hand, allow multiple users to share the same resource (e.g., a time/frequency resource block) and separate the users in other domains with some additional receiver complexity~\cite{8823873}. When the power domain is used to separate the users, it is referred to as the power-domain NOMA (PD-NOMA)~\cite{islam2017power} scheme and is the focus of this survey paper. Alternatively, if the users are separated through non-orthogonal codes, it is referred to as code-domain NOMA (CD-NOMA)~\cite{cai2018modulation, dai2018survey}. PD-NOMA has also been studied in conjunction with CD-NOMA recently~\cite{Sharma2019PDCDNOMA}, where such combination is beyond the scope of this paper.

To support the large explosion of connected users forecast in the massive machine-type communications (mMTC) paradigm of 5G and beyond networks, interest has been growing in academia, industry and even standardization bodies like the 3rd generation partnership project (3GPP) to adopt NOMA schemes~\cite{towardsNomaStandardization}. NOMA and/or hybrid NOMA-OMA schemes tend to have the following attractive merits for 5G~\cite{dai2018survey, vaezi2019multiple} and B5G networks:

\begin{itemize}
    \item Improved spectral efficiency; as multiple users of the networks that adopt NOMA schemes can occupy the same bandwidth which is not applicable in multi-user networks that utilize OMA schemes. Moreover, in OMA schemes, the resource block might be allocated to a user with a poorly received signal strength leading to reduced spectral efficiency.     
    \item Improved user fairness; as NOMA schemes can accommodate multiple users in the same resource block with guaranteed minimum rate requirements. On the contrary, OMA users with poor channel conditions might not be served for a long time by a scheduler trying to optimize the overall spectral efficiency.
    \item Low transmission latency; as the users in the networks that adopt OMA schemes need to wait for an available orthogonal resource block to grant access for transmission. In contrast, NOMA schemes offer more flexible user scheduling opportunities as well as grant-free transmission.
     \item Higher cell-edge throughput; as NOMA schemes allow a base station (BS) to flexibly change the fraction of power allocated to a cell-edge user to support a certain quality of service (QoS), which accordingly enhances its transmission rate.
\end{itemize}

With its promising performance, PD-NOMA is considered as a candidate multiple access scheme in various standardization activities. For long term evolution advanced (LTE-A) systems (3GPP release 13), NOMA was considered under the name of multi-user superposition transmission (MUST)~\cite{3Gpp-release-13-MUST}. Furthermore, in LTE-A Pro (3GPP release 14), the standardization body recognized that at least uplink NOMA schemes should be considered especially for mMTC~\cite{3Gpp-proposal-release-14}. In 5G new radio (NR) phases (1~\&~2) (3GPP release 15 and release 16)~\cite{3Gpp-release-16,3Gpp-release-16-2}, multiple studies on the advancements needed in the transmitter and receiver sides for adopting NOMA schemes have been proposed. Also, some link-level and system-level performance evaluations that show the potential of adopting NOMA schemes have been conducted in these studies.

Further, apart from NOMA schemes, several other PHY schemes, technologies, and network architecture paradigms are brewing in the literature to meet the aforementioned high data-rate demands of future wireless networks. These include massive-MIMO (mMIMO), cell-free mMIMO (CF-mMIMO), reconfigurable antenna systems, large intelligent surfaces (LIS), 3-Dimensional MIMO (3-D MIMO), millimeter-wave (mmWave) and terahertz (THz) communications, coordinated multi-point (CoMP) schemes, cooperative communications, cognitive radio communications systems, visible light communications (VLC) systems, unmanned aerial vehicle (UAV) assisted communications systems, and other enabling schemes and technologies for B5G networks. The realization of the target rates in these networks will require the integration of two or more of these enabling schemes and technologies. Hence, these schemes and technologies can be potentially integrated with PD-NOMA in one form or the other and is not simply the addition of two existing technologies as discussed in detail in~\cite{8792153}. In this paper, we survey the vast body of recent literature on the  optimization of the achievable rates in such a combined setting of PD-NOMA with these enabling schemes and technologies. 

\begin{subtables}
\begin{table*}[!htb]
\centering
\vspace{-2em} 
\caption{A comprehensive list of the existing survey papers and magazine articles that considered the integration of rate-optimal NOMA with enabling technologies of future wireless networks (advanced multi-antenna architectures, include, cell-free mMIMO, reconfigurable antenna systems, large intelligent surfaces, and 3-D MIMO). Notions: \dmark \ \amark \ scattered discussion (i.e., rate-optimal NOMA works were mentioned alongside the NOMA works that considered optimizing other metrics, such as power minimization and energy efficiency maximization as well as the NOMA works that investigated performance analysis metrics, such as BER, SER, and outage probability. Hence, rate-optimal NOMA works were discussed in a scattered fashion within those survey papers). \cmark \ \amark \ partial discussion (i.e., there was at least one dedicated subsection/table for rate-optimal NOMA scheme within those survey papers). \cmark\cmark \ \amark \ detailed discussion.}
\label{table: NOMA surveys and magazine papers}
\resizebox{\textwidth}{!}{%
\begin{tabular}{|c|c|c|c|c|c|c|c|c|c|c|c|c|c|c|c|c|c|}

\hline 
\textbf{[\#]} & \textbf{Year}& \textbf{Type} & \textbf{\begin{tabular}[c]{@{}c@{}}MIMO\\ and\\ mMIMO \end{tabular}}  & \textbf{CoMP} & \textbf{Cooperative} & \textbf{\begin{tabular}[c]{@{}c@{}}Cognitive \\ Radio\end{tabular}} & \textbf{mmWave} & \textbf{VLC} & \textbf{UAV} & \textbf{V2X} & \textbf{BackCom} & \textbf{\begin{tabular}[c]{@{}c@{}}Terrestrial-\\Satellite \end{tabular}} & \textbf{MEC} & \textbf{Underwater} & \textbf{IRS}& \textbf{THz}&
\textbf{\begin{tabular}[c]{@{}c@{}}Advanced\\ multi-antenna\\ architectures\end{tabular}}\\ 
 \hline


\cite{yan2016non} & 2016 & \multirow{18}{*}{\begin{tabular}[c]{@{}c@{}}Survey\\ Papers\end{tabular}} & \dmark &  &  &  &  &  &  &  &  &  &  &  &  &  & \\ 
\cline{1-2} \cline{4-18}

\cite{wei2016survey} & 2016 &  & \dmark &  & \dmark &  &  &  &  & &  &  &  &  &  & & \\
\cline{1-2} \cline{4-18}


\cite{ding2017survey} & 2017 &  & \dmark &  & \dmark &  & \dmark &  &  & &  &  &  &  &   & &  \\ 
\cline{1-2} \cline{4-18}

\cite{liu2017non} & 2017 &  & \cmark & \dmark & \cmark & \dmark & \dmark & \dmark &  & &  &  &  &  & & &  \\ 
\cline{1-2} \cline{4-18}

\cite{basharat2018survey} & 2017 &  &  &  & \dmark &  &  &  &  & &  &  &  &  &  &  &  \\
\cline{1-2} \cline{4-18}

\cite{islam2017power} & 2017 &  & \dmark & \dmark & \dmark &  &  & \dmark &  & &  &  &  &  &  & &\\
\cline{1-2} \cline{4-18}

\cite{cai2018modulation}& 2018 &  & \cmark &  & \cmark & \dmark &  &  &  &  &  &  &  &  &  & &  \\
\cline{1-2} \cline{4-18}

\cite{dai2018survey} & 2018 &  & \dmark & \dmark & \dmark &  &  &  &  &  &  &  &  &  & &  &\\
\cline{1-2} \cline{4-18}

\cite{ding2018embracing} & 2018 &  & \dmark &  & \dmark &  & \dmark & \dmark &  & \dmark &  & \dmark &\dmark &  & & & \\
\cline{1-2} \cline{4-18}

\cite{liaqat2018power}& 2018 &  &  &  & \cmark & \dmark & & \dmark &  & &  &  &  &  &  & &\\
\cline{1-2} \cline{4-18}

\cite{aldababsa2018tutorial} & 2018 &  & \dmark &  & \dmark &  &  &  &  & &  &  &  &  &  & & \\
\cline{1-2} \cline{4-18}





\cite{bawazir2018multiple} & 2018 &  &  &  &  &  &  & \dmark &  & &  &  &  &  &  &  &\\
\cline{1-2} \cline{4-18}

\cite{zeng2018investigation} & 2018 &  & \dmark &  & \dmark &  & \dmark &  &  & &  &  &  &  & & & \\ 
\cline{1-2} \cline{4-18}

\cite{li2018uav} & 2019 &  &  &  &  &  &  &  & \dmark & &  &  &  &  & & &\\
\cline{1-2} \cline{4-18}

\cite{8792153}& 2019 &  & \dmark & \dmark & \dmark & \dmark & \dmark & \dmark &  & &  &  &\dmark &  &  &  &\\
\cline{1-2} \cline{4-18}

\cite{obeed2019optimizing} & 2019 &  &  &  &  &  &  &  \cmark &  & &  &  &  &  &  & & \\
\cline{1-2} \cline{4-18}

\cite{8698841} & 2019 &  & &  &  &  &  & \dmark &  & &  &  &  &  &  & & \\
\cline{1-2} \cline{4-18}

\cite{8764406} & 2019 &  &  &  &  &  & \dmark &  & \dmark  & &  &  &  &  &  & &\\
\cline{1-2} \cline{4-18}


\cite{anwar2019survey} & 2019 &  &  & \dmark &  &  &  &  &  &  &  &  &  &  &  & & \\
\cline{1-2} \cline{4-18}

\cite{memon2019backscatter} & 2019 &  &  &  &  &  &  &  &  & & \dmark &  &  &  &  & & \\
\cline{1-2} \cline{4-18}

\cite{8715508} & 2019 &  &  &  &  &  &  &  &  & &  & \dmark &  &  &  & &\\
\cline{1-2} \cline{4-18}

\cite{tian2019multiple} & 2019 &  & \dmark  &  & \dmark & \dmark  & \dmark &  &  & &  &  &  &  &  & &\\
\cline{1-2} \cline{4-18}

\cite{8798636} & 2019 &  &  &  &  & & \dmark & & & &  &  &  &  &  & & \\
\hline


\cite{dai2015non} & 2015 & \multirow{23}{*}{\begin{tabular}[c]{@{}c@{}}Magazine \\ Articles\end{tabular}} & \dmark &  &  &  &   &  &  &  &  &  &  &  &  & & \\
\cline{1-2} \cline{4-18}

\cite{li2015non}& 2015 &  & \dmark &  &  &  &  &  &  & &  &  &  &  &  & & \\
\cline{1-2} \cline{4-18}

\cite{noma-v2v}& 2017 &  &  &  &  &  &  &  &  & \dmark&  &  &  &  &  & & \\
\cline{1-2} \cline{4-18}

\cite{8088525}& 2017 &  &  &  &  &  & \dmark &  &  & &  &  &  &  & & & \\
\cline{1-2} \cline{4-18}

\cite{ding2017application}& 2017 &  & \dmark &  & \dmark & \dmark &  &  &  & &  &  &  &  & & &\\
\cline{1-2} \cline{4-18}

\cite{shin2017non} & 2017 &  &  & \dmark &  &  &  &  &  &  &  &  &  &  & & &\\
\cline{1-2} \cline{4-18}




\cite{islam2018resource} & 2018 &  & \dmark &  &  &  &  &   &  & &  &  &  &  &  & &\\
\cline{1-2} \cline{4-18}

\cite{liu2018multiple} & 2018 &  & \dmark &  &  &  &  &  &  & &  &  &  &  &  & &\\
\cline{1-2} \cline{4-18}

\cite{huang2018signal} & 2018 &  & \dmark &  &  &  & \dmark &  &  & &  &  &  &  &  & &\\
\cline{1-2} \cline{4-18}

\cite{zhong2018spatial}& 2018 &  & \dmark &  &  &  &  &   &  & &  &  &  &  &  & &\\
\cline{1-2} \cline{4-18}

\cite{wan2018non} & 2018 &  &  &  & \dmark &  &  &  &  &  &  &  &  &  &  & &\\
\cline{1-2} \cline{4-18}


\cite{zhou2018state} & 2018 &  &  &  &  & \dmark &  &  &  & &  &  &  &  &  & &\\
\cline{1-2} \cline{4-18}

\cite{lv2018cognitive} & 2018 &  &  &  & \dmark & \dmark &  &  &  & &  &  &  &  &  & &\\ 
\cline{1-2} \cline{4-18}

\cite{marshoud2018optical} & 2018 &  &  &  &  &  &  & \cmark &  & &  &  &  &  & & & \\
\cline{1-2} \cline{4-18}

\cite{zhou2018energy} & 2018 &  & \dmark &  & \dmark & \dmark & \dmark &  &  & &  &  &  &  &  & & \\ 
\cline{1-2} \cline{4-18}

\cite{bai2018multi} & 2018 &  &  &  &  &  &  &  &  & &  & \dmark &  &  & & &\\
\cline{1-2} \cline{4-18}

\cite{zhang2018heterogeneous} & 2018 &  &  & \dmark &  &  &  &  &  & &  &  &  &  &  & &\\
\cline{1-2} \cline{4-18}

\cite{ali2018coordinated} & 2018 &  &  & \dmark &  &  &  &  &  & &  &  &  &  &  & &\\
\cline{1-2} \cline{4-18}

\cite{chandra2018unveiling} & 2018 &  &  &  &  &  & \dmark &  &  & &  &  &  &  &  & & \\
\cline{1-2} \cline{4-18}

\cite{8403960} & 2018 &  &  &  &  &  &  &  &  & &  &  & \dmark &  &  & & \\
\cline{1-2} \cline{4-18}

\cite{8694790} & 2019 &  &  &  & \dmark & \dmark &  &  &  & &  &  &  &  &  &  & \\
\cline{1-2} \cline{4-18}

\cite{liu2019uav} & 2019 &  &  &  &  &  &  &  & \dmark & &  &  &  &  &  & & \\
\cline{1-2} \cline{4-18}

\cite{8786078} & 2019 &  &  &  & \dmark & &  & & & &  &  &  &  &  & & \\
\cline{1-2} \cline{4-18}

\cite{8930827} & 2019 &  &  &  & &  &  & & & &  & \dmark &  &  &  & & \\
\cline{1-2} \cline{4-18}

\cite{zhu2019cooperative} & 2019 &  &  &  & \dmark &  &  & & & &  & \dmark &  &  & &  & \\
\hline

\multicolumn{3}{|c|}{This survey paper} & \cmark\cmark & \cmark\cmark & \cmark\cmark & \cmark\cmark & \cmark\cmark & \cmark\cmark & \cmark\cmark & \cmark\cmark & \cmark\cmark  & \cmark\cmark & \cmark\cmark & \cmark\cmark & \cmark\cmark & \cmark\cmark & \cmark\cmark \\
\cline{4-18}
\hline
\end{tabular}%
}
\end{table*}

\begin{table*}[!htb]
\centering
\caption{A finer look into the survey papers and magazine articles that partially considered rate-optimal NOMA-enabled schemes and technologies.}
\label{Table: A deep look}
\resizebox{\textwidth}{!}{%
\begin{tabular}{|c|c|c|c|}
\hline
\textbf{[\#]} & \textbf{Year}& \textbf{Type} & \textbf{Similarities}\\ 
\hline %

\cite{liu2017non} & 2017 & \multirow{7}{*}{\begin{tabular}[c]{@{}c@{}}Survey\\ Papers\end{tabular}} & \begin{tabular}[c]{@{}c@{}}For MISO-MIMO NOMA/cooperative NOMA: the authors listed in Table 4/5~\cite{liu2017non} seven/five\\ rate-optimal NOMA works and discussed their main technical contributions in text. In addition,\\ scattered discussion of rate-optimal NOMA scheme for CoMP, cognitive radio, mmWave, and\\ VLC were provided in text. \end{tabular} \\
\cline{1-2} \cline{4-4} 

\cite{cai2018modulation} & 2018 &  & \begin{tabular}[c]{@{}c@{}}For MISO-MIMO NOMA/cooperative NOMA: the authors listed in Table II/III~\cite{cai2018modulation} six/three\\ rate-optimal NOMA works and discussed their main technical contributions in text. \end{tabular} \\
\cline{1-2} \cline{4-4} 

\cite{liaqat2018power} & 2018 &  & \begin{tabular}[c]{@{}c@{}}For cooperative NOMA: the authors listed in Table 3~\cite{liaqat2018power} six rate-optimal NOMA works and\\ discussed their main technical contributions in text.\end{tabular} \\
\cline{1-2} \cline{4-4} 

\cite{obeed2019optimizing} & 2019 &  & \begin{tabular}[c]{@{}c@{}}For VLC-NOMA: the authors listed in Table IV~\cite{obeed2019optimizing} four rate-optimal NOMA works and\\ discussed their main technical contributions in text. \end{tabular} \\
\hline 

\cite{marshoud2018optical} & 2018 & \begin{tabular}[c]{@{}c@{}}Magazine \\ Article\end{tabular} & \begin{tabular}[c]{@{}c@{}} For VLC-NOMA: the authors dedicated one subsection for rate-optimal VLC-NOMA system\\ entitled "Performance of Optical-NOMA" were they discussed two rate-optimal NOMA works.\end{tabular} \\
\hline 
\end{tabular}%
}
\end{table*}

\end{subtables}

\begin{table*}[!htbp]
\centering
\caption{List of abbreviations.}
\label{Table: Abbreviations}
\begin{tabular}{|l|l|l|l|}
\hline
\textbf{Abbreviation} & \textbf{Description} & \textbf{Abbreviation} & \textbf{Description}\\ \hline

3GPP &3rd generation partnership project & LTE-A &Long term evolution advanced\\ \hline

3-D MIMO & 3-Dimensional MIMO & MEC &Mobile edge computing\\ \hline

5G	&Fifth generation & MIMO & Multiple-input-multiple-output \\ \hline

AF	&Amplify-and-forward & MISO &Multiple-input-single-output \\ \hline

AO	&Alternating optimization & ML & Machine learning\\ \hline

AP	&Access point & MMA	&Minorization maximization algorithm\\ \hline

B5G &Beyond fifth generation & mMIMO&Massive multiple-input multiple-output \\ \hline

BA	&Bisection algorithm & mMTC&Massive machine type communications\\ \hline

BackCom	&Backscatter communications & mmWave&Millimeter-wave \\ \hline

BCD	&Block coordinate descent method & MRC & Maximum ratio combining \\ \hline

BD & Backscatter devices & MU-MIMO&Multi-user multiple-input-multiple-output \\ \hline

BER	&Bit error rate & NGDPA&Normalized gain difference power allocation \\ \hline

BF	&Beamforming & NOMA &Non-orthogonal multiple access\\ \hline

BnB	&Branch-and-bound method & NR	&New radio\\ \hline

BR & Backscatter receiver& OFDMA &Orthogonal frequency division multiplexing access \\ \hline

BS	&Base station & OMA	&Orthogonal multiple access  \\ \hline

CB	&Coordinated beamforming & PA	&Power allocation \\ \hline

CCCP& Constrained convex-concave procedure & PD-NOMA	&Power domain NOMA \\ \hline

CD-NOMA& Code-domain NOMA & PHY	&Physical layer\\ \hline

CDMA &Code division multiple access & PU	&Primary user or licensed user \\ \hline

CF	& Compress-and-forward & QoS & Quality of service \\ \hline

CF-mMIMO	& Cell-free mMIMO & RF	&Radio frequency \\ \hline

CoMP	&Coordinated multi-point & RL & Reinforcement learning\\ \hline

CR	&Cognitive radio & SA	&Simulated annealing algorithm \\ \hline

CR-NOMA	&Cognitive radio non-orthogonal multiple access & SCA	&Sequential/Successive convex approximation\\ \hline

CSI	&Channel state information &SCMA	&Sparse code multiple access \\ \hline

CSIT&Channel state information at the transmitter & SDMA	&Spatial division multiple access \\ \hline

D.C.&Difference of convex & SDP	&Semi-definite programming \\\hline

D2D	&Device-to-device & SDR	&Semi-definite relaxation\\ \hline

DE	&Differential evolution algorithm  & SE	&Spectral efficiency   \\ \hline

DF	&Decode-and-forward & SER	&Symbol error rate\\ \hline

DL	&Deep learning & SI & Self-interference \\ \hline

EE  &Energy efficiency & SIC	&Successive interference cancellation  \\ \hline

FDD	&Frequency division duplex & SINR	&Signal-to-noise and interference ratio\\ \hline

FD	&Full-duplex & SISO	&Single-input-single-output\\ \hline

FDMA &Frequency division multiple access & SNR	&Signal-to-noise ratio \\ \hline

FoV	&Field of view & SU	&Secondary user or non-licensed user\\ \hline

FPA	&Fixed power allocation & SWIPT	&Simultaneous wireless information and power transfer\\ \hline

GP	&Geometric program  & TDMA &Time division multiple access \\ \hline

GPA	&Gradient projection algorithm  & TPs	 &Transmission points\\ \hline

GRPA&Gain ratio power allocation & TRPs	&Transmission reception points\\ \hline

HAP	&High altitude platform & UAV	&Unmanned aerial vehicle  \\ \hline

HD	&Half-duplex & UE & User equipment \\ \hline

HetNets &Heterogeneous networks & UL	&Uplink \\ \hline

ICI	&Inter-cell interference   & US/UC	&User selection/User clustering \\ \hline

IoT &Internet of things & V2V	&Vehicle-to-vehicle communications\\ \hline

IRS	&Intelligent reflecting surfaces & V2I	&Vehicle-to-infrastructure communications \\ \hline

JT	&Joint transmission & V2X	&\begin{tabular}[c]{@{}c@{}}Vehicle-to-everything communications subsumes vehicle-\\to-vehicle and vehicle-to-infrastructure communications \end{tabular}\\ \hline

KKT	&Karush-Kuhn-Tucker  & VLC	&Visible light communications  \\ \hline

LAP	&Low altitude platform & WMMSE	&Weighted minimum mean square error\\\hline

LED	&Light emitting diode & WSR & Weighted sum rate \\\hline

LIS	&Large intelligent surfaces & ZF  & Zero-forcing  \\\hline

LoS	&Line-of-sight & ZF-BF	& Zero-forcing beamforming\\\hline

\end{tabular}%
\end{table*}

In this paper, our contribution is to present a comprehensive up-to-date review on the integration of PD-NOMA~\footnote{In this paper, we refer to NOMA and PD-NOMA interchangeably. Hence, where ever NOMA is used after this point, it always refers to the PD-NOMA.} with advanced antenna architectures, in sub 6 GHz, mmWave, and THz bands, CoMP, cooperative communications, cognitive radio, VLC, UAV, and the other emerging communications schemes to maximize the achievable rates and hence the spectral efficiency in future wireless networks. The rate optimization problems in such NOMA-enabled schemes or technologies are typically non-convex and their level of complexity increase as the number of system parameters get large. Hence, in this survey, we present the adopted system models and specify optimization techniques utilized for each category in the above list of NOMA-enabled schemes and technologies. In each category, we highlight the main findings of the surveyed papers while offering insights into the common themes of the rate optimization methods in each category. Finally, we offer future research directions for the integration of PD-NOMA with each of these schemes/technologies and their possible combinations.  

\subsection{Related Work and Existing Surveys} 

The up-to-date list of surveys~\cite{yan2016non,wei2016survey, ding2017survey,liu2017non,islam2017power, basharat2018survey,cai2018modulation,ding2018embracing,dai2018survey,liaqat2018power,aldababsa2018tutorial, bawazir2018multiple,li2018uav,8792153,zeng2018investigation,obeed2019optimizing,8698841,8764406,anwar2019survey,memon2019backscatter,8715508,tian2019multiple,8798636} and magazine articles~\cite{dai2015non,li2015non,noma-v2v,ding2017application,8088525,shin2017non, islam2018resource,liu2018multiple,huang2018signal,zhong2018spatial,wan2018non,zhou2018state,lv2018cognitive,marshoud2018optical,zhou2018energy,bai2018multi,zhang2018heterogeneous,ali2018coordinated,chandra2018unveiling,8403960,8694790,liu2019uav,8786078,8930827, zhu2019cooperative} that have appeared on PD-NOMA scheme are shown in Table~\ref{table: NOMA surveys and magazine papers}. As the table captures, many of these survey papers go beyond the typical NOMA setting and discuss the integration with multiple antennas and other technologies. Of these, the paper in~\cite{8792153} is the most related in looking into the interplay between NOMA and other enabling technologies. However, we differentiate our survey from the work in~\cite{8792153} and others by focusing on the rate-optimization problems that arise from the integration of these several technologies with PD-NOMA. As discussed before in this Section, the central theme of this survey surrounds the system models, design variables, constraints, etc. involved in solving the typically non-convex and possibly combinatorial rate optimization problems that arise from the integration of the system parameters of both NOMA and the enabling technologies. In this way, we distinguish our work from~\cite{8792153} which is merely a general discussion of how would NOMA enables some of these technologies. Some other surveys identified in Table~\ref{Table: A deep look} discuss rate-optimal NOMA schemes in certain settings or with some enabling technologies. However, to the best of the authors' knowledge, no survey yet exists that is dedicated to the investigation of the rate optimization of PD-NOMA with the main PHY-layer enabling schemes and technologies for high data rate future wireless communications networks. 

It is also worth highlighting that achievable rate optimization problems form the vast majority of 
existing work in these NOMA-integrated enabling B5G schemes studied in the literature. Hence, we limit the scope of the survey to the rate optimization problems to ensure a good focus for the discussion, but it still captures a vast majority of the work in this area. As a result, papers that investigated performance analysis metrics such as symbol error rate (SER), bit error rate (BER), and outage probability in such PD-NOMA are beyond the scope of this survey. The interested reader may refer, for example, to the work published in~\cite{ding2014performance,ding2015impact,yang2016general,8717609} and references therein to explore the performance of the PD-NOMA scheme for the aforementioned metrics. In addition, works that investigated the power minimization, for example,~\cite{chen2016optimal,zhou2018artificial,sun2019robust,mao2019power}, the maximization of secrecy rate, for example,~\cite{7426798,chen2018exploiting,sun2018robust,8809566}, the maximization of energy efficiency (EE), for example,~\cite{7523951,7488207,fang2017joint,8648507}, minimizing transmission delay or latency, for example,~\cite{8488481,8611381,8666055,8486633}, and multi-objective optimization problems that involve the achievable rates and other performance metrics such as joint optimization of the spectral efficiency and energy efficiency and its trade-offs, for example,~\cite{8214974,8357581,8854318,8247249} in PD-NOMA  are also beyond the scope of this survey paper. 

\begin{figure*}[hp!]
\centering
\vspace{-4.5em}
\includegraphics[width=0.7\textwidth]{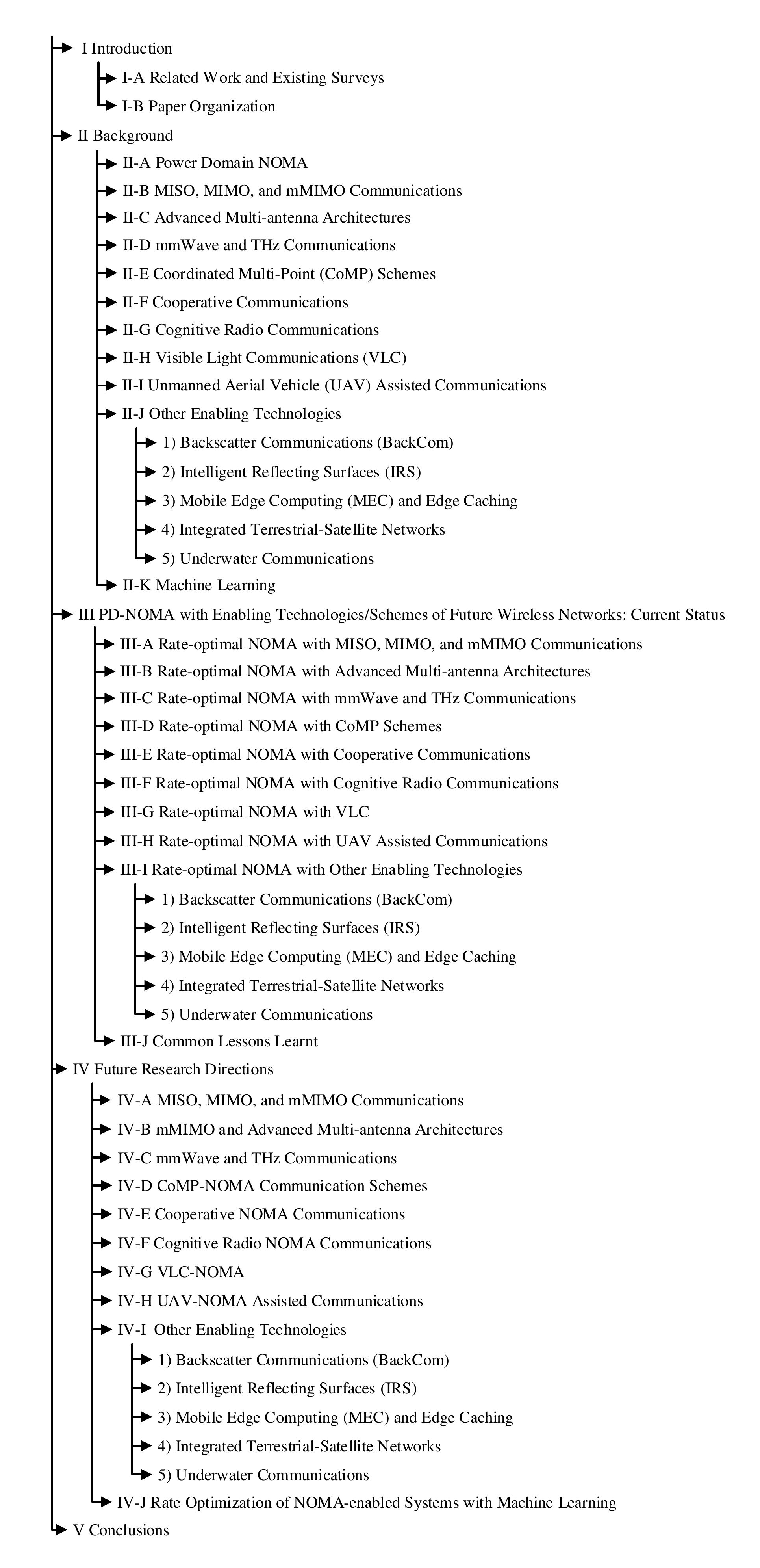}
\caption{The structure of this paper.}
\label{fig: structure}
\vspace{-6.5em}
\end{figure*}

So, the main contributions of this survey paper are:    
\begin{itemize}
    \item A thorough up-to-date survey of the integration of PD-NOMA with each of the candidate enabling technologies of future B5G networks where the adopted system models, the utilized optimization techniques, and the main results are fully investigated. The existing related surveys, at best, have partially discussed the rate optimal integration of some of these enabling technologies with PD-NOMA as explained before in detail in Table~\ref{table: NOMA surveys and magazine papers} and Table~\ref{Table: A deep look}. 
    \item The integration of rate-optimal PD-NOMA  with combinations of two or more of these technologies and/or schemes is also investigated in a similar manner. Up to the authors' knowledge, this has not been carried out in open literature before.
    \item A thorough up-to-date survey of the integration of PD-NOMA with advanced multi-antenna architectures (including cell-free mMIMO, reconfigurable antenna systems, large intelligent surfaces, 3-D MIMO), THz communications, intelligent reflecting surfaces (IRS), and underwater communications. Up to the authors' knowledge, this has not been carried out in open literature before.
    \item Both of the lessons learnt obtained from the up-to-date literature and the directions for future research on such integration of PD-NOMA with one or more of these technologies and/or schemes are presented along with highlights of the associated merits and/or trade-offs.  
    \item An elaborate investigation of the role of machine learning in these NOMA-enabled schemes is presented and the possible directions for future work on ML-assisted NOMA-enabled schemes are stated.    
    \end{itemize}

\subsection{Paper Organization}

In Section~\ref{Section: Background}, we first provide a brief description of PD-NOMA itself, followed by a description of the main enabling communications schemes and technologies for future wireless networks that are discussed in this paper. Section~\ref{Section: existing work} surveys the integration of PD-NOMA with these enabling technologies, namely, MISO, MIMO, mMIMO, advanced multi-antenna architectures (including CF-mMIMO, reconfigurable antenna systems, LIS, and 3-D MIMO), mmWave and THz, CoMP, cooperative communications, cognitive radio, VLC, UAV communications, BackCom, IRS, mobile edge computing (MEC) and edge caching, integrated terrestrial-satellite networks, and underwater communications. Also, it states the main learnt lessons on such integration. In Section~\ref{Section: Conclusions and Open Research Problems}, we provide a list of possible directions for future research for each of the aforementioned technologies and their combinations. The structure of this article is shown in Fig.~\ref{fig: structure} at a glance. Also, the list of abbreviations used in this survey paper is presented in Table~\ref{Table: Abbreviations}.

\section{Background} 
\label{Section: Background}
\begin{figure*}[htp!]
\centering
\includegraphics[width=1.0\textwidth]{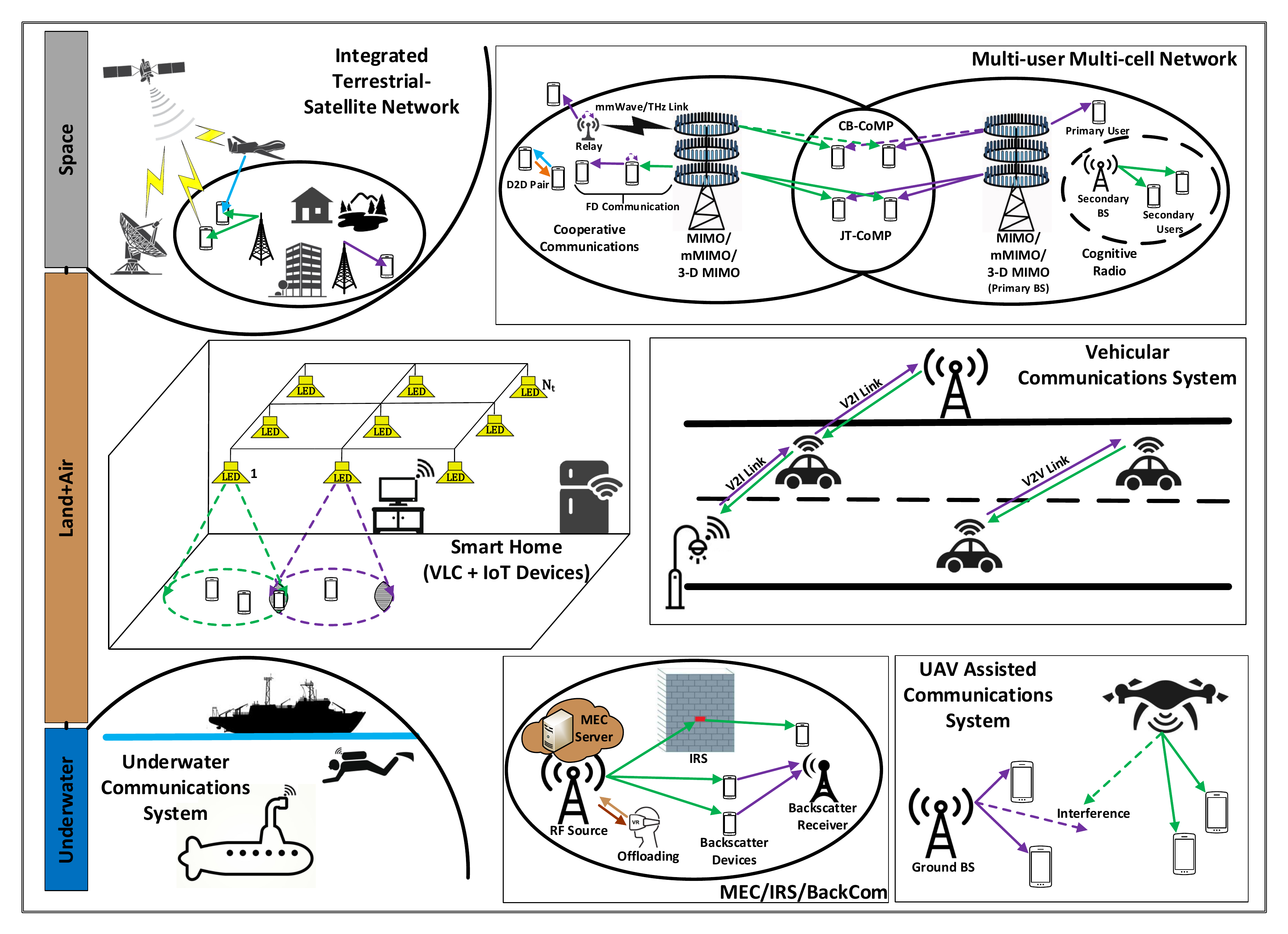}
\caption{The enabling schemes and technologies of future
wireless networks, including MIMO, mMIMO, advanced multi-antenna architectures, mmWave and THz, CoMP, cooperative communications, cognitive radio, VLC, UAV communications, BackCom, intelligent reflecting surfaces, mobile edge computing and edge caching, integrated terrestrial-satellite networks, and underwater communications.}
\label{fig: All technologies bkgd}
\end{figure*}

In this section, a brief introduction to PD-NOMA as well as to some of the main PHY-layer enabling schemes and technologies for future communications networks is presented. The objective of this section is to familiarize the reader with the fundamentals of each of these communications schemes or technologies. Hence, relevant references are provided in each sub-section where further details about each scheme or technology  can be found. A graphical representation of the main PHY-layer enabling schemes and technologies of future wireless networks is depicted in Fig.~\ref{fig: All technologies bkgd}.

\subsection{Power Domain NOMA} \label{sec:PD_NOMA}

Power-domain NOMA concept was first introduced in~\cite{saito2013non} to improve the spectral efficiency of wireless networks by allowing multiple users to simultaneously share both the time and frequency resources. The theoretical roots of PD-NOMA scheme lie in multi-user information theory. These include scalar and vector multiple access and broadcast channels along with the superposition coding, successive interference cancellation (SIC), joint decoding, iterative water-filling, dirty paper coding, and other related transmission and reception schemes~\cite{liu2017non}. However, NOMA schemes add additional "comm-theoretic" constraints on the users' target rates, in addition to the typical transmit power constraint in the info-theoretic models, to guarantee the fairness among the served users. 

\begin{figure*}[htp!]
\centering
\vspace{-1em}
\includegraphics[width=0.58\textwidth, height=7.2cm]{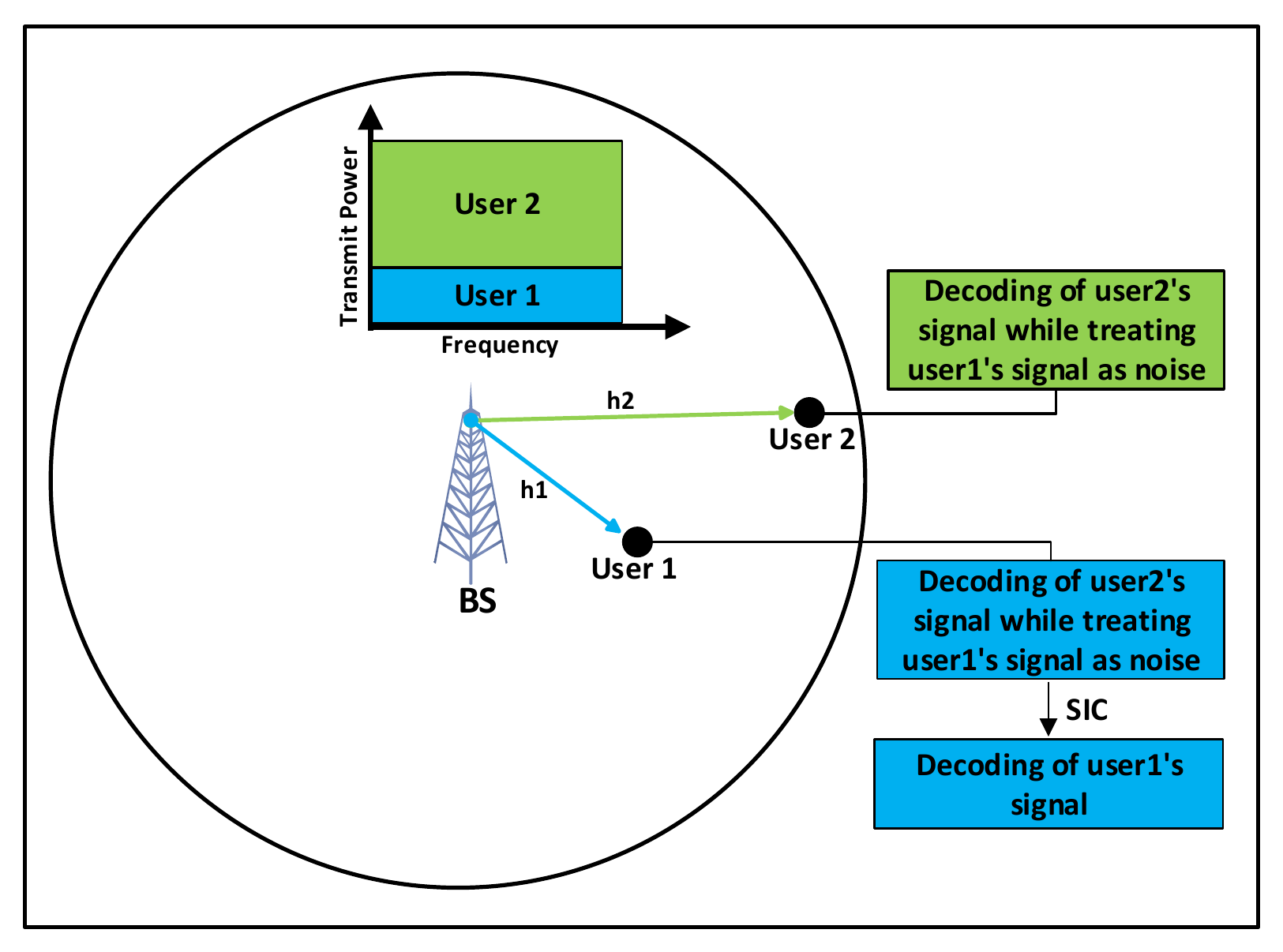}
\caption{An illustration of a two-user downlink power-domain NOMA scheme with superposition coding and successive interference cancellation decoding~\cite{al2018multi}.}
\label{fig: NOMA_2users (SISO)}
\vspace{-1em}
\end{figure*}

A basic two-user downlink NOMA scheme is shown in Fig.~\ref{fig: NOMA_2users (SISO)}. User 1 (strong user) is close to the BS with strong channel gain and User 2 (weak user) is farther away from the BS with weaker channel gain. At the transmitter, both signals for the weak and the strong users are superimposed upon each other with different power allocation. The transmitter tends to allocate more power to the weak user as it has a larger path loss, as compared to the strong user (note that the size of the rectangles for User 1 and User 2 are different indicating the different powers allocated for each user). At the strong user receiver, the signal of the weak user has a high signal-to-noise ratio (SNR) which implies that the strong user can successfully decode and subtract the weak user signal before decoding its own signal (i.e., performing SIC). On the other hand, at the weak user receiver, the strong user signal is considered as noise as its transmission power is lower than the weak user signal. Subsequently, the weak user can decode its signal directly without SIC~\cite{choi2017noma}. In a basic two-user uplink (UL) NOMA scheme, both users transmit their signals to the BS simultaneously (i.e., in the same time slot and frequency channel). For practical reasons~\cite{yang2016general}, both users need to have some sort of cooperation to transmit both signals under a total transmit power constraint. The power share of both users is distributed based on their channel conditions with respect to the BS. After the transmission process, the BS receives a superimposed signal that contains both users' signals. Subsequently, the BS performs SIC with a decoding order from the strongest user to the weakest user to decode both individual users' signals. Noting that user equipments' (UEs') interference cancellation capability is already incorporated in an earlier wireless standard, namely, network-assisted interference cancellation and suppression (NAICS) in LTE (3GPP release 12)~\cite{3Gpp-release-12-NAICS}.

The grouping or clustering of the users to be served in the same resource block is essential in PD-NOMA and is typically carried out as two users per cluster (also known as {\it user pairing}) or multiple users per cluster. The selection in basic NOMA scheme, with single-antenna BS/access point (AP) and single-antenna users, is usually based on the instantaneous scalar channel gains and the users are ranked accordingly to allow proper SIC decoding which tends to improve as the channel gain disparity increases unless the channel gain of the weaker user is very small which may render NOMA inefficient. In general, the optimal user clustering requires an exhaustive search and may not be affordable for practical systems and networks with a large number of users~\cite{ali2016_optUCandPA}. For this reason, researchers resort to low complexity solutions to solve the user clustering problem through heuristic algorithms which may lead to unpredictable results.

Alternatively, the other approaches for handling the NOMA clustering problem are~\cite{liu2017non}: 
\begin{itemize}
    \item The monotonic optimization approach as in~\cite{7812683}; the monotonic nature of the objective function and constraints allows researchers to limit the exhaustive search to a much smaller region of the feasible set; this accordingly simplifies the problem and can lead to the optimal solution but still with relatively high complexity.
    \item The combinatorial relaxation approach; the selection and association of a particular user to a particular cluster is realized by a binary variable; by relaxing this variable to a continuous one, the original non-convex problem can be transformed to a convex one and solved optimally through classic Lagrangian dual method~\cite{boyd2004convex}, but this relaxation results in a performance gap between the original problem and the relaxed one.
    \item The matching theory approach; which is often utilized to deal with problems that have beneficial relationship parties, such as the many-to-one bipartite matching problem~\cite{cui2017user}, the many-to-one two-sided matching problem~\cite{fan2019channel}, and the many-to-many two-sided matching problem~\cite{zhang2017sub,xu2018joint}.
    \item The game theory approaches; which transform the clustering problem to a game, such as the coalition game and improved coalition game proposed in~\cite{7994888,8809359,8767908} and in~\cite{ding2019improved}, respectively, and the Stackelberg game proposed in~\cite{8684765}.
\end{itemize}

Early work on PD-NOMA started with basic single-cell scenarios and then evolved to multi-cell and multi-carrier scenarios. Recently, PD-NOMA scheme has been integrated with more practical and sophisticated system models alongside other promising communications schemes and technologies to meet the target performance of future wireless networks.

\subsection{MISO, MIMO, and mMIMO Communications}
\label{sec:bkd Conventional MISO/MIMO Communication}

The use of multiple antennas at the transmitter, receiver, or both has helped significantly drive up data rates and been an integral part of cellular systems right from 4G~\cite{tse2005fundamentals}. When multiple antennas are available at the BS, but only a single antenna at the user, it is referred to as a multiple-input-single-output (MISO) system; while if multiple antennas are also available at the receiver, it is referred to as a multiple-input-multiple-output (MIMO) system. These multiple antennas at the transmitter and receiver are used to realize beamforming or spatial multiplexing gains in single-user scenarios. In a multi-user environment, the multiple antennas can be used to separate the users in the space domain, creating the so-called spatial division multiple access~\cite{tse2005fundamentals}. The use of large-scale antenna arrays at the BS, where the number of transmit antennas far exceeds the number of users in the system, is referred to as a massive-MIMO (mMIMO) system. In such mMIMO systems, when the number of transmit antennas approach infinity, it creates favorable propagation conditions whereby a unique beam can be formed for each user and perfect separation in the space domain is possible~\cite{zhang2018favourableprop}. When a very large number of antennas are used for MIMO, it is sometimes referred to as ultra mMIMO.

\subsection{Advanced Multi-antenna Architectures}  \label{sec: bkd advMimoNoma}

The high spectral efficiency demands in B5G networks (100 b/s/Hz or more) require advanced multi-antenna configurations beyond the conventional MIMO and mMIMO in 4G and 5G networks, respectively. Some of these advanced multi-antenna architectural schemes are outlined in the subsequent sub-sections.

\subsubsection{Cell-free mMIMO (CF-mMIMO)}
\label{sec:bkd Cell-free mMIMO}
With B5G required to support different use-cases, large scale antenna arrays can be used for other engineering challenges such as providing cost and energy-efficient solutions. One such low-cost solution that is an extension of the current mMIMO, is the concept of cell-free mMIMO (CF-mMIMO)~\cite{zhangAccess2019CellFree}. With CF-mMIMO, a subset of UEs are served by a wide geographic distribution of a large number of individually controllable antenna elements. The key idea with CF-mMIMO is to use all the antennas available in the network to serve a subset of UEs, without the typical constraints of having a defined cell coverage area. This allows the served UEs to access a much larger pool of antennas, moving closer to theoretical mMIMO of having infinite antennas at the BS.

\subsubsection{Reconfigurable Antenna Systems}
\label{sec:bkd Reconfigurable Antenna Systems}
Reconfigurable antennas offer the opportunity to dynamically change the radiation patterns by altering the physical configuration of the antenna~\cite{6474484}. In other words, a reconfigurable antenna modifies the antenna\textquotesingle s pattern, polarization, or other physical attributes that help achieve the desired antenna beam-steering. This provides an extra degree of freedom to the beamforming problem and has been successfully applied to both sub-6 GHz~\cite{6474484} as well as mmWave and above communications~\cite{7917231}.

\subsubsection{Large Intelligent Surfaces}
\label{sec:bkd Large Intelligent Surfaces}

Contrasted with CF-mMIMO, another emerging extension of mMIMO that is on the other end of the cost spectrum, is the large intelligent surfaces (LIS)~\cite{Hu2018activeLIS}. An LIS is like a large contiguous surface of antennas, capable of generating power to send the desired electromagnetic signals. It can be thought of as another way of realizing theoretical mMIMO with infinite antennas, but unlike CF-mMIMO, LIS is not constrained to support a smaller number of UEs. On the contrary, LIS can support an even larger number of connected devices, ideal for the massive connectivity requirements for B5G networks in indoor scenarios.

\subsubsection{3-D MIMO}
\label{sec:bkd 3-D MIMO} 
In 5G NR systems, 3-D separation of users is possible in both the azimuth and vertical direction, sometimes called 3-D MIMO~\cite{zhang20183dmimo}. In~\cite{bjornson2019massive}, many future directions for mMIMO systems are outlined for B5G networks, including extremely large aperture arrays, 3-D MIMO, holographic mMIMO, 6-D positioning, large-scale MIMO radar and smart mMIMO integrated with artificial intelligence.

\subsection{mmWave and THz Communications}
\label{sec:bkd mmWave Communications and THz communications}

In contrast to the sub-6 GHz frequency range, a large amount of bandwidth is available in the mmWave frequency bands and beyond. Hence, they  are seen as a key enabler of high data rates in 5G and beyond cellular networks. However, unlike in the sub-6 GHz spectrum, the propagation of electromagnetic waves in the mmWave bands exhibits high path loss and is highly directional in nature~\cite{fan2019channel}. As a result, the beam gain offered by large-scale antenna arrays, i.e., mMIMO systems, is typically seen as a requirement to enable successful deployment of networks in these mmWave-frequency bands~\cite{han2015MagmmWave}. Additionally, the short wavelengths of the electromagnetic waves in the mmWave frequency band make it practical to deploy large antenna arrays in a compact area, as required by a mMIMO system~\cite{xiao2019user}. 

While mmWave communications in the 30 GHz spectrum is already part of the 5G standard, the academic literature is growing in even higher frequency bands such as mmWave for 100-300 GHz and THz bands~\cite{han2019terahertz}. The sheer amount of bandwidth available in this region of the frequency spectrum makes it very appealing to meet the large data requirements of 100 Gbps and more for B5G networks, but it comes at the cost of very high propagation loss. The study by Rappaport~\textit{et al.} in~\cite{8732419} highlights several promising applications for communications above 100 GHz in 6G networks, including wireless fiber for back-haul and information showers where a blast of data is sent to the user in a short amount of time as the user passes through an area of coverage at these high frequencies. The same study in~\cite{8732419} also highlights the additional challenges for wave propagation at these frequencies, particularly that the free-space path loss and penetration loss increases significantly as we go to higher frequencies, leading to shorter coverage areas. However, at these frequency bands, the wavelength and consequently the size of the antennas are very small, allowing for the use of highly directional antennas for ultra-precise beamforming, multi-reflector antennas, lens-integrated antenna arrays and other such innovative solutions~\cite{han2019terahertz}. Hence, B5G networks will likely employ ultra mMIMO systems to solve the propagation challenges at these bands.

\subsection{Coordinated Multi-Point (CoMP) Schemes} 
\label{Subsection: background: CoMP}

 The practical considerations in interference-limited heterogeneous multi-cellular environments as well as the promising info-theoretic benchmarks, where a network with fully cooperating cells reduces to a simple broadcast channel, have led to the introduction of network cooperation schemes known as network MIMO or CoMP schemes. This is to allow the base stations or the access points to cooperate in order to mitigate or reduce the inter-cell interference (ICI) and to enhance the received signal quality~\cite{comp-1, comp-2}. The degree of cooperation may vary depending on the availability of users' data and/or the channel state information (CSI) as itemized below: 
 
 \begin{itemize}
  \item  Both the users' data and CSI are available at all the transmission and reception points (TRPs). This availability enables full joint coordination or processing. Here, the cooperating BSs or APs (also known as transmission points (TPs) or TRPs or CoMP-cells) jointly transmit to the served users (typically the cell-edge users) or TP selection where only one or a subset of the cooperating TPs, usually the TP with the highest received signal-to-noise and interference ratio (SINR), sends to the desired user(s). 
  
  \item  Only the CSI is available at all the TRPs. This relaxed and less-demanding requirement of having only the CSI at the cooperating TPs allows reducing ICI through coordinated scheduling of the users and joint beamforming.  
\end{itemize}
   
 As can be seen, the TP selection and coordinated beamforming schemes are less complex. This is primarily because of its less overhead requirement, less stringent requirement on back-haul performance, and less synchronization requirements, as compared to the full joint coordination scheme. Moreover, different requirements or priorities by the operators would lead to different CoMP implementations such as cell-centric, user-centric, and hybrid clustering algorithms~\cite{comp-2}.  

Early deployment of CoMP was in 3GPP LTE-A standard where all the CoMP variants above were investigated and evaluated~\cite{rcomp3gpp}. In B5G networks, the ultra-dense deployment of small cells or small BSs in heterogeneous cellular networks would result in more of both co-tier and cross-tier inter-cell interference and hand-off requests which necessitates the use of advanced AP/BS CoMP schemes to efficiently combat the inter-cell interference and/or to enable "cell-less" networks to improve network performance, especially for cell-edge users.          

\subsection{Cooperative Communications}

Cooperative communications systems allow some communication terminals in the network to interact with each other for the sake of improving reliability, enhancing spectrum efficiency as well as power efficiency, and improving network connectivity by taking advantage of the broadcast nature of the wireless links~\cite{nosratinia2004cooperative}. B5G networks are expected to comprise of billions of heterogeneous nodes, each with different processing capabilities, transmit power, etc. As a result, a network where the nodes co-operate with each other for common goals offers the potential for great enhancements in overall spectral efficiency. In what follows, we discuss different realizations for cooperative communications schemes. 

There are three possible ways to realize the cooperation among the nodes in the network; (i) by allowing the communication terminals to hear-and-forward messages to other terminals, (ii) by using a dedicated relaying node to help in information and/or energy transfer between the source(s) and their corresponding destination(s), and (iii) by allowing some user pair(s)/group(s) to interchange signals without traversing the BS or the core network. The first type of cooperation is often referred to as user-assisted communications, the second one is often referred to as relay-assisted communications, and the third type is often referred to as device-to-device (D2D) communications~\cite{li2012cooperative,6805125}. 

D2D communications can be considered as an enabler for vehicle-to-everything (V2X) applications. In 3GPP Release 14, an initial cellular V2X standard has been completed as a part of expanding the LTE platform to new services~\cite{3Gpp-V2X-release-14}. Later in 3GPP Release 15, in V2X phase 2, some service requirements to enhance the support for V2X scenarios have been identified~\cite{3Gpp-V2X-release-15}. In 3GPP Release 16, in NR-V2X, further architecture enhancements to the 5G system have been specified in order to facilitate vehicular communications for V2X services~\cite{3Gpp-V2X-release-16}. In 3GPP Release 17, some enhancements to the application architecture to support V2X services have been specified~\cite{3Gpp-V2X-release-17}. In V2X applications, vehicles can exchange data with each other and with the infrastructure/network units. The former is often referred to as vehicle-to-vehicle (V2V) communications and the latter is often referred to as vehicle-to-infrastructure (V2I) communications. 

In addition to the above, there are two types of relaying strategies, namely, half-duplex and full-duplex. In the half-duplex strategy, the communication between the source and the destination takes place in two phases, where in the first phase, the BSs or APs broadcast the signals to the nearby receivers (i.e., users/relays). Consequently, in the second stage, the nearby receivers forward the messages of far receivers. On the other hand, in the full-duplex relaying strategy, the relays/users simultaneously receive and forward the signals to their intended destinations. Although the full-duplex strategy introduces some self-interference (SI) caused by the signal leakage from the transceiver output to the input~\cite{riihonen2011mitigation}, which degrades the performance of the full-duplex system, still full-duplex technology has the capability of utilizing the total bandwidth of the system as it hears-and-forwards the signals on the same frequency. Recent advancements in SI cancellation techniques and the move towards short-range networks such as small-cell cellular networks, which have lower path loss than  the traditional cellular networks, have reduced the severity of SI and paved the way for the full-duplex strategy to be an enabling technique for future wireless networks~\cite{liu2015band}.

Moreover, there are three main relaying protocols: (i) decode-and-forward (DF) relaying, (ii) amplify-and-forward (AF) relaying, and (iii) compress-and-forward (CF) relaying. In DF relaying protocol, the received signal is completely decoded and then re-encoded and sent to the intended receiver(s). In AF relaying protocol, the received signal with its embedded noise is amplified and forwarded to the designated receiver. While in CF relaying protocol, the relay quantizes/compresses the overheard signal and then sends the resultant signal to the designated receiver~\cite{liaqat2018power}.

\subsection{Cognitive Radio Communications}

Most of the radio spectrum has been allocated to certain wireless applications because of its increasing demand, and the need for more bandwidth for new applications. The Federal Communications Commission (FCC) reported that the utilization of this allocated spectrum varies from 15\% to 85\% with high variance in time and space~\cite{fcc2003docket}. To address the low spectrum utilization at different times and locations, the cognitive radio (CR) idea appeared as a proposed solution~\cite{mitola1999cognitive}. The main idea of CR is to enhance the utilization of the pre-allocated spectrum by allowing the non-licensed users, called secondary users (SUs), to either coexist with primary users (PUs) or exploit the white spaces of the spectrum in the absence of PUs~\cite{biglieri2013principles}. B5G networks are expected to support a wide variety of users and use-cases whose demands fluctuate rapidly in time and space. Consequently, dynamic spectrum access solutions like CR can play a key role to meet these demands by increasing network spectral efficiency. Applications of CR in B5G networks include the opportunistic use of an unlicensed spectrum as well as the use of a licensed spectrum for a vast number of heterogeneous users. The CR network needs to support such applications through a priority-based PU-SU system.

CR schemes can be classified into three paradigms, namely, underlay, overlay, and interweave~\cite{goldsmith2009breaking}. In the underlay CR paradigm, the SUs are allowed to access the PUs networks as long as the interference power constraint of the PU is not violated~\cite{zhao2007survey}. While in the overlay CR paradigm, the PUs and SUs are capable of transmitting signals simultaneously. In this paradigm, the PUs' SNR will be affected by the interference caused by the SUs' transmission. In order to compensate for that decrease in the PUs' SNR, the SUs are forced to divide their transmission power between (i) secondary transmissions and (ii) the remaining power is used to relay (assist) primary transmissions~\cite{4197718}. Furthermore, in the interweave CR paradigm, the SUs are allowed to utilize the PUs' spectrum only if the spectrum is not occupied by the PUs. Note that, the interweave CR paradigm does not authorize any concurrent transmission between SUs and PUs, hence it is known to be an interference-limited spectrum sharing mode~\cite{8703419}.

\subsection{Visible Light Communications (VLC)}

Visible light communications (VLC) has recently emerged as a potential technology for complementing and/or off-loading radio frequency (RF) communications systems for various indoor user-dense scenarios such as homes, office rooms, conference and exhibition halls, airplanes and train cabins as well as some outdoor and V2X applications in 5G and 6G networks. VLC is based on the principle of modulating light emitted by  diodes (LEDs), without any adverse effects on the human eye  and the required illumination levels, which gives an opportunity to exploit the existing illumination infrastructure for wireless communication purposes. VLC technology is expected to play an important role in enabling very high data-rate short-range communications as well as V2V and in general V2X applications in B5G networks. 

A generic model for a multi-user VLC system is shown in  Fig.~\ref{fig: All technologies bkgd} where a set of transmitting $N_t$ LEDs communicate with a set of users. This is a typical indoor downlink scenario where the LEDs on the ceiling transmit the downlink data to the VLC receivers that are typically equipped with photo-detectors~\cite{pratama2018bandwidth}. The field of view (FoV) of the LEDs and the photo-diodes are designed to satisfy both the illumination and communication requirements. Each shaded area represents the interference or overlap region imposed by the signals carrying different information and arriving from the neighboring LEDs. While there is negligible signal fading experienced in VLC channels, the SINR significantly drops at the boundary of interference region and by obstruction of the line-of-sight (LoS) path due to user mobility, especially for outdoor applications. 

\subsection{Unmanned Aerial Vehicle (UAV) Assisted Communications} \label{sec:UAV_bkgd}

UAV-assisted communication networks is an emerging paradigm in the B5G wireless networks to support high transmission rates, to provide ubiquitous connectivity to a diverse set of end-users, and to facilitate wireless broadcast~\cite{irem2019drones, andreevHalim2019UAVmag, li2018uav}. UAV can play several roles in this paradigm, e.g., act as a flying BS, a relay in flying ad-hoc networks, or even as user equipment (UE)~\cite{mozaffari2019tutorial}. The UAVs can be deployed as a high altitude platform (HAP) for long-term use or as a low altitude platform (LAP) for fast and flexible deployment when used as a BS~\cite{caoHalim2018UavSurvey, mozaffari2019tutorial}.

We now focus on the use-case where the UAV acts as a LAP BS to serve users in an ad-hoc manner. This is illustrated in Fig.~\ref{fig: All technologies bkgd}, where the UAV complements the ground BS and serves some users in conjunction with the ground BS. The main idea in this type of deployment is to bring the supply of wireless networks to where the demand is in both space and time, acting as a complement to the terrestrial communications network~\cite{irem2016UAVmag}. As illustrated in Fig.~\ref{fig: All technologies bkgd}, like in any multi-cell network, the UAV will cause interference to the users served by the ground BS and vice-versa. However, the problem can be addressed with additional degrees of freedom that a flying BS allows which does not exist in traditional terrestrial BSs. These include designing the height and placement of the UAV BS and the high probability of LoS links between the UAV and the served users~\cite{li2018uav}.

\subsection{Other Enabling Technologies} \label{sec:Other_bkd}

\subsubsection{Backscatter Communications (BackCom)} \label{sec:backscatter}

BackCom enables low-power transmission of information by backscatter devices (BD) through the modulation of information over incident RF carriers, without the BDs having to generate any power by themselves.  An illustration is shown as part of Fig.~\ref{fig: All technologies bkgd}, where the RF source is the power-consuming device that generates carrier RF waves. The low-cost BDs simply modulate their information over these incident carriers, which is decoded by the backscatter receiver (BR). Hence, it is a technology that primarily targets the internet of things (IoT) deployment, a key use-case in B5G networks, where the BD could be low-cost sensors. 

\subsubsection{Intelligent Reflecting Surfaces (IRS)}\label{sec:irs_bkgd}

Another emerging technology to provide low-cost massive connectivity solutions in B5G networks, particularly in indoor environments, is the intelligent reflecting surfaces illustrated as part of Fig.~\ref{fig: All technologies bkgd}. IRS solutions typically consist of a planar array of passive phase shifters that can collaborate to steer the propagation path of the transmitted signals in the wanted direction~\cite{8811733}. In this way, the IRS acts as a smart reflecting surface to help steer the transmitted signal towards the user. Since the IRS only reflects signals and does not require antennas that generate any power for transmission, it is a low-cost solution to enhance spectral efficiency and support massive connectivity. For this reason that the IRS devices are merely passive reflectors, they are different from the active LIS solution discussed as part of the advanced multi-antenna architectures. While the LIS  can be viewed as an extension to mMIMO as it involves a surface full of active transmitting antennas~\cite{Hu2018activeLIS}, the IRS is a different technology that does not use antennas at the surface, and hence we do not consider IRS as part of the advanced multi-antenna architectures. IRS also has similarities with backscatter communications, but unlike backscatter devices which are the sources of information, IRS devices are merely helpers that reflect the transmitted wave in the intended direction.

\subsubsection{Mobile Edge Computing (MEC) and Edge Caching} \label{Sec: MEC bkgd}

Mobile edge computing can offer remote computation for low computing-capability IoT devices that may need  intensive computing-capability in some latency-sensitive applications like interactive networked gaming, real-time control, and virtual reality. In the MEC network, the BS usually equipped with a MEC server. The MEC server can receive and execute some offloaded tasks from the IoT devices. Later, the results are sent to the IoT devices after execution. There are two offloading scenarios in  the MEC network, namely, partial and binary offloading. In the former scenario, a task can be divided into two parts: a local computing part and an offloading part. While in the latter scenario, a task can be either locally computed by the IoT device or remotely accomplished by the MEC server, i.e., cannot be divided~\cite{7879258}.

Local content servers can be placed alongside the MEC servers to enables edge caching. In edge caching, some of the most popular content files are pre-stored in local content servers at the BS that is in close physical proximity to the UEs. The UEs within the coverage of that BS can have fast access to these files without burdening the back-haul links of the network. Edge caching has mainly two phases, namely, content pushing and content delivery. In the content pushing phase, the BS fetch the most popular content files from the back-haul and push these files to the local content servers. While in the content delivery phase, the network users access these pre-stored files directly. Subsequently, edge caching provides a capacity enhancement for 5G networks~\cite{wong2017key}.

\subsubsection{Integrated Terrestrial-Satellite Networks} 
In this framework, satellites are used in conjunction with the typical terrestrial ground BS-based network to provide enhanced coverage and rates to the network users. Satellite communications is viewed as one of the key enablers to achieve the goal of truly ubiquitous coverage in B5G networks~\cite{8088533}. The satellite and terrestrial networks can co-operate with each other similar to a CoMP multi-cell setting~\cite{zhu2017nonSat}. Alternatively, the terrestrial BS can act as an AF relay to signals transmitted by the satellite network~\cite{7230282}. Also, similar to the case of UAV-enabled systems, satellites could also be users in the system, e.g., satellites dynamically access the spectrum through a CR approach~\cite{7336495}.

\begin{figure*}[ht!]
\centering
\includegraphics[width=1.0\textwidth]{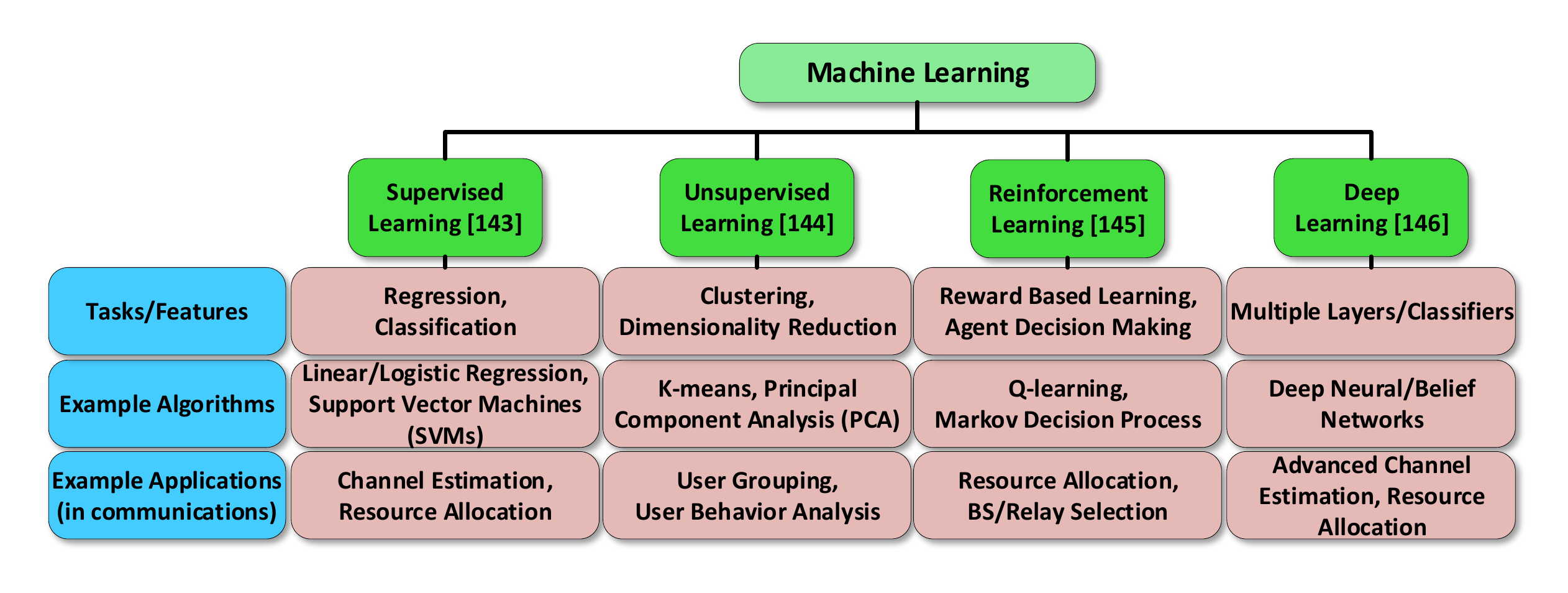}
\caption{The different categories of machine learning algorithms with the main tasks/features, some example algorithms, and some example applications in wireless communications~\cite{zhang2019thirty}.~\nocite{suthaharan2016supervised,hastie2009unsupervised,kaelbling1996reinforcement,schmidhuber2015deep}}
\label{fig:ML_tree}
\end{figure*}

\subsubsection{Underwater Communications}

Underwater wireless communications play an important role in oil and gas exploration, water-pollution monitoring, and marine-life observation through the use of IoT devices like sensors. With B5G networks expected to provide ubiquitous coverage to support such sensors, underwater wireless communications is an emerging technology for B5G networks. At present, the three common transmission technologies for underwater wireless communications are acoustic, wireless optical, RF communications. Each of the aforementioned transmission technologies has its own characteristics, challenges, and adopted set of applications~\cite{Ali2019}. Generally, underwater acoustic transmission prevails the underwater long-range (several kilometers) wireless communications but suffers from low data rates due to acoustic signals' limited bandwidth and the highly variable delay due to acoustic signals' low propagation speed. Underwater optical wireless communications has a medium communication range ($50-100\mbox{m}$) and can achieve high data rates (on the order of gigabits per second) with low transmission latency due to the high propagation speed of light in the aqua~($\approx 2.55 \times 10^8 \ \mbox{m/s}$). Underwater RF communications is suitable for short-distance propagation ($10 \mbox{m}$) and immune to harsh conditions and noises, particularly, in shallow water at the expanse of bulky RF modules with large antennas that are costly and power-hungry~\cite{9071997}. It is worthy to note that, in 2017, the North Atlantic Treaty Organization (NATO) has issued the first international standard for communication between underwater nodes using acoustic signals that can achieve a transmission range up to 10 Km at 11.5 kHz~\cite{8012224}.

\begin{figure*}[ht!]
\centering
\vspace{-1em}
\includegraphics[width=1.0\textwidth]{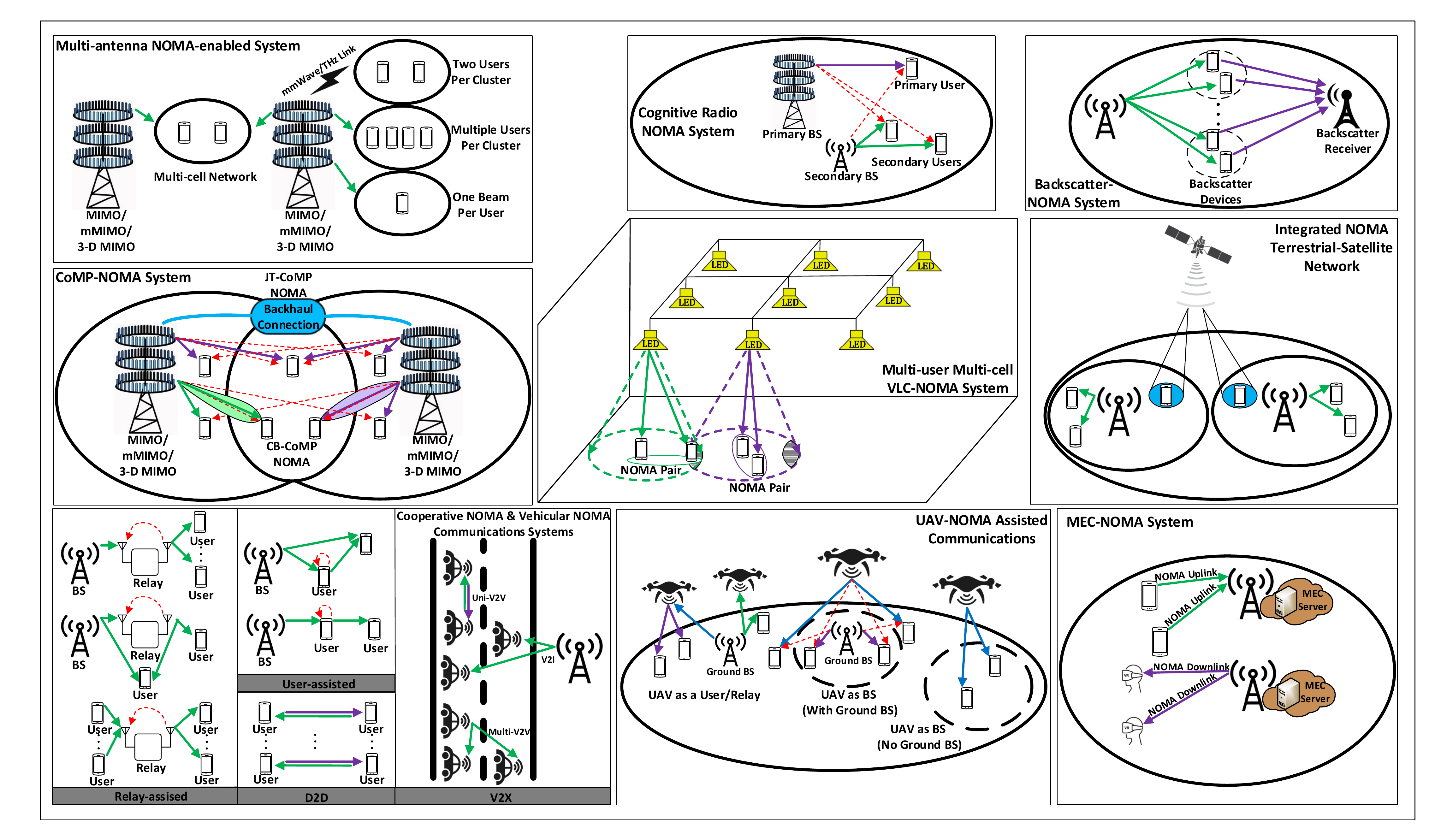}
\caption{NOMA-enabled schemes and technologies for future wireless networks as outlined before in the caption of Fig.~\ref{fig: All technologies bkgd}.}
\label{fig: All technologies}
\vspace{-1em}
\end{figure*}

\subsection {Machine Learning}\label{sec:Bkgd_ML}

Another developing trend for next-generation wireless networks, such as B5G cellular networks, is the advent of a data-driven approach using machine learning (ML) to complement or even replace the traditional model-driven approach~\cite{zappone2019WirelessML, jiang2017ML} used in communications systems. Relevant to the discussion in this paper, applying ML algorithms to resource optimization problems has certain advantages over traditional optimization techniques. Primarily, optimization approaches suffer from high cost and complexity when the number of parameters to be configured becomes large. Optimization algorithms are often sensitive to the parameter selection and heuristics have to be re-run from scratch every time there is a small change in the system model, e.g., the arrival of new users. In other words, the entire algorithm has to be run every time there is a small change in any system parameter. All these limitations of the traditional optimization tools have motivated researchers to explore the use of ML techniques for resource optimization in communications systems~\cite{zhang2019thirty}.  Additionally, when multi-objective optimization problems are framed, the goal is to find Pareto-optimal solutions, i.e., a solution space where the improvement of one metric necessarily degrades some other metric. Due to the large search space involved, ML is an attractive solution for finding such Pareto-optimal solutions~\cite{zhang2019thirty}.

ML algorithms can be broadly classified into three main categories - supervised, unsupervised and reinforcement learning, as illustrated in Fig.~\ref{fig:ML_tree}. Supervised learning algorithms use labeled training data for tasks like regression and classification. Unsupervised ML algorithms do not use training data and can be used for tasks like clustering and dimensionality reduction. Reinforcement learning algorithms refer to the set of ML algorithms that have an agent that learns an optimal set of actions by interacting with the environment. Recently, a more powerful form of learning algorithms called deep learning (DL) has emerged. DL algorithms aim to mimic the neurons in the brain, by forming artificial neural networks that comprise multiple hidden layers between the input and output. Unlike classic supervised ML algorithms, DL is more powerful as it can first extract a set of features from the data and then use that for classification or prediction. Such neural networks can also be used as an agent in an RL system, forming a Deep Reinforcement Learning algorithm (DRL). DRL algorithms can be used to solve complex non-convex network optimization problems, particularly in modern communications systems where ad-hoc, autonomous decisions need to be taken~\cite{luong2019ApplicationsML}. As illustrated in Fig.~\ref{fig:ML_tree}, all these ML algorithms have been used to solve different problems in communications systems. For example, resource allocation is one such problem often tackled by all classes of these ML algorithms, each tackling the problem from a different angle. A detailed description of this and its potential applications to NOMA communications systems are provided in Section~\ref{sec: future ML_NOMA} of the future research directions.

\section{PD-NOMA with Enabling Technologies/Schemes of Future Wireless Networks: Current Status}
\label{Section: existing work}

In this part of the paper, a survey on the integration of PD-NOMA with MISO, MIMO, mMIMO, advanced multi-antenna architectures (cell-free mMIMO, and reconfigurable antenna systems), mmWave, THz, CoMP, cooperative, cognitive radio, VLC, UAV communications, and others is conducted through the up-to-date lists provided in Table~\ref{tab:misoMimoNoma} to Table~\ref{tab:otherNoma}, of the representative work for each of the aforementioned schemes and technologies. Besides, an illustration of the commonly utilized system models in NOMA-enabled schemes and technologies of future wireless networks is depicted in Fig.~\ref{fig: All technologies}.

\begin{table*}[!htbp]
\centering
\vspace{-2em} 
\caption{MISO-NOMA and MIMO-NOMA communications systems: current status of rate optimization schemes.} 
\label{tab:misoMimoNoma} 
\resizebox{\textwidth}{!}{%
}%
\end{table*}

\begin{table*}[!htbp]
\centering
\vspace{-2.5em} 
\caption{mMIMO-NOMA, advanced  multi-antenna  architectures (cell-free mMIMO-NOMA, and reconfigurable antenna NOMA-enabled scheme), mmWave-NOMA and THz-NOMA systems: current status of rate optimization schemes.}
\vspace{-0.7em}
\label{tab:massiveMimoNoma1}
\resizebox{\textwidth}{!}{%
%
\\
\hline 

\end{tabular}%
}
\end{table*}

\subsection{Rate-optimal NOMA with MISO, MIMO, and mMIMO Communications} \label{sec:listBasicMimoNoma}

In this section, we first present the representative work for conventional multi-antenna NOMA systems, sometimes referred to as MIMO-NOMA systems in Table~\ref{tab:misoMimoNoma} as it forms the base for rate optimization works for all other multi-antenna NOMA-enabled technologies. The representative work on MISO and MIMO integrated NOMA systems are presented in Table~\ref{tab:misoMimoNoma} for both the single-cell and multi-cell scenarios. Compared to the basic single-input-single-output NOMA (SISO-NOMA) system described in Section~\ref{sec:PD_NOMA}, the use of multiple antennas at either the transmitter, receiver, or both means that the channel is now described by a vector or a matrix. Hence, unlike in a SISO-NOMA system, the ordering of users according to their channel conditions is no longer trivial. Hence, the MIMO-NOMA system is usually broken down into a SISO-NOMA form that is easier to work with~\cite{ding2017survey}. This is typically achieved by using the multiple antennas at the transmitter for beamforming, such that users can be grouped into beams, often referred to as clusters. Within each cluster, the problem breaks down into a typical SISO-NOMA setting and the traditional PD-NOMA scheme as described in Section~\ref{sec:PD_NOMA} can be used. This approach is classified as NOMA-BF in~\cite{ding2017application} and cluster-based MIMO-NOMA in~\cite{liu2017non}. Alternatively, the multiple antennas at the base station can also be used to form one beam per user like in MU-MIMO, and then the beamforming weights are designed to create enough difference between the users' channel conditions such that the NOMA principles can be applied~\cite{hanif2016minorization}.

With the clustering-based approach, the two users should have sufficient difference in the magnitude of the channel gain coefficients for the PD-NOMA scheme to work well~\cite{islam2017power}. Thus, user selection (US) is a key parameter in such schemes. The objective here is to pair users who have sufficiently different channel conditions and  also fit within a cluster, i.e., a beam. This design variable (or parameter) is often optimized in such schemes along with the power allocation (PA) coefficients assigned to each user in the cluster, e.g.,~\cite{kimy2013milcom, islam2018resource}. However, with randomly located users, good user pairing algorithms are not guaranteed to find the ideal user clustering even with an exhaustive search of all possible solutions~\cite{ding2016signalalignment} because acceptable solutions may not exist. Hence, user-specific beamforming weights are also considered as a parameter in these sum-rate optimization problems~\cite{seo2018beam, zhang2016robust, cai2017low, ali2017non, ding2016signalalignment}. The addition of the optimization of the per-user beamforming (BF) weights to the clustering-based MIMO-NOMA rate optimization problem allows the schemes to either create more channel gain difference among the users in a cluster, or to better separate the different clusters from each other.

With the one-beam-per-user model, the user ordering is predefined and the optimization schemes design a set of precoder weights to meet the given user ordering~\cite{hanif2016minorization, zhu2018beamforming}. The sum rate of the users is maximized through a minorization maximization algorithm (MMA) for the non-convex problem proposed by Hanif~\textit{et al.}~\cite{hanif2016minorization}; while in the technique suggested by Zhu~\textit{et al.}~\cite{zhu2018beamforming}, it is shown that forming a weighted sum-rate maximization problem instead has hidden convexity that can be solved using the KKT optimality conditions. While this one-beam-per-user approach generally requires full channel state information at the transmitter (CSIT) for the required beamforming design, the study in~\cite{tian2018robust} framed an optimization problem to maximize the worst-case achievable rate in order to offer robustness to channel uncertainties.

The availability of CSIT also affects the formulation of the MIMO-NOMA rate optimization problems considered in the literature. If the full CSIT is available, the design objective is to maximize one of the overall sum-rate~\cite{kimy2013milcom, ali2017non, hanif2016minorization}, weighted sum-rate~\cite{yalcin2019downlink, choi2017performance}, or, the worst user-rate~\cite{seo2018beam, liu2016fairness}. On the other hand, if only statistical CSIT is available, then the design objective is to maximize the ergodic sum-rate~\cite{sun2018precoder, sun2015ergodic} rather than the actual sum-rate. While obtaining full CSIT helps maximize the system throughput, it comes with additional cost and overhead, especially in frequency division duplex (FDD) systems where reciprocity cannot be used. This motivates the need for system design assuming only limited feedback or partial CSIT~\cite{ding2016limitedfeedback}. In the work by Yang~\textit{et al.}~\cite{yang2017noma}, the number of feedback bits for CSIT is optimized and it is demonstrated that the system throughput (i.e., sum rate) can be increased as the number of feedback bits for CSIT is increased.

Further, the representative works on mMIMO integrated NOMA systems are presented in Table~\ref{tab:massiveMimoNoma1}. While the favorable propagation conditions in mMIMO systems say that an individual beam can be formed per user when the number of antennas at the BS is infinite, users with highly correlated user channels are hard to separate through SDMA with a finite number of antennas. However, as with the cluster-based MIMO-NOMA systems, such users are ideal to be grouped in a cluster if there is sufficient difference in their large-scale fading coefficients~\cite{senel2019_mMimovsNOMA_bjornson}. This leads to similar US and PA optimization algorithms as the cluster-based MIMO-NOMA systems. As such, sum-rate optimization problems in such a mMIMO-NOMA setting have been studied in single-cell~\cite{chen2018fully, chen2018upas} and multi-cell~\cite{kudathanthirige2019_earlyaccess} frameworks using this clustering-based approach.

\begin{itemize}[label=$\blacksquare$]
\item Lessons Learnt 
    \begin{itemize}[label=$\bullet$]
    \item The literature of NOMA integrated with conventional MIMO forms the base for integrating NOMA with almost every other technology surveyed in this paper, as multiple antennas are typically involved at either end of the communication system. The typical approach in most of the literature is to form effective clusters of users after considering the effective channel gains that account for the beamforming weight multiplication. Typically, once the clusters are formed and user ordering established, optimal power allocation coefficients are derived to maximize the sum rate of the users in the cluster.
    \item In mMIMO systems that serve multiple users, cluster formation for NOMA and SDMA can be seen as complementary approaches to jointly serve multiple users and improve the overall spectral efficiency. Users with channels that are highly correlated are easy to separate through NOMA but hard to separate through SDMA and vice-versa. This property of SDMA and NOMA to favor opposing characteristics in the channel conditions, make them well suited to be used in conjunction with each other.
    \end{itemize}
\end{itemize}

\subsection{Rate-optimal NOMA with Advanced Multi-antenna Architectures}
\label{sec:Survey advanced Mimo Noma}

NOMA has recently been integrated with some of the other advanced multi-antenna architectures, presented in Section~\ref{sec: bkd advMimoNoma} and the representative work in this area is captured in Table~\ref{tab:massiveMimoNoma1}. Only those technologies from the list in Section~\ref{sec: bkd advMimoNoma} that have representative NOMA-integrated rate optimization works are included here, while the integration of NOMA with other advanced multi-architecture schemes, namely, LIS and 3-D MIMO are discussed in the future work.

\subsubsection{Cell-free mMIMO (CF-mMIMO)}
\label{sec:body Cell-free mMIMO}

The integration of NOMA in CF-mMIMO is an emerging trend in the literature as it captures two promising B5G technologies that offer high spectral efficiency~\cite{8957708}. Instead of considering cells, the system models in such a setting involve distributed access points (AP's) in a geographical area trying to serve a set of users. The AP's are connected to a central processing unit that does the bulk of the baseband processing. Like in other MIMO-NOMA settings, the integration of NOMA means that the distributed antennas are used to form clusters of users that can be jointly served in the same orthogonal resource by PD-NOMA. Hence, compared to CF-mMIMO-OMA, the additional design complexity comes in the user pairing problem as the right sets of users have to be identified to be grouped in the cluster. The introduction of NOMA however brings spectral efficiency gains to a CF-mMIMO system, both in the DL~\cite{8368267} and UL~\cite{8756881}. In~\cite{8756881}, a sum-rate optimization problem similar to the theme of this paper for the UL direction is solved using the SCA method.

\subsubsection{Reconfigurable Antenna Systems}
\label{sec:body Reconfigurable Antenna Systems}

The integration of NOMA with the concept of reconfigurable antennas has recently been explored in the literature~\cite{8756960, 8935164}. In~\cite{8756960}, reconfigurable antennas and NOMA are used as complementary multiple access schemes and decisions are made based on the grouping of users as to which scheme is selected for which users. In NOMA, particularly in mmWave systems where the LoS path dominates, the best clusters of users are those that have a similar angle of departure (AoD), but different channel gains. The scheme in~\cite{8756960} then serves users who meet such criteria through PD-NOMA, while users who have not only a similar AoD, but also similar channel gains are served with a reconfigurable antenna based multiple access scheme. On the other hand, in~\cite{8935164}, reconfigurable antennas are used to mitigate the inter-cluster interference in multi-user MISO-NOMA settings. Through only knowledge of the large scale channel properties, the proposed scheme is shown to effectively cancel out the inter-cluster interference. This allows a large number of users to be served in the same orthogonal resource element, through a combination of PD-NOMA SIC for intra-cluster interference management and reconfigurable antennas for inter-cluster interference.

\begin{itemize}[label=$\blacksquare$]
\item Lessons Learnt
    \begin{itemize}[label=$\bullet$]
    \item While work in this area is new and still limited, the common theme of integrating NOMA with these advanced multi-antenna architecture schemes is that NOMA adds an additional degree of flexibility that enables further spectral efficiency gains compared to when these techniques are used in conjunction with OMA. The main design challenge compared to OMA in each of these techniques is the user clustering problem, as an effective beam to cover a set of users in a NOMA cluster has to be formed with distributed AP's in CF-mMIMO, reconfigurable antennas or with the extremely thin pencil beams in ultra mMIMO systems operating at higher frequencies.
    \end{itemize}
\end{itemize}

\subsection{Rate-optimal NOMA with mmWave and THz Communications}
\label{sec:body mmwave and THz Communications}

When operating in these higher frequency bands, the system model needs to consider the low-rank channels and spatial correlation characteristics that are typical for the mmWave communications~\cite{cui2017user, fan2019channel}. While the high spatial correlation of users is typically a constraint for user separation in MU-MIMO systems, MIMO-NOMA systems operating in the mmWave band can use the clustering approach to exploit this correlation to form a cluster and separate the users in the power domain\cite{fan2019channel, cui2018unsupervised}. Another practical limitation of systems operating in the mmWave bands with mMIMO large scale antenna arrays is that scaling the number of transceivers with the number of antennas is often unfeasible. Hence, unlike with regular MIMO-NOMA systems, the system models studied in the literature  often use either analog BF with a single RF chain~\cite{xiao2019user} or a hybrid BF design with a reduced number of RF chains~\cite{dai2019Hybrid, fan2019channel, wei2018multi}. With these additional constraints in the system model, sum-rate optimization schemes in the mmWave bands have similar US and PA design objectives as those with other MIMO-NOMA schemes, e.g.,~\cite{cui2017user}. In multi-cell mmWave-NOMA settings, the authors in~\cite{8967149} propose an angle-domain NOMA scheme that schedules one cell-center and one cell-edge user in a NOMA pair, for each beam in each cell. The sum rate is then maximized by optimizing the precoder and decoder BF design along with the selection of paired users, in settings where the cells co-operate as well as when they do not.For the US sub-problem, while it is typically tackled through traditional optimization schemes, in~\cite{cui2018unsupervised} and~\cite{8856221}, the US problem is solved using an unsupervised clustering ML approach. Both these works exploit the high correlation amongst users' channels and the fact that mmWave propagation is dominated by the LoS path to effectively employ K-means clustering. Since the effects of multipath propagation are limited in mmWave spectrum, the user clustering in mmWave-NOMA systems comes down to finding spatially correlated users with the available CSI at the BS. This is precisely what unsupervised clustering algorithms are capable of achieving without any labeled training data.

The appeal of integrating NOMA in even higher parts of the spectrum, e.g., THz communications, that offer orders of magnitude more bandwidth than even the mmWave spectrum is obvious for B5G networks, as it offers opportunities to cluster more users in a very wide band. Through the joint optimization of beam, bandwidth, and power allocation, a THz-NOMA system allows opportunities for meeting both massive connectivity and extremely high data rates in B5G systems~\cite{han2019terahertz}. In~\cite{8824971}, the authors design a THz-NOMA system that accounts for the distance and frequency selective properties of the THz band. In traditional THz bands, the Long-User-Central-Window (LUCW) principle is usually used to assign the central sub-band to long users, while the side sub-bands are allocated to the short users~\cite{han2019terahertz}. Through the integration of NOMA, the authors in~\cite{8824971} enhance this principle to NOMA clusters of four users each, rather than individual users. Through a hybrid BF design in an ultra mMIMO system that forms 4-user clusters, followed by power allocation and sub-band assignment, a rate optimization that outperforms THz-OMA is demonstrated in~\cite{8824971}.

\begin{itemize}[label=$\blacksquare$]
\item Lessons Learnt 
    \begin{itemize}[label=$\bullet$]
    \item In mmWave systems, due to the high correlation amongst user channels, particularly when users are geographically clustered, it offers great opportunities for integrating NOMA. Users that are hard to separate with individual beams using MU-MIMO techniques can now easily be grouped together in a NOMA cluster. Through the use of mMIMO, beams are formed to serve clusters of users and separated from other beams (clusters) using BF techniques. Since the mmWave channels are usually dominated by the most dominant path, usually the LoS path, the US problem breaks down to finding users with highly correlated channels in the angle domain, often captured through the cosine similarity metric. Unsupervised machine learning techniques that automatically identify such clusters of correlated users is an emerging trend in the mmWave-NOMA literature.
    \end{itemize}
\end{itemize}

\begin{table*}[!htbp]
\centering
\vspace{-1.5em} 
\caption{NOMA-enabled CoMP communications schemes: current status of rate optimization schemes.}
\label{Table: CoMP-NOMA}
\resizebox{\textwidth}{!}{%
\begin{tabular}{|c|c|c|c|c|c|c|c|c|}
\hline
\textbf{[\#]} & \textbf{CoMP Type} & \textbf{System Model} & \textbf{\begin{tabular}[c]{@{}c@{}}Design Objective\end{tabular}} & \textbf{Optimization Method}& \textbf{Main Finding(s)}\\ 
\hline %
\cite{sun2018joint} & Coordinated Beamforming (CB) & \begin{tabular}[c]{@{}c@{}} Two users per cluster \\ model in a multi-cell \\ multi-user MIMO network \end{tabular} &\begin{tabular}[c]{@{}c@{}} Max. sum rate \end{tabular}& \begin{tabular}[c]{@{}c@{}} SDP and SCA technique \\  in each iteration \end{tabular} & \begin{tabular}[c]{@{}c@{}} The proposed CoMP-NOMA\\ scheme outperforms both \\CoMP-NOMA scheme with fixed \\power and NOMA  without CoMP \\for medium  to high SNR \end{tabular}\\
\hline

\cite{ali2018downlink} & \multirow{8}{*}{Joint Transmission (JT)} & \begin{tabular}[c]{@{}c@{}} Two/multiple users per \\ cluster in a two-tier \\HetNet multi-cell network  \end{tabular} & \begin{tabular}[c]{@{}c@{}} Max. sum rate\end{tabular}&  \begin{tabular}[c]{@{}c@{}} A suboptimal distributed \\ power optimization at \\ each CoMP-BS and KKT \\ optimality conditions \end{tabular}  &  \begin{tabular}[c]{@{}c@{}} The JT-CoMP-NOMA scheme \\significantly outperforms the \\JT-CoMP-OMA scheme with\\ more energy requirements\end{tabular}\\  
\cline{1-1} \cline{3-6} 


\cite{noma-comp-4} &  &  \begin{tabular}[c]{@{}c@{}} A large-scale interference-\\limited  HetNet with \\randomly distributed users \end{tabular}  & \begin{tabular}[c]{@{}c@{}} Max. sum rate \end{tabular} & Heuristic algorithms & \begin{tabular}[c]{@{}c@{}} The Coordinated JT-NOMA \\scheme significantly increases \\the throughput in dense networks \end{tabular} \\
\cline{1-1} \cline{3-6} 

\cite{8675425} &  &  \begin{tabular}[c]{@{}c@{}} A user-aided cooperative\\ CoMP-NOMA system \end{tabular}  & Max. sum rate &  {\begin{tabular}[c]{@{}c@{}} BnB method and\\ low complexity\\ greedy algorithm \end{tabular}} & \begin{tabular}[c]{@{}c@{}} The sum-rate performance of the \\proposed scheme outperforms both\\ NOMA and OMA schemes \end{tabular} \\
\hline 

\cite{8781867} & \begin{tabular}[c]{@{}c@{}} Generalized Joint\\ Transmission CoMP (GCoMP) \end{tabular} & \begin{tabular}[c]{@{}c@{}} Multiple users per cluster \\model in downlink\\ CoMP-NOMA system \end{tabular} & \begin{tabular}[c]{@{}c@{}} Max. normalized \\ sum rate \end{tabular} & \begin{tabular}[c]{@{}c@{}} KKT optimality \\ conditions \end{tabular} & \begin{tabular}[c]{@{}c@{}} The sum-rate performance\\ of the proposed scheme\\ outperforms both CoMP-NOMA\\ and GCoMP-OMA schemes \end{tabular}\\
\hline 

\end{tabular}%
}
\end{table*}

\subsection{Rate-optimal NOMA with CoMP Schemes}

The promising performance gains of the PD-NOMA scheme in terms of spectral efficiency and fairness in the single-cell scenarios has motivated extending it to multi-cell scenarios. This is especially true  for the weaker cell-edge users who would also increase their achievable rates through CoMP schemes leading to CoMP-NOMA systems resulting in enhancement of both spectral efficiency and network throughput. A summary of the up-to-date work on the rate-optimal CoMP-NOMA systems is provided in Table~\ref{Table: CoMP-NOMA}. Early work in~\cite{sun2018joint} has considered a downlink CoMP-NOMA where beamforming vectors and power allocation coefficients are jointly optimized through transforming the sum-rate optimization problem into an equivalent rank-constrained SDP problem and then utilize SCA to allow the use of convex solvers. The obtained results have shown a performance gain trade-off of CoMP-NOMA and CoMP-OMA as highlighted in the table. The sum-rate maximization in heterogeneous networks was carried out in~\cite{ali2018downlink} and~\cite{noma-comp-4} where the significant improvement of spectral efficiency gain for joint transmission CoMP-NOMA for various access distances was demonstrated in~\cite{ali2018downlink} while the JT-CoMP-NOMA scheme was shown to significantly enhance the coverage and throughput in tier-m networks as compared to the non-coordinated NOMA scheme in~\cite{noma-comp-4}. The achievable sum rate of two-BS JT NOMA scheme with multiple user pairs, to serve the cell-edge users, was compared in~\cite{8675425} to a proposed relay-aided multiple access (RAMA) scheme and it was shown that JT-CoMP-NOMA, whenever feasible power-wise, outperforms the RAMA scheme. While the previous papers have considered the use of CoMP-NOMA to mainly serve the cell-edge users, a generalized JT CoMP (GCoMP)-enabled NOMA scheme for all the users in the network was proposed in~\cite{8781867} and it was demonstrated that its sum rate is significantly higher than that of CoMP-NOMA. However, the additional complexity of the proposed generalized scheme was not addressed.        

In addition to this list, the performance of two hybrid CoMP-NOMA and CoMP-OMA schemes in heterogeneous ultra-dense networks was investigated in~\cite{zhang2018heterogeneous}. This was conducted  using system-level simulations and the results indicated interesting trade-offs in these networks depending on the roles of the macro- and micro-BSs in the coordinated transmissions and the density of the micro-BSs. This is further highlighted in Section~\ref{sec:futureCompNoma}.     

\begin{itemize}[label=$\blacksquare$]
\item Lessons Learnt
    \begin{itemize}[label=$\bullet$]
     \item The integration of CoMP with NOMA has shown performance gain relative to CoMP-OMA; however, that gain decreases as the number of the cooperating BSs decreases due to the interference made by the non-cooperating BSs. This suggests targeting a trade-off between the extent of cooperation among the networks BSs and stringent requirements of CSI availability and exchange as well as synchronization among the cooperating BSs.     
    
    \item In general, it can be seen for interference-free and weak interference environments such as single-cell scenarios and low-user-density multi-cell scenarios, most of the reported work that the performance gains of NOMA as compared to OMA schemes. On the other hand, for strong-interference environments such as multi-cell scenarios with strong inter-cell interference and heterogeneous networks with high-density micro-BSs, some work reported that the achievable rates by OMA schemes outperform the NOMA scheme. Since the sum-rate performance of the system is deteriorated by the demanding minimum-rate user's constraint which is used for achieving user fairness (i.e., we need to allocate larger power for users that have weak channel conditions to achieve user fairness). This would require the use of hybrid NOMA-OMA and/or other variants of PD-NOMA that are less interference-limited as discussed in Section~\ref{sec:futureCompNoma} on future research topics. 
    
    \item The papers in Table~\ref{Table: CoMP-NOMA} assumed the availability of full CSI (instantaneous channel gains) which would be demanding in dense networks. Such requirements in CoMP-NOMA schemes may result in large signaling overhead and the performance of these schemes may degrade with imperfect or partial CSI as transmit pre-coding is known to be dependent on the availability of full CSI at the transmitter.    
    
\end{itemize}
\end{itemize}

\begin{subtables}
\begin{table*}[!htbp]
\vspace{-2.0em}
\centering
\caption{Cooperative NOMA communications systems: current status of rate optimization schemes.}
\vspace{-0.7em}
\label{Table: Cooperative NOMA (Part A)}
\resizebox{\textwidth}{!}{%
%
}
\end{table*}

\begin{table*}[!htb]
\centering
\vspace{-2em}
\caption{Cooperative NOMA communications systems: current status of rate optimization schemes.}
\label{Table: Cooperative NOMA (Part B)}
\resizebox{\textwidth}{!}{%
%
\\ 

\end{tabular}%
}
\end{table*}
\end{subtables}

\subsection{Rate-optimal NOMA with Cooperative Communications}
\label{Cooperative- existing work}

The merits of cooperative transmission schemes lie in the ability of extending the coverage area of the network and of improving the users' SINR, consequently, the users' rate is improved. This happens by providing some users with multiple copies of the same message while utilizing some form of combining techniques such as the maximum ratio combining (MRC). Recently, the adoption of the cooperative communications systems in conjunction with NOMA has received a lot of interest~\cite{7812683,zhang2017sub, luo2017adaptive,xue2017joint,gau2018optimal, wu2018optimal, cao2018online,zhao2018robust, wan2019achievable, 8653277, wang2019joint,chen2019optimal, zhang2019resource,  agarwal2018performance,zhang2018joint, sun2016transceiver, 7946258,alsaba2018full,yuan2019joint,su2019robust,li2018cooperative,8713856,liu2018hybrid,8667101, zhang2017performance, pan2018resource,8753486,zhao2017joint, liu2019cooperative,chen2017performance, 8713527,8438868,8624380,8707106,8398200,8906057, chen2018interference,xiao2019power,wang2019interference}. This is due to the advantage of the spatial diversity gain in improving the users' rate while enhancing the spectral efficiency of the network~\cite{zhang2017performance}.  

A summary of the up-to-date work on the rate-optimal NOMA schemes in relay-assisted, user-assisted, and device-to-device (D2D) cooperative communications systems is provided in Table~\ref{Table: Cooperative NOMA (Part A)} and Table~\ref{Table: Cooperative NOMA (Part B)}. From the outlined work provided in both Table~\ref{Table: Cooperative NOMA (Part A)} and Table~\ref{Table: Cooperative NOMA (Part B)}, one can observe that researchers have investigated cooperative NOMA schemes under three types of cooperation, namely, relay-assisted, user-assisted, and D2D transmission as follows:
\begin{itemize}
    \item [(i).]  Relay-assisted cooperative NOMA schemes: 
        \begin{enumerate}
         \item In this variant, a BS sends a superimposed signal to a relay, then the relay forwards the received signal to two or $K$ users through DF or AF modes. Subsequently, the users perform SIC procedures to decode the signal sent from the relay as described in Section~\ref{sec:PD_NOMA}. This system model is adopted by several researchers~\cite{luo2017adaptive,xue2017joint,gau2018optimal, cao2018online,zhao2018robust, 8653277, wang2019joint}.
         \item In this case, the system model is similar  to the previous one, but there exists an additional communication link between the BS and the strong user. Usually, in this setup, the objective of the optimization is to maximize the strong user-rate while assuring an acceptable rate to the weak user as proposed in~\cite{chen2019optimal}.
         \item Alternately, a two-way relaying model,  where uplink users transmit signals to a relay using uplink NOMA scheme and then the relay broadcasts the superimposed signal to other downlink users. This model is usually deployed in decentralized networks that aim to reduce the signaling overhead and the deployment overhead of the centralized networks as proposed in~\cite{agarwal2018performance,zhang2018joint}.
       \end{enumerate}
    \item [(ii).] User-assisted cooperative NOMA schemes:
        \begin{enumerate}
         \item In this case,  a BS transmits a superimposed signal through the NOMA scheme to two users that have different channel gains. The strong user uses DF or AF modes to forward the signal to the weak user which employs SIC to decode its own message. Such a model was  studied in~\cite{7946258,alsaba2018full,yuan2019joint,su2019robust,li2018cooperative,liu2018hybrid,8667101}.
         \item Another system model is similar to the previous one but without a direct link between the BS and the weak user. This model is useful to extend the network coverage and was proposed by Zhang~\textit{et al.}~\cite{zhang2017performance}.
       \end{enumerate}
    \item [(iii).] D2D cooperative communications system models~\cite{pan2018resource,8753486,zhao2017joint}; such system models have been considered to improve the spectral efficiency and to reduce the end-to-end latency. 
    \item [(iv).] V2X cooperative communications system models~\cite{liu2019cooperative,chen2017performance, chen2018interference,xiao2019power,wang2019interference}. These models cover all the previous possibilities which are (i) relay-assisted, (ii) user-assisted, (iii) V2V cooperation types. Such models were proposed to enhance connectivity and to improve spectrum efficiency.
\end{itemize}

\begin{itemize}[label=$\blacksquare$]
\item Lessons Learnt
    \begin{itemize}[label=$\bullet$]
    \item The vast majority of the representative works in NOMA-enabled cooperative communications systems considered half-duplex communication mode~\cite{zhang2017sub, luo2017adaptive, xue2017joint,gau2018optimal, wu2018optimal, cao2018online,zhao2018robust, wan2019achievable, 8653277, wang2019joint, agarwal2018performance,zhang2018joint, sun2016transceiver, 7946258, yuan2019joint,su2019robust,li2018cooperative,8713856, pan2018resource,8753486,zhao2017joint, chen2017performance, 8713527,8438868,8624380,8707106,8398200,8906057, chen2018interference,xiao2019power,wang2019interference,chen2019optimal,liu2019cooperative,liu2018hybrid}; since the half-duplex mode yields lower implementation complexity and inherently avoids the self-interference as compared to FD mode. However, the recent advancements in SI cancellation technologies, the move towards short-range networks, and the low spectral efficiency of HD mode paved the way for the adoption of FD mode in NOMA-enabled cooperative communications systems~\cite{7812683, chen2019optimal, zhang2019resource, liu2019cooperative, zhang2017performance, 8667101, alsaba2018full, liu2018hybrid}. In order to further improve the system performance, by exploiting the tradeoff between the efficiency loss of the HD mode and the inherent self-interference of the FD mode, a hybrid half/full-duplex cooperative NOMA that dynamically switch between and HD and FD modes has been proposed by Chen~\textit{et al.}~\cite{chen2019optimal} and Liu~\textit{et al.}~\cite{liu2018hybrid}. Overall, since NOMA and full-duplex are complementary in principle toward improving the SE of a network without the need for additional bandwidth resources~\cite{8694790}, we expect that the research in B5G networks will focus more on the integration of NOMA with full-duplex communication as compared to the integration of NOMA with half-duplex communication.
    \item As highlighted in the previous point, to enable the use of FD relaying some advanced SI cancellation techniques should be adopted to reduce the SI to a tolerable level. This will increase the system complexity and require more signal processing power from the nodes. With FD relaying, it is intuitive from a practical point of view, to use relay-assisted cooperative communications as relay nodes have a powerful signal processing capabilities. However, using a dedicated node for relaying increases the system cost. To avoid such cost and to utilize the network infrastructure, it is advisable, whenever applicable, to use user-assisted communications while equipping the users with energy harvesting technology to avoid draining user\textquotesingle s batteries.
    \end{itemize}
\end{itemize}

\begin{table*}[!htbp]
\centering
\vspace{-2em} 
\caption{Cognitive Radio NOMA communications systems: current status of rate optimization schemes.}
\vspace{-0.5em}
\label{Table: CR-NOMA}
\resizebox{\textwidth}{!}{%
\begin{tabular}{|c|c|c|c|c|c|c|c|}
\hline
\textbf{[\#]} & \begin{tabular}[c]{@{}c@{}} \textbf{Cognitive} \\ \textbf{Paradigm} \end{tabular}  & \begin{tabular}[c]{@{}c@{}} \textbf{System Model} \\ Note: the notion "nodes" here \\ includes BS + users \end{tabular} & \textbf{Design Objective} & \textbf{Optimization Method}& \textbf{Main Finding(s)}\\ 
\hline 

\cite{xu2018joint} & \multirow{26}{*}{Underlay} & \begin{tabular}[c]{@{}c@{}} Half-duplex and full-duplex \\ cognitive OFDM-NOMA \\systems, single-antenna nodes \end{tabular} & \begin{tabular}[c]{@{}c@{}}Max. weighted \\ sum rate of\\  secondary network
\end{tabular} & \begin{tabular}[c]{@{}c@{}}The three decomposed problems \\ are solved based on the bisection \\algorithm, matching theory, and \\Lagrangian dual method together \\with Newton\textquotesingle s method. Also, an \\alternate iteration framework has \\been proposed for joint optimization\end{tabular} & \begin{tabular}[c]{@{}c@{}} The both proposed HD and FD\\ cognitive OFDM-NOMA \\ systems outperform the traditional \\ cognitive OFDMA system \end{tabular} \\ 
\cline{1-1} \cline{3-6} 


\cite{xu2018max} &  & \begin{tabular}[c]{@{}c@{}}Video transmission in NOMA-\\ based cognitive wireless \\ networks, single-antenna nodes\end{tabular} & \begin{tabular}[c]{@{}c@{}}Maxmin rate of \\ secondary users
\end{tabular} & \begin{tabular}[c]{@{}c@{}} SCA technique, binary search, \\ dual decomposition methods, and \\ a heuristic-based algorithm \end{tabular} & \begin{tabular}[c]{@{}c@{}} The proposed algorithm  improves\\ the minimum video transmission\\ quality and achieves fairness \\ among different secondary users \end{tabular}\\ 
\cline{1-1} \cline{3-6} 


\cite{budhiraja2019cross} &  & \begin{tabular}[c]{@{}c@{}} NOMA assisted cognitive\\ radio  femtocell networks, \\ single-antenna nodes  \end{tabular} & \begin{tabular}[c]{@{}c@{}}Max. sum rate \\ of secondary network
\end{tabular} & \begin{tabular}[c]{@{}c@{}} SCA technique, \\KKT optimality conditions, \\ and a greedy algorithm \end{tabular} & \begin{tabular}[c]{@{}c@{}} The proposed scheme enhances \\ the sum rate of the femto-cell \\ users as compared with the \\ conventional cognitive radio \\OMA-based femtocell scheme \end{tabular} \\ 
\cline{1-1} \cline{3-6} 

\cite{chen2018novel} &  & \begin{tabular}[c]{@{}c@{}} Relay-assisted spectrum \\sharing scheme with user-\\assisted cooperation in \\cognitive radio network, \\single-antenna nodes \end{tabular} & \begin{tabular}[c]{@{}c@{}}Max. sum rate of\\ secondary receivers\end{tabular} & \begin{tabular}[c]{@{}c@{}} The optimization problem is \\transformed to a convex \\problem through reformulation, \\then the closed-form solution \\is obtained through relaxation\end{tabular} & \begin{tabular}[c]{@{}c@{}}The proposed scheme has been\\ compared with an equivalent \\scheme but without user-assisted \\cooperation. The proposed scheme \\reduces the required transmit \\power while achieving same rate\\ performance results\end{tabular}\\
\cline{1-1} \cline{3-6} 

\cite{mohammadi2018beamforming} &  & \begin{tabular}[c]{@{}c@{}} CR network with full-duplex\\ relay-assisted transmission, \\  relay: multiple antennas, \\ BS: single antenna, \\ and users: single antenna\end{tabular} & \begin{tabular}[c]{@{}c@{}}Max. the rate of \\ near secondary user
\end{tabular} & \begin{tabular}[c]{@{}c@{}} SDR in conjunction \\with line-search approach \end{tabular} & \begin{tabular}[c]{@{}c@{}} The proposed scheme improves \\both the near and far secondary \\users rates as  compared to the\\ HD mode \end{tabular}\\ 
\cline{1-1} \cline{3-6} 

\cite{9088119} & &\begin{tabular}[c]{@{}c@{}} An industrial cognitive IoT \\over multi-homing-based\\ cognitive NOMA HetNets,\\ single-antenna nodes  \end{tabular} & Max. sum rate & \begin{tabular}[c]{@{}c@{}} SCA technique and\\dual decomposition method \end{tabular} & \begin{tabular}[c]{@{}c@{}} The sum-rate performance of the\\ proposed scheme outperforms \\OMA scheme at the cost of \\increasing the receiver\textquotesingle s complexity \end{tabular}\\ 
\hline

\cite{8753517} & \begin{tabular}[c]{@{}c@{}} Underlay  \\ and\\ Overlay \end{tabular}   & \begin{tabular}[c]{@{}c@{}} A NOMA-based SWIPT \\scheme, single-antenna nodes \end{tabular} & Max. sum rate & Dichotomy method & \begin{tabular}[c]{@{}c@{}} The proposed scheme can achieve \\ a maximum sum rate via setting \\ an optimal sensing time for\\ the secondary network  \end{tabular}\\
\hline

\cite{li2018optimized} &  Overlay & \begin{tabular}[c]{@{}c@{}} A joint OMA and NOMA\\  transmission protocol in an \\industrial  cooperative-cognitive \\network, single-antenna nodes  \end{tabular} & \begin{tabular}[c]{@{}c@{}}Max. secondary\\  transmitter rate 
\end{tabular} & A linear search algorithm
& \begin{tabular}[c]{@{}c@{}} The performance of the proposed \\ joint OMA and NOMA protocol \\ outperforms OMA schemes in \\certain scenarios \end{tabular} \\
\hline

\cite{8703419} & Interweave & \begin{tabular}[c]{@{}c@{}} A two-tier cognitive radio\\  heterogeneous NOMA \\scheme, single-antenna nodes\end{tabular} & \begin{tabular}[c]{@{}c@{}} Max. the sum rate of \\ second-tier small- \\ cells network \end{tabular} & \begin{tabular}[c]{@{}c@{}} D.C. programming \\ transformation and KKT \\ optimality conditions \end{tabular} & \begin{tabular}[c]{@{}c@{}} The sum-rate performance of \\ the proposed small-cell scheme \\ outperforms its counterpart of \\NOMA scheme with equal power\\ allocation and with OFDMA \end{tabular}\\
\hline

\end{tabular}%
}
\end{table*}

\subsection{Rate-optimal NOMA with Cognitive Radio Communications}
\label{Rate-optimal CR-NOMA}

The main aim of cognitive radio communications systems is to improve the utilization of the available spectrum. Therefore, PD-NOMA is adopted in conjunction with CR  to further enhance the system spectral efficiency. A summary of the up-to-date work on the rate-optimal CR-NOMA schemes is provided in Table~\ref{Table: CR-NOMA}. One can notice from the outlined work in Table~\ref{Table: CR-NOMA}, that researchers have investigated three  main system models, namely: (i) conventional cognitive radio~\cite{xu2018joint, 8753517,xu2018max}, (ii) cognitive radio in heterogeneous networks (i.e., with macro-BS(s) and femto-BS(s) setting)~\cite{budhiraja2019cross,8703419,9088119}, and (iii) cognitive radio with cooperative communications systems, namely, relay-assisted~\cite{mohammadi2018beamforming},  user-assisted~\cite{li2018optimized}, and relay-assisted alongside with user-assisted~\cite{chen2018novel}.

In conventional CR-NOMA systems, the early work in~\cite{xu2018joint} has considered the weighted-sum capacity maximization problem for half-duplex/full-duplex cognitive OFDM-NOMA systems with two/multiple users on each subcarrier. A joint optimization problem of sensing duration, user assignment, and power allocation has been formulated and iteratively optimized using an alternate iteration framework. The former optimization problem, due to its intractability, is decoupled into three subproblems and separately optimized using the bisection algorithm, matching theory, and Lagrangian dual method together with Newton\textquotesingle s method, respectively. Through simulations, the authors demonstrated that the proposed joint solution can achieve a weighted-sum capacity near to the one achieved by employing the exhaustive-search algorithm with 120 times complexity reduction. Later, in~\cite{xu2018max}, a maxmin rate optimization problem for optimizing the secondary user scheduling, power allocation, and video packet scheduling in CR-NOMA-based network has been proposed. In this work, the authors also decouple the former optimization problem into two sub-problems, namely, a joint secondary user scheduling and power allocation subproblem that is solved through SCA technique, binary search, and dual decomposition methods and a video packet scheduling subproblem that is solved through a heuristic-based algorithm. Through simulations, the authors demonstrated that the proposed solution can improve the minimum video transmission quality with a reduced computational complexity as compared to a solution that depends on the earliest deadline first approach (EDFA). Afterward, in~\cite{8753517}, a sum-throughput maximization problem that jointly optimizes the sensing time and the power allocation for an overlay/underlay CR-NOMA-based SWIPT network has been proposed. The joint optimization problem is solved through an algorithm based on the Dichotomy method. This study indicated that the maximal system throughput can be achieved through optimizing the sensing time for the secondary network.

In CR-NOMA HetNets, the early work in~\cite{8703419} has considered the sum-rate maximization problem for optimizing the bandwidth resource allocation, secondary users clustering, and power allocation in interweave CR-NOMA HetNets. In this work, the authors decouple the former optimization problem and separately optimize (i) the bandwidth resource allocation to efficiently utilize the limited idle bandwidths, (ii) the secondary-users clustering to improve the sum rate of secondary users, and (iii) the power allocation to maximize sum throughput within a NOMA cluster. The authors demonstrated that the proposed secondary-users clustering scheme has a much lower complexity compared to the exhaustive search algorithm and the sum-rate performance of the proposed D.C. programming power allocation outperforms both the NOMA scheme with equal power allocation and OFDMA scheme. Later, in~\cite{budhiraja2019cross}, a sum-rate maximization problem that jointly optimizes the channel allocation and power allocation in CR-NOMA with femtocell users has been proposed. The channel allocation is optimized to mitigate co-tier and cross-tier interferences while the power allocation is optimized to ensure guaranteed QoS for femtocell users. Through simulations, the authors demonstrated the performance gain in the average sum-rate of the proposed scheme as compared with the CR-OMA scheme. Afterward, in~\cite{9088119}, a sum-rate maximization problem that jointly optimizes secondary IoT device scheduling and power allocation with imperfect CSI and imperfect spectrum sensing in CR-NOMA HetNets has been proposed. The proposed joint optimization problem has been solved through the dual decomposition method and the SCA technique. The proposed scheme improves the sum-rate performance of the secondary devices in industrial cognitive IoT as compared with the OMA scheme.

In Table~\ref{Table: CR-NOMA}, three works considered cognitive radio with cooperative communications systems. In~\cite{mohammadi2018beamforming}, a rate maximization of the near-secondary user that jointly optimizes the power allocation and beamforming through employing SDR with the line-search approach in FD relay-assisted CR-NOMA has been analyzed. Through simulations, the authors showed the ability of the proposed scheme in reducing the self-interference impact at the relay and the inter-user interference at the near user. In~\cite{li2018optimized}, a rate maximization of the secondary transmitter with a joint OMA-NOMA scheme in user-aided overlay industrial CR networks has been analyzed. The former optimization problem has considered the optimization of the time-sharing between primary and secondary systems and the transmit-power allocation at the secondary transmitter and solved iteratively through a linear search algorithm. In this study, at the secondary transmitter, the authors proposed two relaying NOMA schemes, namely, decode-and-forward based NOMA (DF-NOMA) and Analog-Network-Coding based NOMA (ANC-NOMA). Through simulations, the authors revealed the superiority of the ANC-NOMA scheme over the DF-NOMA scheme. In~\cite{chen2018novel}, a sum-rate maximization problem of secondary receivers in relay assisted NOMA spectrum sharing scheme has been analyzed. The optimization problem in this study optimizes the power allocation at the NOMA relay and solved through relaxation. Through simulations, the authors articulated the importance of the user-assisted cooperation link in improving the sum-rate performance of the secondary receivers.

\begin{itemize}[label=$\blacksquare$]
\item Lessons Learnt
    \begin{itemize}[label=$\bullet$]
    \item The majority of the CR-NOMA representative works in Table~\ref{Table: CR-NOMA} have considered the underlay paradigm because the interference power constraint of the PU in CR networks is directly related to the minimum rate constraint of the weak user in NOMA pair. It is worthy to note that Ding~\textit{et al.}~\cite{ding2015impact} has proposed a novel PA scheme in downlink NOMA called cognitive radio inspired NOMA (CR-inspired-NOMA) that depends on users\textquotesingle \ QoS requirements. NOMA scheme here is considered as a special case of underlay CR networks. In this scheme, the weak/strong user in the NOMA-pair can be viewed as a primary/secondary user. Particularly, the serving BS allocates enough power to the weak user (PU) to meet his QoS requirement (rate constraint), then the strong user (SU) is being served with the remaining BS power. However, this scheme does not guarantee the QoS requirement for the strong user in the NOMA-pair and even can but the strong user in an outage if the weak user\textquotesingle s rate requirement is very large.
    \end{itemize}
\end{itemize}

\begin{table*}[!htbp]
\centering
\vspace{-2em} 
\caption{VLC-NOMA systems: 
current status of rate optimization schemes.}
\vspace{-0.5em}
\label{Table: NOMA-VLC}
\resizebox{\textwidth}{!}{%
\begin{tabular}{|c|c|c|c|c|c|c|}
\hline
\textbf{[\#]} & \textbf{Classification} & \textbf{System Model} & \textbf{Design Objective} & \textbf{Optimization Method}& \textbf{Main Finding(s)}\\ 
\hline

\cite{yang2017fair} &  \multirow{22}{*}{\begin{tabular}[c]{@{}c@{}} Single-cell\\(single-carrier) \end{tabular}} &\begin{tabular}[c]{@{}c@{}} Downlink VLC-NOMA \\scheme, one LED luminary+ \\$K$ photo-diodes\end{tabular} & Max. sum rate & \begin{tabular}[c]{@{}c@{}}KKT optimality  \\ conditions\end{tabular} & \begin{tabular}[c]{@{}c@{}}The sum-rate performance \\of the proposed scheme \\ outperforms the OFDMA scheme\\ with some additional computations\end{tabular}\\
\cline{1-1} \cline{3-6} 

\cite{tahira2019optimization} & & \begin{tabular}[c]{@{}c@{}}Downlink VLC-NOMA \\ scheme,  one LED luminary+ \\two photo-diodes\end{tabular} & \begin{tabular}[c]{@{}c@{}}Max. sum rate \end{tabular} & SCA technique & \begin{tabular}[c]{@{}c@{}}The sum-rate performance of the \\proposed optimized VLC-NOMA \\scheme outperforms the conventional \\ VLC-NOMA scheme for both a LoS \\ and a LoS+NLoS systems \end{tabular}\\
\cline{1-1} \cline{3-6}


\cite{8742679} & &\begin{tabular}[c]{@{}c@{}} Downlink VLC-NOMA \\scheme, one LED luminary+ \\ $K$ photo-diodes\end{tabular} & Max. sum rate & \begin{tabular}[c]{@{}c@{}} Standard interior point \\method  \end{tabular} & \begin{tabular}[c]{@{}c@{}} The proposed scheme achieves higher \\system sum-rate performance as\\ compared with equivalent schemes of \\ the gain ratio power allocation (GRPA)\\ and of the fixed power allocation (FPA)\end{tabular}\\ 
\cline{1-1} \cline{3-6} 

\cite{8936365} & &\begin{tabular}[c]{@{}c@{}} Downlink VLC-NOMA \\scheme  with beam steering, \\one LED luminary + \\ $K$ photo-diodes\end{tabular} & Max. sum rate & \begin{tabular}[c]{@{}c@{}} An iterative algorithm\\ based on majorization-\\minimization procedure \end{tabular} & \begin{tabular}[c]{@{}c@{}} The proposed scheme achieves an\\ additional 10 Mbps sum-rate gain\\ for each NOMA user pair as\\ compared to the TDMA scheme \end{tabular}\\ 
\cline{1-1} \cline{3-6} 

\cite{9075277} & & \begin{tabular}[c]{@{}c@{}}A downlink  UAV-assisted\\ VLC-NOMA system,\\ one LED luminary + \\$K$ photo-diodes\end{tabular} & Max. sum rate & \begin{tabular}[c]{@{}c@{}} Harris Hawks\\ optimization with\\ artificial neural\\ networks \end{tabular} & \begin{tabular}[c]{@{}c@{}} The sum-rate performance of the\\ proposed scheme that jointly optimizes\\ UAV\textquotesingle s placement and PA outperforms\\ the schemes that individually optimize\\ the PA (i.e., GRPA) and the UAV\textquotesingle s\\ placement (i.e., random placement)\\ as well as the OFDMA scheme \end{tabular}\\ 
\hline

\cite{fu2018enhancedo} &\multirow{6}{*}{\begin{tabular}[c]{@{}c@{}} Single-cell\\(multi-carrier) \end{tabular}} & \begin{tabular}[c]{@{}c@{}}Downlink OFDM-based \\ VLC-NOMA scheme, \\  four LED luminaries + \\$K$ photo-diodes\end{tabular} & Max. sum rate & Analytical method 
& \begin{tabular}[c]{@{}c@{}} The proposed power allocation \\algorithm outperforms the fixed \\ power allocation (FPA) and the \\ gain ratio power allocation (GRPA) \\ algorithms in terms of sum rate \end{tabular}\\
\cline{1-1} \cline{3-6}

\cite{8669970} & &\begin{tabular}[c]{@{}c@{}} A downlink power-line-fed \\ VLC-NOMA scheme, \\ one LED luminary + \\ $K$ photo-diodes\end{tabular} & Max. sum rate & \begin{tabular}[c]{@{}c@{}} KKT optimality \\ conditions \end{tabular} & \begin{tabular}[c]{@{}c@{}}The proposed scheme achieves higher \\system sum-rate performance as\\ compared with equivalent schemes of \\ the normalized gain difference power \\ allocation (NGDPA) and FPA  \end{tabular}\\ 
\hline


\cite{feng2018multiple} & \multirow{8}{*}{Multi-cell}  & \begin{tabular}[c]{@{}c@{}} Ultra-dense VLC hybrid\\ OMA and NOMA scheme,\\ eight LED luminaries + \\six photo-diodes\end{tabular} & \begin{tabular}[c]{@{}c@{}}Max. sum \\ throughput \end{tabular} & \begin{tabular}[c]{@{}c@{}}A dynamic programming-\\ based $l$th layer-recursion \\model \end{tabular} & \begin{tabular}[c]{@{}c@{}} The proposed hybrid NOMA-OMA \\scheme outperforms the conventional \\ TDMA and NOMA schemes in \\ terms of the achievable throughput\end{tabular} \\ 
\cline{1-1} \cline{3-6} 


\cite{zhang2017user} &  &\begin{tabular}[c]{@{}c@{}}A user grouping-based \\ VLC-NOMA scheme,\\ four LED luminaries + \\$K$ photo-diodes\end{tabular} & \begin{tabular}[c]{@{}c@{}} Both max. sum rate \\ and maxmin user-rate\end{tabular} & \begin{tabular}[c]{@{}c@{}}Gradient projection \\ algorithm (GPA) \end{tabular} & \begin{tabular}[c]{@{}c@{}} The proposed NOMA scheme \\ achieves a higher sum-rate \\ performance than the OMA scheme\end{tabular}\\ 
\cline{1-1} \cline{3-6} 

\cite{9062301} &  &\begin{tabular}[c]{@{}c@{}} A multiuser multi-cell \\VLC-NOMA scheme, \\four LED luminaries + \\eight photo-diodes \end{tabular} & \begin{tabular}[c]{@{}c@{}} Both max. throughput\\ and maxmin rate \\ of cell-edge users \end{tabular} & \begin{tabular}[c]{@{}c@{}}Interior point method \end{tabular} & \begin{tabular}[c]{@{}c@{}} The proposed ZF-pre-coding NOMA\\ scheme outperforms both conventional\\ ZF and conventional NOMA \end{tabular}\\ 
\hline

\end{tabular}%
}
\end{table*}

\subsection{Rate-optimal NOMA with VLC}

Although VLC systems have abundant free bandwidth, the current off-the-shelf LEDs have limited bandwidth which necessitates the adoption of spectrally efficient schemes like NOMA-enabled schemes to attain the desired high data rates. Moreover, indoor and low mobility VLC systems have the attractive features of quasi-static nature of the propagation channel and relatively high SNR conditions, especially for short-distance communications and clear LoS conditions, as these features allow more reliable estimation of the channel gains and SIC, respectively. A summary of the up-to-date work on the rate-optimal NOMA-enabled VLC systems is provided in Table~\ref{Table: NOMA-VLC} for different trans-receive configurations. 

The early work in~\cite{yang2017fair} has developed a convexified power allocation problem for an arrangement of one LED transmitter and $M$ receiving photo-diodes and introduced an optimal power allocation algorithm that was shown to outperform both the gain ratio power allocation (GRPA) and the fixed power allocation schemes. A similar approach was used in~\cite{8742679} where the convexified  multiple factors power allocation problem was solved using CVX solvers. In~\cite{tahira2019optimization}, the Cuckoo Search (CS) meta-heuristic optimization algorithm was utilized to solve a multi-objective optimization problem in terms of the achievable rate and the users received power and the optimized scheme was again shown to achieve higher larger sum-rate than fixed power VLC-NOMA scheme. However, that gain was relatively small and the complexity of the used algorithm was not addressed. In~\cite{8936365}, an iterative  algorithm for the selection of the optimal NOMA pairs and then the determination of their allocated powers using the majorization-minimization (MM) procedure was utilized to maximize the sum-rate for VLC-NOMA with steerable beams. The rate maximization for integration of UAVs in VLC-NOMA was considered in~\cite{9075277} where the Harris Hawks Optimizer (HHO), as a swarm intelligence technique, was utilized to jointly optimize the UAV\textquotesingle s placement and power allocation.      

For multi-carrier VLC-NOMA single-cell scenarios, in~\cite{fu2018enhancedo}, an enhanced power allocation algorithm, as compared to fixed power allocation and gain ratio power allocation, for an OFDM-based VLC-NOMA system was shown to outperform these power allocation schemes in terms of sum rate and subcarrier loss rate. In~\cite{8669970}, a joint power allocation strategy to maximize the sum-rate for a combined power line communications (PLC)-VLC network, where the data is fed by the PLC into the VLC network, was proposed and compared to the other power allocation schemes. 

In multi-cell scenarios, a graph theory-based clustering strategy was proposed in~\cite{feng2018multiple} for layered asymmetrically clipped optical OFDM (LACO-OFDM) aided ultra-dense VLC networks and a hybrid NOMA-OMA scheme is adopted to reduce the excessive interference in such dense networks. A gradient projection (GP) algorithm was adopted in~\cite{zhang2017user} for optimizing the power allocation for sum-rate maximization where the users are assigned to the APs so that no inter-cell interference is there. The inter-cell interference was addressed in~\cite{9062301} where two variants of zero-forcing-based pre-coding in NOMA-VLC, where the cell-edge user is either associated with all the neighboring APs or associated with only the AP with the strongest channel gain, were introduced. The proposed schemes were shown to outperform both conventional ZF pre-coding and conventional NOMA schemes and the computational complexity of the different schemes considered in the paper was addressed in details.

\begin{itemize}[label=$\blacksquare$]
\item Lessons Learnt
    \begin{itemize}[label=$\bullet$]
    \item The optimization of VLC systems is in general more involved than RF systems due to the additional non-negative input signal and peak power constraints. The capacity-achieving inputs for amplitude-constrained multi-user multiple-access and broadcast VLC and optical channels are still, in general, open problems~\cite{optical-capacity-1, optical-capacity-2,optical-capacity-3} and hence the capacity bounds and/or the modulation schemes such as pulse-amplitude modulation, on-off keying, and pulse-position modulation are widely used in optimizing the achievable rates in such channels.
    \item Although NOMA schemes are attractive for VLC systems and networks due to the typical high SNR conditions and the quasi-static channel characteristics especially for the LoS indoor scenarios, the additional signal design constraints, as highlighted in the item before, in VLC result in more involved rate optimization problems as compared to RF systems. So, the computational complexity of the utilized optimization methods and algorithms becomes an issue and should have been addressed in all the papers in the table.     
\end{itemize}
\end{itemize}

\begin{table*}[!htb]
\centering
\vspace{-2em} 
\caption{UAV-NOMA assisted communications systems: current status of rate optimization schemes.}
\vspace{-0.5em}
\label{tab:uavNoma}
\resizebox{\textwidth}{!}{%
\begin{tabular}{|c|c|c|c|c|c|}
\hline
\textbf{[\#]} & \textbf{Category} & \textbf{System Model} & \textbf{Design Objective} & \textbf{Optimization Method}& \textbf{Main Finding(s)}\\ 
\hline
\cite{sohail2018non} & \multirow{22}{*}{ \begin{tabular}[c]{@{}c@{}}UAV as BS \\(no ground BS)\end{tabular}} & \begin{tabular}[c]{@{}c@{}}One UAV + two users, \\ single-antenna UAV, users \end{tabular} & \begin{tabular}[c]{@{}c@{}}Max. sum rate \end{tabular} & Exhaustive search & \begin{tabular}[c]{@{}c@{}}Proposed UAV altitude and PA\\ scheme outperforms OMA in \\a similar system model setting \end{tabular}\\ 

\cline{1-1} \cline{3-6} 
\cite{liu2019placement} & &\begin{tabular}[c]{@{}c@{}}One UAV + multiple users, \\ single-antenna UAV, users \end{tabular}& \begin{tabular}[c]{@{}c@{}} Max. sum rate\end{tabular} & \begin{tabular}[c]{@{}c@{}} Break down non-convex\\problem into convex\\ problems with KKT \\optimality conditions \end{tabular} & \begin{tabular}[c]{@{}c@{}}Proposed UAV placement and\\ PA scheme outperforms OMA\\ and fixed PA NOMA schemes \end{tabular}\\  

\cline{1-1} \cline{3-6} 
\cite{sun2019cyclical} & &\begin{tabular}[c]{@{}c@{}}One UAV + multiple users, \\ single-antenna UAV, users \end{tabular} & \begin{tabular}[c]{@{}c@{}}Maxmin user-rate \end{tabular}& \begin{tabular}[c]{@{}c@{}} BCD method\end{tabular} &\begin{tabular}[c]{@{}c@{}} The proposed joint optimization \\of UAV placement and user \\scheduling can double the \\sum rate compared to \\cyclical TDMA scheduling\end{tabular}\\   

\cline{1-1} \cline{3-6} 
\cite{nasir2019uav} & &\begin{tabular}[c]{@{}c@{}}One UAV + multiple users, \\ single-antenna UAV, users \end{tabular} &\begin{tabular}[c]{@{}c@{}} Maxmin user-rate \end{tabular} & \begin{tabular}[c]{@{}c@{}}Path following algorithm \end{tabular}& \begin{tabular}[c]{@{}c@{}} The proposed joint optimization \\of multiple design variables \\improved the sum rate \\significantly compared to OMA\end{tabular}\\ 

\cline{1-1} \cline{3-6} 
\cite{8685130} &  &\begin{tabular}[c]{@{}c@{}} One UAV + multiple users,\\single-antenna UAV, users\\Joint OMA + NOMA \\transmission modes \end{tabular}  &\begin{tabular}[c]{@{}c@{}} Maxmin user-rate\\ \end{tabular} & \begin{tabular}[c]{@{}c@{}} Penalty dual \\ decomposition method \end{tabular} &\begin{tabular}[c]{@{}c@{}} Proposed method outperformed \\benchmarks for OMA only\\ or NOMA only based user\\ scheduling\end{tabular} \\ 
\cline{1-1} \cline{3-6} 
\cite{8848428} & &\begin{tabular}[c]{@{}c@{}} One UAV + multiple users, \\ mobile users with GPS, \\ UAV at fixed height\end{tabular}  &Max. sum rate & Interior point method & \begin{tabular}[c]{@{}c@{}} Given a fixed decoding order, the\\ proposed scheme finds the optimal \\ UAV position and outperforms OMA \\ and NOMA with fixed UAV position\end{tabular} \\ 
\hline

\cite{zhao2019joint} & \multirow{7}{*}{\begin{tabular}[c]{@{}c@{}}UAV as BS\\(with ground BS)\end{tabular}} &\begin{tabular}[c]{@{}c@{}}One BS + one UAV \\+ multiple users, \\multi-antenna BS, single\\-antenna UAV and users \end{tabular}& \begin{tabular}[c]{@{}c@{}}Max. sum rate\end{tabular}& \begin{tabular}[c]{@{}c@{}} Iterative algorithm using\\ BCD method\end{tabular}& \begin{tabular}[c]{@{}c@{}} The proposed low complexity\\ algorithms that steer the UAV\\ close to its users and away\\ from BS served users \end{tabular}\\ 
\cline{1-1} \cline{3-6} 

\cite{8700188} &  &\begin{tabular}[c]{@{}c@{}}One BS + multiple UAVs\\ + multiple IoT nodes,\\ single-antenna UAV and\\ IoT nodes \end{tabular} & \begin{tabular}[c]{@{}c@{}}Max. system\\ capacity\end{tabular} & \begin{tabular}[c]{@{}c@{}} An algorithm based\\ on K-means clustering\\ method and matching\\ theory as well as an\\alternative optimization\\ algorithm \end{tabular} & \begin{tabular}[c]{@{}c@{}}The total system capacity of the\\proposed scheme outperforms\\ both NOMA scheme with fixed\\ height UAV and OMA scheme \end{tabular}\\
\hline

\cite{mei2019uplink} & \multirow{10}{*}{ \begin{tabular}[c]{@{}c@{}}UAV as a user\end{tabular}}&\begin{tabular}[c]{@{}c@{}} One BS + UAV as a user \\+ ground-users, uplink \\communication with \\ground BSs that co-operate\\ with each other \end{tabular}  &\begin{tabular}[c]{@{}c@{}} Max. WSR of \\UAV and\\ ground users\end{tabular} & \begin{tabular}[c]{@{}c@{}}Alternating algorithm\\ and SCA technique \end{tabular} &\begin{tabular}[c]{@{}c@{}} The proposed scheme achieved\\ significant sum-rate gains \\compared to OMA and\\  non-cooperative schemes\end{tabular} \\ 
\cline{1-1} \cline{3-6} 

\cite{8906143} & &\begin{tabular}[c]{@{}c@{}}Multiple BSs + UAV as \\a user + ground-users, \\uplink communication links \\between users-BSs and \\between the UAV-BSs \end{tabular}  & Max. sum rate & \begin{tabular}[c]{@{}c@{}}A low complexity\\ suboptimal\\ iterative algorithm \end{tabular} &\begin{tabular}[c]{@{}c@{}}The proposed NOMA scheme \\outperforms TDMA scheme in \\term of the sum rate \end{tabular} \\ 
\cline{1-1} \cline{3-6} 

\cite{9070200} &   & \begin{tabular}[c]{@{}c@{}}One BS + one UAV \\+ one user,\\single-antenna UAV, user \end{tabular}  &  \begin{tabular}[c]{@{}c@{}}Max. ground-\\user rate \end{tabular} & \begin{tabular}[c]{@{}c@{}}One-dimensional\\ searching method \end{tabular} &\begin{tabular}[c]{@{}c@{}}The proposed aerial-ground\\ NOMA scheme with both perfect\\ and partial CSI achieved a\\ significant rate-improvement as\\ compared to OMA\end{tabular} \\ 
\hline

\cite{mu2020joint} & UAV as a relay&\begin{tabular}[c]{@{}c@{}} One BS + UAV as a FD\\ relay + two ground users,\\ multi-antenna BS, UAV, \\ and single-antenna users \end{tabular}  &\begin{tabular}[c]{@{}c@{}}Max. sum\\ throughput \end{tabular}  &\begin{tabular}[c]{@{}c@{}}Inner approximation \\method \end{tabular} &\begin{tabular}[c]{@{}c@{}}The sum-throughput performance\\ of the proposed FD-NOMA-SWIPT\\ scheme outperforms both HD-NOMA-\\SWIPT and FD-OMA-SWIPT schemes\end{tabular} \\ 
\hline

\end{tabular}%
}
\end{table*}

\subsection{Rate-optimal NOMA with UAV Assisted Communications}
\label{sec:litsuvUAV}

As described in Section~\ref{sec:UAV_bkgd}, the use of UAVs in 5G and beyond communications systems is fast growing. The typical use-case studied in UAV-NOMA systems is when the UAV acts as a flying BS and provides a capacity boost to an existing terrestrial network~\cite{liu2019uav}. The goal is to serve more users by scaling up the number of BSs in an ad-hoc fashion to meet any unprecedented traffic needs. PD-NOMA scheme also aims to serve more users concurrently. This is accomplished  via non-orthogonal spectrum sharing. For further spectral efficiencies gains, the two technologies can be combined to have the flying BSs to serve users through the NOMA scheme. This UAV-NOMA framework adds an extra degree of freedom to the rate optimization problem by allowing for the choice of UAV placement, altitude, and trajectory. In this section, we survey how this additional degree of freedom has been exploited by researchers in such UAV-NOMA systems~\cite{sohail2018non, liu2019placement, sun2019cyclical, nasir2019uav, zhao2019joint}. 

Table~\ref{tab:uavNoma} summarizes the existing literature on rate optimization schemes in UAV-NOMA systems. The system models typically involve a UAV acting as a flying BS and multiple users served by the UAV~\cite{sun2019cyclical, liu2019placement, nasir2019uav}. As described in earlier sections, the design variables in the rate optimization problems for NOMA schemes typically involve power allocation, user scheduling, or even beam design if multiple antennas are available. In UAV-NOMA systems, there is the additional design variable of the UAV placement (the location in the 3-D space). In the work by Liu~\textit{et al.}~\cite{liu2019placement} the sum rate is maximized through an algorithm that first finds an optimal placement for the UAV, followed by the PA coefficients. In another work by Sun~\textit{et al.} and Nasir~\textit{et al.}~\cite{sun2019cyclical,nasir2019uav}, a joint optimization of the UAV placement with other design variables is considered for a max-min rate optimization problem. In~\cite{sun2019cyclical}, it is only the UAV placement and PA that is jointly optimized; while in~\cite{nasir2019uav}, the authors jointly optimize several design variables including the UAV altitude, user scheduling, PA, and the transmit antenna beam-width. While allowing for flexible UAV placement can help increase the spectral efficiency, the energy efficiency is inversely related to the UAV altitude. Hence, the authors in~\cite{sohail2018non} studied a scheme with fixed UAV altitude and compared it with a scheme where the UAV altitude was allowed to be optimized by the algorithm. The authors showed that there are significant spectral efficiency gains from allowing the UAV altitude to be optimized, but at the cost of energy efficiency.

As the UAV is used as an add-on to a terrestrial network, multiple users will be served partly by a ground BS and partly by the UAV acting as a flying BS. In such a model, the UAV transmissions will cause interference to the BS-served users. The study in~\cite{zhao2019joint} addresses such a scenario  with a two-part strategy: (i)~in the first, a joint user scheduling and UAV trajectory iterative optimization scheme is designed for the UAV-served users; and, (ii)~in the  second, a NOMA precoding scheme is developed for the BS-served users to cancel the interference from the UAV-served users. This is accomplished by exploiting the multiple antennas available at the BS.    

In the work by Mei~\textit{et al.}~\cite{mei2019uplink}, an uplink system model of BSs that co-operate with each other, one UAV and multiple ground users are considered. However, in this case, the UAV is considered as an uplink user rather than a flying BS. Hence, the ground BS has to serve the UAV as well as the ground users. In this model, the higher the uplink rate for the UAV, the more the interference to the ground users. As a result, the weighted sum rate of the UAV and ground users is optimized through a user scheduling and power allocation algorithm.

\begin{itemize}[label=$\blacksquare$]
\item Lessons Learnt
    \begin{itemize}[label=$\bullet$]
    \item The introduction of UAVs to NOMA systems allows for an extra degree of freedom to serve. The UAV altitude and placement can be optimized in conjunction with the power allocation coefficients and user clustering to maximize the sum rate. Such a setting creates a system where several design variables can be optimized to improve the performance, e.g., the UAV altitude, user scheduling, PA, and the transmit antenna beam-width in~\cite{nasir2019uav}.
    \item Typically, UAVs will be used in conjunction with terrestrial BSs to jointly serve users on the ground. The interference caused by the UAV BSs can affect some NOMA clusters, served by different BSs in multi-cell settings.
    \end{itemize}
\end{itemize}

\begin{table*}[!p]
\centering
\caption{NOMA communications schemes with other enabling technologies: current status of rate optimization schemes.}
\label{tab:otherNoma}
\resizebox{\textwidth}{!}{%
\begin{tabular}{|c|c|c|c|c|c|}
\hline
\textbf{[\#]} & \textbf{Classification} & \textbf{System Model} & \textbf{Design Objective} & \textbf{Optimization Method}& \textbf{Main Finding(s)}\\ 
\hline

\cite{lyu2017optimal} & \multirow{15}{*}{\begin{tabular}[c]{@{}c@{}} Backscatter\\ Communications\\(BackCom) \end{tabular}} & \begin{tabular}[c]{@{}c@{}}A TDMA-based and\\ NOMA-based backscatter\\ assisted wireless powered\\ communication network \\(WPCN), with BackCom\\ mode and  harvest-then-\\transmit (HTT) mode,\\ single antenna nodes \end{tabular} & Max. sum rate & Interior-point method & \begin{tabular}[c]{@{}c@{}}In HTT mode, the sum rate of\\TDMA scheme outperforms NOMA \\as TDMA scheme can harvest more\\ energy and consequently achieves\\higher transmission rate. In BackCom,\\ NOMA outperforms TDMA\end{tabular}\\ 
\cline{1-1} \cline{3-6} 



\cite{8851217} &  &\begin{tabular}[c]{@{}c@{}}A NOMA-dynamic\\TDMA scheme in bi-static\\ backscatter WPCN,\\ single antenna backscatter\\ devices and a multi-antenna\\ backscatter receiver \end{tabular} & \begin{tabular}[c]{@{}c@{}}Maxmin\\backscatter\\devices\\ throughput \end{tabular} & \begin{tabular}[c]{@{}c@{}}A suboptimal iterative\\ algorithm based on\\ BCD method \\and SCA technique \end{tabular} & \begin{tabular}[c]{@{}c@{}} The sum-throughput performance\\ of the proposed scheme outperforms\\ the traditional TDMA scheme;\\ because of the radio-resource\\ sharing capability in NOMA systems\end{tabular}\\
\cline{1-1} \cline{3-6}

\cite{8962090} &  &\begin{tabular}[c]{@{}c@{}}A NOMA-dynamic \\ TDMA scheme in FD\\ symbiotic radio network,\\ single antenna backscatter\\ devices and a two-antenna\\ full-duplex access point\end{tabular} & \begin{tabular}[c]{@{}c@{}} Maxmin\\backscatter\\devices\\ throughput \end{tabular} & \begin{tabular}[c]{@{}c@{}}A suboptimal iterative\\ algorithm based on\\ BCD method \\and SCA technique \end{tabular} & \begin{tabular}[c]{@{}c@{}} The addition of NOMA to \\dynamic TDMA enhanced \\overall throughput and BD fairness \\compared to pure dynamic TDMA\end{tabular}\\
\hline

\cite{zhu2017nonSat} & \multirow{16}{*}{\begin{tabular}[c]{@{}c@{}} Integrated\\ Terrestrial-\\Satellite\\ Networks \end{tabular}} & \begin{tabular}[c]{@{}c@{}}A downlink integrated\\ terrestrial-satellite\\NOMA system,\\ single antenna users \end{tabular} & \begin{tabular}[c]{@{}c@{}}Max. system\\ capacity \end{tabular} & \begin{tabular}[c]{@{}c@{}}Hungarian method\\ and SCA technique \end{tabular} & \begin{tabular}[c]{@{}c@{}} The proposed integrated \\network can achieve higher \\system capacity and can serve\\ more users as compared with\\ the case of no satellite  \end{tabular}\\ 
\cline{1-1} \cline{3-6} 

\cite{8642812} &  &\begin{tabular}[c]{@{}c@{}}A downlink mmWave \\integrated satellite-\\terrestrial NOMA system,\\ single antenna users \end{tabular} & Max. sum rate & \begin{tabular}[c]{@{}c@{}}S-procedure and Taylor \\expansion as well as \\an iterative penalty\\ function (IPF) algorithm \end{tabular} & \begin{tabular}[c]{@{}c@{}}The sum-rate performance of\\ the proposed IPF beamforming\\ NOMA scheme outperforms both \\the fixed beamforming NOMA \\ and the traditional OMA schemes \end{tabular}\\
\cline{1-1} \cline{3-6}



\cite{8951059} & &\begin{tabular}[c]{@{}c@{}} A hybrid satellite/UAV\\ terrestrial NOMA system,\\ the UAV acts as a HD-DF \\relay, single antenna nodes \end{tabular} & Max. sum rate & \begin{tabular}[c]{@{}c@{}} First order\\ optimality conditions \end{tabular} & \begin{tabular}[c]{@{}c@{}} The sum-rate performance of the\\ proposed optimized UAV-location\\ NOMA scheme outperforms an\\ equivalent scheme with random\\ UAV deployment \end{tabular}\\ 
\cline{1-1} \cline{3-6}

\cite{8892479} & &\begin{tabular}[c]{@{}c@{}} A multi-beam satellite-\\ground Industrial IoT\\ that employs NOMA,\\ operating in Ka-band, \\single antenna nodes\end{tabular} & Max. sum rate  &  \begin{tabular}[c]{@{}c@{}}  KKT optimality\\conditions \end{tabular} & \begin{tabular}[c]{@{}c@{}} The proposed NOMA integrated \\scheme achieves significantly \\higher rates than OMA as all\\ nodes use the whole
frequency\\ band to transmit \end{tabular}\\ 
 \hline
 

\cite{8761991} & \multirow{14}{*}{\begin{tabular}[c]{@{}c@{}} Mobile Edge\\ Computing\\ (MEC) \\and Edge \\ Caching \end{tabular}} &\begin{tabular}[c]{@{}c@{}}A downlink NOMA-\\enabled Fog radio\\ access network,\\single antenna nodes \end{tabular} & \begin{tabular}[c]{@{}c@{}}Max. weighted \\sum rate \end{tabular} & \begin{tabular}[c]{@{}c@{}}A suboptimal solution: \\ many-to-one \\matching algorithm\\ and SCA technique\end{tabular} & \begin{tabular}[c]{@{}c@{}} The WSR performance of the\\ proposed NOMA scheme\\ outperforms OMA scheme \end{tabular}\\
\cline{1-1} \cline{3-6} 


\cite{8891423} & & Similar to~\cite{8761991} & \begin{tabular}[c]{@{}c@{}}Max. weighted \\sum rate \end{tabular} & \begin{tabular}[c]{@{}c@{}} The optimal solution: \\monotonic optimization\\ and polyblock \\approximation method \end{tabular} & \begin{tabular}[c]{@{}c@{}}The WSR performance of the\\ proposed optimal NOMA scheme\\ outperforms the suboptimal NOMA\\ scheme~\cite{8761991} and OMA scheme \end{tabular}\\
\cline{1-1} \cline{3-6} 


\cite{8951269} & &  \begin{tabular}[c]{@{}c@{}}A user-assisted HD \\cooperative MEC-\\enabled NOMA system, \\ single antenna nodes\end{tabular} & \begin{tabular}[c]{@{}c@{}}Max. sum \\offloaded\\ data rate\end{tabular} & \begin{tabular}[c]{@{}c@{}}Some mathematical\\ simplification methods\end{tabular} & \begin{tabular}[c]{@{}c@{}}The average sum-rate performance\\ of the proposed scheme outperforms\\ both NOMA scheme without \\cooperation and TDMA scheme \end{tabular}\\
\cline{1-1} \cline{3-6} 

\cite{8675357} &  & \begin{tabular}[c]{@{}c@{}}A downlink cache-aided\\ NOMA scheme with\\ one BS + two users,\\ single antenna nodes \end{tabular} & \begin{tabular}[c]{@{}c@{}}Max. sum rate \\and min. file \\delivery time \end{tabular} & \begin{tabular}[c]{@{}c@{}}Bisection search and \\ an analytical method \\based on the optimality\\ conditions \end{tabular} & \begin{tabular}[c]{@{}c@{}} The proposed cache-aided NOMA\\ with UE side-caching outperforms\\ both NOMA without caching and\\ cache-aided OMA schemes in term \\of sum rate and file delivery time\end{tabular}\\ 
\hline


\cite{mu2019exploiting} & \multirow{7}{*}{\begin{tabular}[c]{@{}c@{}} Intelligent\\ Reflecting \\ Surfaces \\ (IRS) \end{tabular}} & \begin{tabular}[c]{@{}c@{}}A downlink MISO\\ IRS-NOMA system,\\ One BS+multiple passive \\ reflecting
elements+ \\ multiple users \end{tabular} & Max. sum rate & \begin{tabular}[c]{@{}c@{}}SCA technique and\\ sequential rank-one\\ constraint relaxation\\ approach \end{tabular} & \begin{tabular}[c]{@{}c@{}}The sum-rate performance of the\\ proposed IRS-NOMA system \\outperforms both the conventional \\system without the IRS and \\the IRS-OMA system  \end{tabular}\\ 
\cline{1-1} \cline{3-6} 

\cite{yang2019intelligent} &  &\begin{tabular}[c]{@{}c@{}}A downlink SISO/MISO\\ IRS-NOMA system,\\ One BS+multiple passive \\ reflecting
elements+ \\ multiple users \end{tabular} & Maxmin SINR & \begin{tabular}[c]{@{}c@{}}BCD method and\\ SDR technique \end{tabular} & \begin{tabular}[c]{@{}c@{}}The sum-rate performance of the\\ proposed IRS-NOMA outperforms\\ three benchmark schemes including\\ the traditional NOMA, IRS-OMA\\ and traditional OMA \end{tabular}\\
\hline 

\cite{8604904} & Underwater & \begin{tabular}[c]{@{}c@{}}A relay-aided FD \\ underwater acoustic\\ NOMA system,\\single antenna nodes \end{tabular} & Max. sum rate & \begin{tabular}[c]{@{}c@{}}A low-complexity\\ iterative algorithm \end{tabular} & \begin{tabular}[c]{@{}c@{}}The sum-rate performance of the\\ proposed relay-aided FD-NOMA\\ (R-FD-NOMA) scheme (i) always\\ outperforms the non-relay aided FD-\\NOMA scheme. (ii) is inferior to the\\ relay-aided HD-OMA scheme at low\\ SIC efficiency\end{tabular}\\ 
\hline  

\end{tabular}%
}
\end{table*}

\subsection{Rate-optimal NOMA with Other Enabling Technologies} \label{sec:Other_existing}

In this subsection, we survey the integration of PD-NOMA scheme with BackCom, IRS, mobile edge computing and edge caching, integrated terrestrial-satellite networks, and underwater communications. A summary of the up-to-date work on the rate-optimal PD-NOMA scheme with those enabling schemes and technologies is provided in Table~\ref{tab:otherNoma}.

\subsubsection{Backscatter Communications (BackCom)}

As described in Section~\ref{sec:backscatter}, BackCom is a technology that is ideally suited for low-cost IoT devices, as they do not need to generate their own power for transmission when they act as backscatter devices (BD's). NOMA can then potentially be a great enabler to BackCom to support a large number of BD's through power domain multiplexing with different backscattered power levels~\cite{guoYanikomeroglu2018BackComNOMA}. In this way, BackCom-NOMA systems can help meet both the low-cost and massive connectivity needs of IoT systems. Further, NOMA-enhanced backscatter symbiotic cellular and IoT networks are a promising B5G deployment to meet both the cellular and IoT use-cases. In such deployments, the BD's transmit their signal on top of incident cellular signals~\cite{zhang2019Backscatter}. A transmission protocol that incorporates both NOMA and TDMA has been proposed in symbiotic cellular-IoT systems in~\cite{8962090} and non-symbiotic systems in~\cite{8851217, lyu2017optimal}. Rate optimization problems aim to maximize the minimum throughput among BDs~\cite{8851217} or the overall sum throughput of all the BDs~\cite{lyu2017optimal}. In such a setting, the power reflection coefficients from the BDs need to be selected as opposed to the typical PA coefficients as in other NOMA-enabled systems. It is shown in these works that in symbiotic cellular-IoT systems or any such setting where the BD's can transmit their signal without first harvesting the energy from the source, by jointly optimizing the power reflection coefficients as well as the BD's time allocation slices, a NOMA-enabled BackCom can achieve higher rates than OMA, e.g., pure TDMA. However, if the BD's operate in a harvest-then-transmit (HTT) mode by harvesting energy from the power source before transmitting their information using the harvested energy, it is shown in~\cite{lyu2017optimal} that the pure TDMA scheme outperforms NOMA as it can harvest more energy. This brings the interesting design challenges in BackCom-NOMA systems of how much power from the source the BDs can use and if they are relying on HTT mode, ensuring that NOMA protocols allow the end-users sufficient time allocation to harvest the required energy. 

\begin{itemize}[label=$\blacksquare$]
\item Lessons Learnt
    \begin{itemize}[label=$\bullet$]
    \item The massive machine type connectivity requirements of 5G and beyond communications systems typically require serving a very large number of devices with low-cost solutions. BackCom-NOMA systems are generating interest in the literature as a large number of low-cost backscatter  devices can be served in a single orthogonal resource through power domain multiplexing, using NOMA. 
    \end{itemize}
\end{itemize}

\subsubsection{Intelligent Reflecting Surfaces (IRS)}
\label{Subsection: body-Intelligent Reflecting Surfaces (IRS)}

The addition of NOMA to the IRS networks described in Section~\ref{sec:irs_bkgd} has recently been proposed in the literature~\cite{ding2019simple}, where the ability of the IRS to align the user's channel vectors is shown to be a great enabler for NOMA cluster formation. Similar to traditional MIMO-NOMA settings where BF techniques are used to create clusters of users that are served through PD-NOMA techniques, the passive phase shifters in IRS systems add an additional degree of flexibility to steer the transmitted signals in the direction of the NOMA user clusters. In two recent works~\cite{mu2019exploiting} and~\cite{yang2019intelligent}, the active BF at the BS and the passive BF at the IRS is optimized to achieve the maximum sum rate for the users in the system and demonstrate a superior performance to OMA-IRS systems. The joint optimization of both the active BF vectors at the BS and the passive reflectors at the IRS poses considerable design challenge as it needs to consider several constraints such as the BS transmit power, the IRS phase-shift constraints along with other users QoS and NOMA SINR constraints. Additionally, the user ordering becomes a challenge in these settings as it depends on the effective channel gain computed after multiplying the BF weights from both the active and passive BF, from the BS and IRS respectively.

\begin{itemize}[label=$\blacksquare$]
\item Lessons Learnt
    \begin{itemize}[label=$\bullet$]
    \item The addition of NOMA to IRS to form IRS-NOMA systems, offers spectral efficiency gains from the enhanced NOMA cluster formation possibilities through active and passive BF techniques. However, this poses additional design challenges from more complicated user orderings to additional constraints from the passive reflectors that need to be considered in rate optimization problems.
    \end{itemize}
\end{itemize}

\subsubsection{Mobile Edge Computing (MEC) and Edge Caching} \label{Subsection: body-MEC}

IoT devices usually have restricted memory and computational capabilities. Hence, offloading tasks in computational-intensive applications to a centralized location (such as base station) can be of great interest to maintain the IoT-enabled network. The transfer of these tasks to the MEC servers still has many challenges. The first one is that a large number of IoT devices can share the same MEC servers;  therefore, ensuring the availability of computational resources in MEC servers that meet the processing requirements of the tasks is not an easy job. Secondly, the IoT devices have limited battery life, the computation of tasks locally on the device or the transfer of data to a MEC server consume energy that could lead to quick depletion of the battery. Subsequently, the optimization of resource allocation in MEC-based networks is essential in real-time applications to ensure timely completion of the computational tasks and long-lasting use of the batteries of the IoT devices. In~\cite{8761991}, a resource allocation problem in a downlink NOMA-enabled heterogeneous cloud radio access network (H-CRAN) that maximizes the weighted sum rate of NOMA users has been proposed. Later an extension to~\cite{8761991} that jointly optimizes the resource allocation and power allocation has been analyzed in~\cite{8891423}. In both works employing NOMA scheme achieve better system weighted sum-rate performance as compared to equivalent systems that employ the OMA scheme; since NOMA allows multiple users to access the same subcarrier simultaneously. Both works have employed multiple fog APs/remote radio heads (RRHs) while enabling the reuse of the same subcarriers among the fog APs and between the fog APs and the RRHs. Hence, NOMA superiority comes at the expense of introducing (i) intra-cell interference (co-tier interference) among different fog APs' users and (ii) inter-cell interference (cross-tier interference) between the fog APs' users and the RRHs' users that caused by frequency reusing. Subsequently, both works utilized a mechanism to control this interference. This mechanism contributes in increasing the system complexity. 

In~\cite{8951269}, a sum-rate maximization problem for offloaded data that jointly optimizes the transmit power, the time allocation, and the task partition for users in a cooperative user-assisted MEC-enabled NOMA system has been proposed.  The proposed system contains three nodes, namely, user, helper, and an AP integrated with a MEC server. The user can offload tasks to the helper and AP simultaneously using NOMA to exploit the idle computational resources in the helper. The helper can only offload tasks to AP and located between the user and the AP. Through simulation, the author demonstrated that the proposed scheme can achieve higher sum-rate performance for the offloaded data compared to the equivalent NOMA scheme without cooperation and TDMA scheme no matter where the physical location of the helper.

Conventional edge cashing assumes that the content can be pushed in off-peak hours in an error-free manner~\cite{8368286}. However, this assumption is valid only if the content files vary slowly (e.g., windows software updates). On the other hand, if content files vary quickly (e.g., trending news), then the edge caches need to be updated during on-peak hours. To efficiently facilitate pushing content during on-peak hours, Ding~\textit{et al.}~\cite{8368286} proposed two NOMA assisted caching strategies, namely, \textit{push-then-deliver} strategy and \textit{push-and-deliver} strategy. In the former strategy, a short time interval is reserved for content pushing during on-peak hours, and the NOMA scheme is utilized to push more content files to the local caching servers as well as to provide more connections in the content delivery stage as compared to OMA scheme. While the latter strategy occurs when a user request cannot be found in the local content servers. Subsequently, the user needs to be served directly by the BS and not by the local content servers. In that case, the BS fetches the requested content file from the back-haul and utilizes the NOMA scheme in order to both serve the user directly and push this new content file to the local content servers through one transmission.  Consequently, this strategy increases the frequency of updating the local content servers. Another emerging application area of caching and NOMA was proposed by Xiang~\textit{et al.}~\cite{8675357}. In~\cite{8675357}, the cached data at the users were considered as side information and exploited for interference cancellation in NOMA transmission.

\begin{itemize}[label=$\blacksquare$]
\item Lessons Learnt
   \begin{itemize}[label=$\bullet$]
    \item As mentioned in Section~\ref{Sec: MEC bkgd} and in Section~\ref{Subsection: body-MEC}, the main goal in MEC-enabled networks with latency-sensitive applications is to support efficient task offloading for low-cost users. Hence, designing a suitable multiple-access scheme is crucial in such networks. With OMA schemes, the users sometimes need to wait for an available orthogonal resource block to grant access which is not acceptable in latency-sensitive applications. In contrast, NOMA schemes offer more flexible user scheduling opportunities while accommodating more users in each resource block. Hence, NOMA can be considered as an enabler for IoT-enabled networks at the expense of increasing the system complexity.
    \end{itemize}
\end{itemize}

\subsubsection{Integrated Terrestrial-Satellite Networks} 
NOMA is used as an enabler in integrated terrestrial-satellite networks to bring spectral efficiency gains that complement the ubiquitous coverage gains from satellite communications. For example, in~\cite{8892479}, a multibeam satellite Industrial IoT (IIoT) is designed to provide coverage to users in remote areas. The addition of NOMA is shown to improve the transmission rate of IIoT under the limited spectrum resources.  In~\cite{zhu2017nonSat}, the ground BS and satellites co-operate to jointly serve a set of users, but only the terrestrial network adopts NOMA as a multiple access scheme. In this work, the authors seek to complement the spectral efficiency gains of a NOMA-based ground BS with an integrated satellite communication system for enhanced coverage. From a rate optimization problem perspective, a user can camp to either a terrestrial or satellite network based on the channel conditions and this adds an additional degree of freedom to the design.

Operating in the same frequency band, the ground and satellite networks will cause interference to each other and is an important design challenge in such networks. Managing the interference caused by a large number of users in a NOMA ground cluster along with the interference caused by satellite communication plus the other ground NOMA clusters can be a significant challenge that can present different tradeoffs between spectral efficiency gains and complexity introduced. In~\cite{zhu2017nonSat}, a joint iteration algorithm for BF vectors and PA coefficients of both networks is proposed  to control the level of interference from one network to another.  On the other hand, in~\cite{8642812}, the satellite uses multicast communication to serve the earth BSs, which in turn serve individual users through NOMA. A sum-rate optimization problem is then formed with per-antenna power constraints and QoS constraints at both the earth BS and end-users considered.

\begin{itemize}[label=$\blacksquare$]
\item Lessons Learnt
    \begin{itemize}[label=$\bullet$]
    \item In integrated terrestrial-satellite communications systems that adopt NOMA, the spectral efficiency gains of NOMA can be complemented with the ubiquitous coverage gains of satellite communications. NOMA typically is used by the ground BS to ground users with good channel conditions through NOMA clusters and offloading users with weak channels to satellite communications. However, when operating in the same frequency band, significant design challenges emerge from mitigating the interference between NOMA clusters on the ground and the users served through satellite beams.
    \end{itemize}
\end{itemize}

\subsubsection{Underwater Communications}
\label{Subsection: body-Underwater Communication}

With the current trend of the internet-of-everything (IoE) that aims to provide everything electronic with an IP address and connect it to the internet, a recent development in underwater communications networks termed as internet-of-underwater-things (IoUT) has emerged~\cite{amoakoh2020wireless}. IoUT seek to enable the communications between different heterogeneous underwater nodes (e.g., submarines, autonomous underwater vehicles,  movable sensors, divers, tracked animals, etc.). In such a multiuser heterogeneous network, to serve multiple IoUT nodes simultaneously, selecting a suitable multiple access scheme is critical. For example, TDMA and FDMA schemes are not suitable for wireless underwater communications because of their long-time guards and limited bandwidth, respectively. On the other hand, the integration of NOMA in underwater wireless communications can improve the system spectral efficiency without additional resources for both acoustic~\cite{8604904} and wireless optical communications~\cite{9049718}. For underwater acoustic  communications~\cite{8604904}, a power allocation optimization problem that aims to maximize the system sum-rate in a relay-aided full-duplex underwater acoustic NOMA system has been proposed. A suboptimal solution has been obtained for the former optimization problem through a low-complexity iterative algorithm. It is shown, through numerical simulations, that the sum-rate performance of the proposed relay-aided full-duplex NOMA scheme outperforms both the non-relay full-duplex NOMA scheme and relay-aided half-duplex OMA scheme under efficient interference cancellation conditions. While, for underwater wireless optical communications~\cite{9049718}, an initial study that compared the performance of NOMA-enabled and OFDMA-enabled in the presence of the oceanic turbulence fading has been analyzed. This study revealed the superiority of the NOMA-enabled scheme in terms of data rate at the expanse of achieving lower coverage performance compared to the OFDMA-enabled scheme.

\begin{itemize}[label=$\blacksquare$]
\item Lessons Learnt
    \begin{itemize}[label=$\bullet$]
    \item In an underwater environment, simultaneous uplink and downlink communications are needed among different entities at the sea-level, in the shallow-water, at the deep-water, and at the ground-level under the sea. Relay-assisted FD-NOMA system can be an enabler for such a scenario as depicted in~\cite{8604904}. In~\cite{8604904}, through employing the FD-NOMA scheme, multiple downlink and uplink transmissions occur simultaneously using the same radio resources. To manage the interference between the transmissions, successive-interference and self-interference cancellation techniques are adopted at the cost of increasing the system computational complexity.
    \end{itemize}
\end{itemize}

\begin{figure*}[ht!]
\centering
\includegraphics[width=1.0\textwidth]{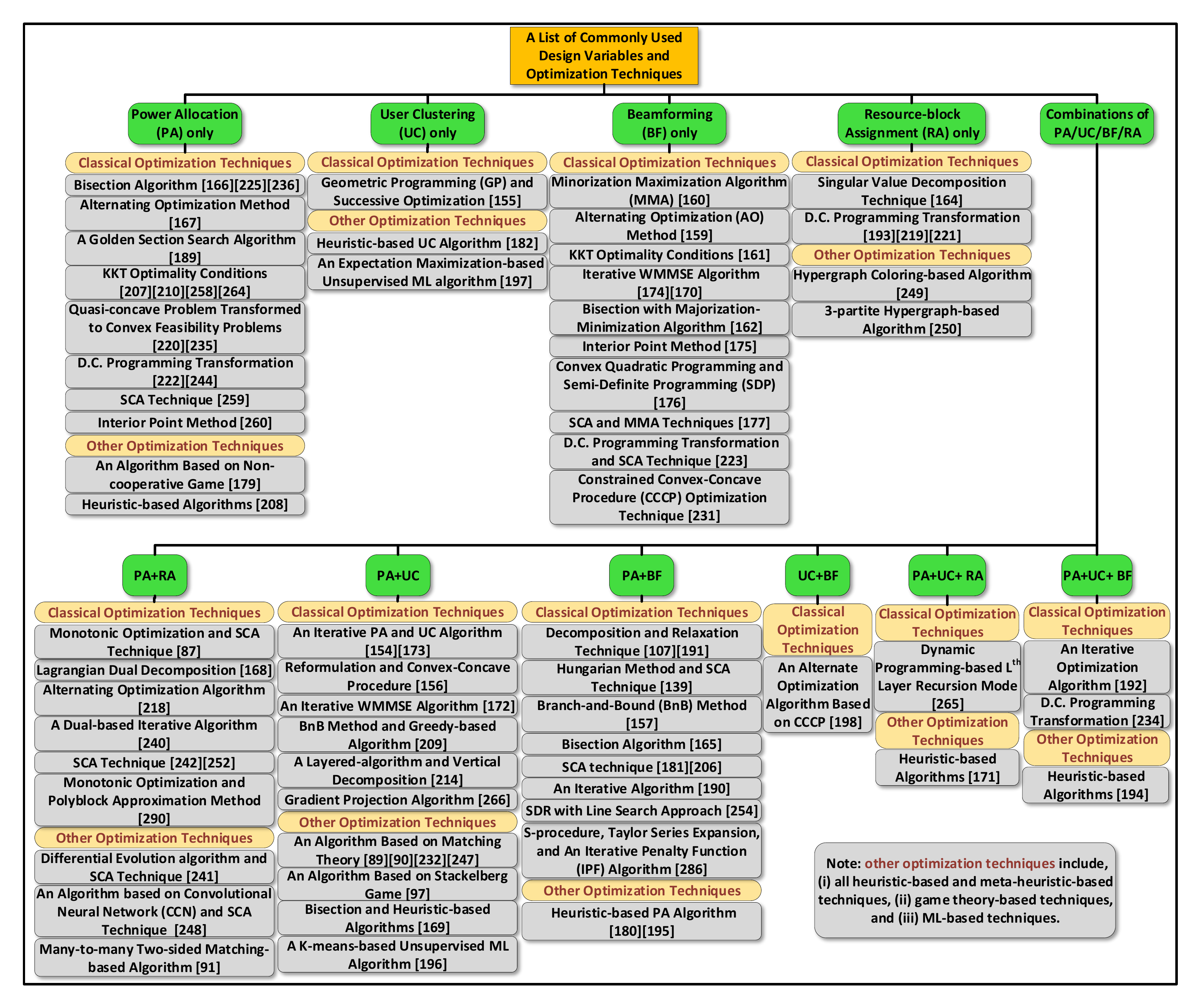}
\caption{A list of the common design variables and the different optimization techniques used in rate-optimal NOMA-enabled schemes and technologies for future wireless networks.} 
\label{fig:lessons_learnt_tree}
\end{figure*}

\begin{figure*}[ht!]
\centering
\includegraphics[width=1.0\textwidth]{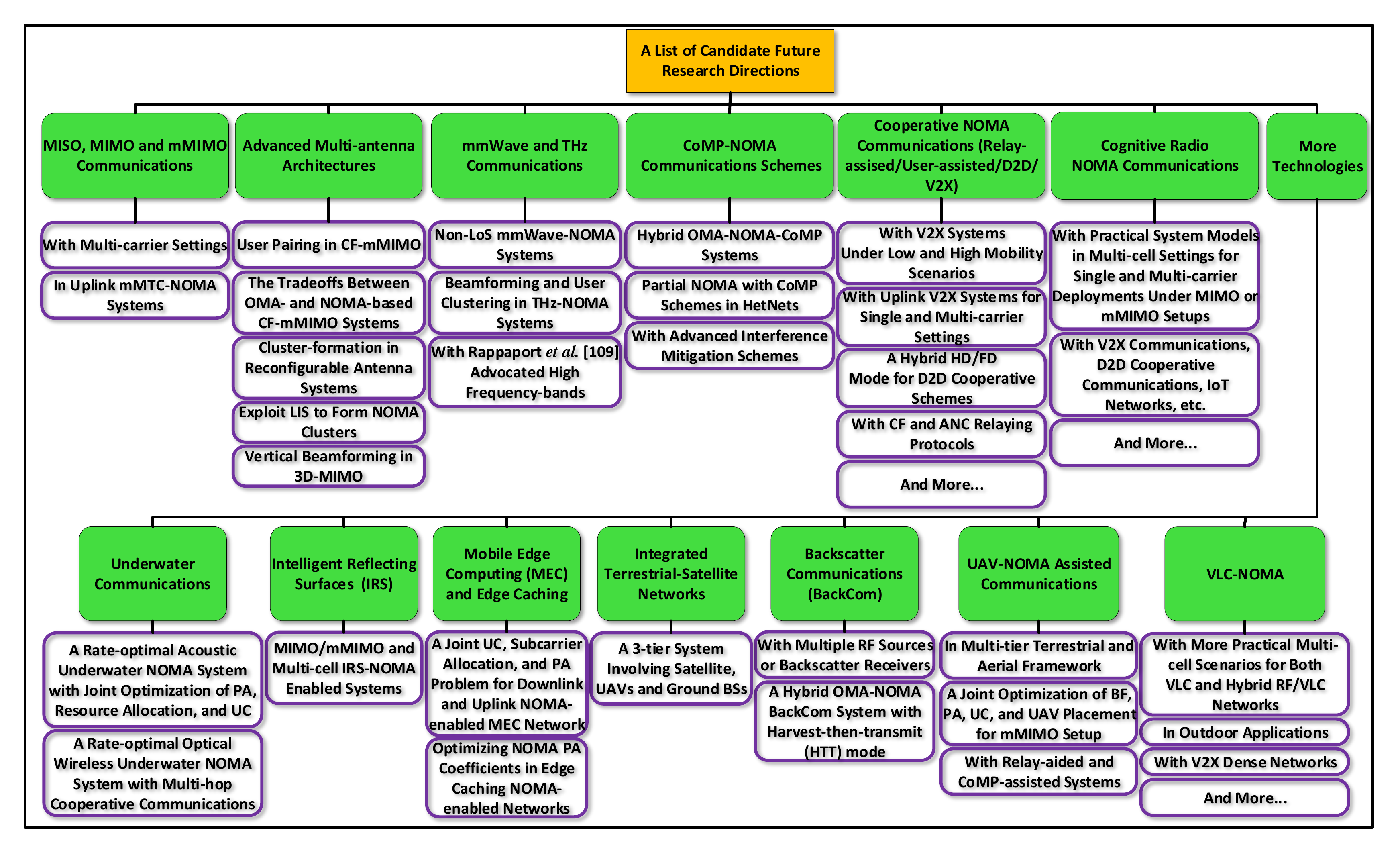}
\caption{A list of candidate future research directions for NOMA-enabled schemes and technologies that are discussed in detail throughout this section.} 
\label{fig:Future Research Directions}
\end{figure*}

\subsection{Common Lessons Learnt}

The general main learnt lessons are listed as follows:
\begin{itemize}

\item The rate optimization problem of NOMA schemes is a joint optimization problem that typically involves optimal power allocation, optimal user-clustering/user-pairing, optimal BF weights design in multi-antenna systems, and possibly considers optimal channel allocation. This would typically lead to a non-convex and combinatorial non-polynomial (NP)-hard optimization problems and finding the optimal solution in most cases needs exhaustive search which is not practical. Consequently, researchers have adopted different approaches to circumvent the difficulty associated with these problems through optimization techniques as listed in Fig.~\ref{fig:lessons_learnt_tree} were the following conclusions may be deduced:
For power allocation only (where a certain suboptimal user clustering is utilized and no channel allocation is considered), the optimization problem is typically non-convex but not combinatorial and the classical relaxation-optimization methods, for example, successive convex approximation (SCA)~\cite{tahira2019optimization}, difference of convex (D.C.) programming~\cite{agarwal2018performance,8438868}, or even using game theory as a way to find a near-optimum solution while relaxing the original non-convex problem, for example, to do this in~\cite{8719970} the PA non-convex problem is relaxed and modeled as a non-cooperative competitive game. Differently, for user clustering and power allocation (no channel allocation is considered), as UC is included the problem becomes typically combinatorial and more computationally-involved optimization methods are been utilized, for example, gradient projection algorithm~\cite{zhang2017user}, some iterative PA and UC algorithms~\cite{kimy2013milcom,7974731}, or even algorithms that are based on matching theory~\cite{cui2017user,fan2019channel,8713856,8398200}, game theory~\cite{8684765}, heuristic search~\cite{liu2016fairness}, and machine learning~\cite{cui2018unsupervised}. A similar observation applies when resource allocation is considered with combinations of PA, BF, and UC as depicted in Fig.~\ref{fig:lessons_learnt_tree}. Additionally, the rate optimization problem of NOMA schemes might include other considerations that indirectly affect system rate-performance, such as the users' classification in CoMP schemes~\cite{ali2018downlink}, optimizing the number of admitted users~\cite{7974731} and developing efficient BS-antenna selection algorithm~\cite{liu2017_3DRA} in multi-antenna NOMA systems, UAV altitude/location/position/trajectory optimization in unmanned aerial vehicles assisted communications~\cite{sohail2018non,sun2019cyclical,nasir2019uav, 8685130,8848428,zhao2019joint}, optimizing SIC decoding order for users; particularly in multi-cell NOMA-HetNets where the near user is not necessary a user with good channel quality~\cite{8387207}.

\item For NOMA-enabled systems operating in higher frequency ranges, e.g., mmWave communications, the nature of the wave propagation means that the waves do not travel as far, the LoS path wave dominates and multipath propagation is not as much as at lower frequencies. The short propagation distance means that large scale antenna arrays are often used for beamforming gains. All these factors influence the design of NOMA-enabled systems at higher frequencies. The lack of strength in multipath signals implies there is a stronger correlation among users' channels. This is a condition favorable to NOMA, as it aids user clustering. Researchers exploit this characteristic for user pairing in mmWave systems, along with the large antenna arrays that aid in placing users in beams, i.e., clusters.

\item ML has emerged as an approach to solve aspects of the rate optimization problems in NOMA-enabled systems. For example, in mmWave systems, exploiting the high correlation amongst users' channels with LoS dominated communication links, unsupervised clustering ML algorithms have been used to tackle the user selection sub-problem~\cite{cui2018unsupervised}. Neural networks have also been employed in D2D cooperative networks for resource allocation~\cite{8906057}. A detailed analysis of these and other ML approaches to rate optimization problems is provided later as part of the future research directions in Section~\ref{sec: future ML_NOMA}.
 
\item For the integration of PD-NOMA with more than one of the enabling technologies or schemes, the existing literature has considered (i) the integration of CoMP-NOMA with user-assisted uplink cooperative communication to enable cell-edge users to transmit signals at low power and to achieve high sum-rate system performance~\cite{8675425}, (ii) UAV-NOMA and integrated terrestrial-satellite communications~\cite{8951059} are two closely linked technologies, where the UAV or satellite can act as a BS to complement the traditional ground BS's. When NOMA is integrated to a UAV acting as a BS or a high altitude satellite, the propagation is largely dominated by the LoS path. This makes co-located users easy to group into clusters and serve them using PD-NOMA. In this regard, these technologies share a similarity with mmWave and above communications, as they are also LoS dominated and make co-located users good candidates for user clustering in NOMA pairs. The interested reader can look at other noticeable integrations of D2D underlaying NOMA with UAV-assisted communications~\cite{8438868,8624380} as well as full-duplex cooperative communication with (i) cognitive radio network~\cite{mohammadi2018beamforming}, (ii) backscatter communication system~\cite{8962090}, and (iii) underwater acoustic network~\cite{8604904}.  

\end{itemize}

\section{Future Research Directions}
\label{Section: Conclusions and Open Research Problems}

The  survey documented  in this article  has revealed performance gains and/or trade-offs that motivate further investigations of the aforementioned NOMA-enabled schemes and technologies. In the following paragraphs, we highlight some of the possible directions of future research on the integration of PD-NOMA with MISO, MIMO, mMIMO, advanced multi-antenna architectures, mmWave and THz communications, CoMP, cooperative communications, cognitive radio, VLC, UAV communications, BackCom, IRS, MEC and edge caching, integrated terrestrial-satellite networks, underwater communications, and their combinations. To make this section easier to traverse, a list of candidate future research directions is provided in Fig.~\ref{fig:Future Research Directions}.

\subsection{MISO, MIMO, and mMIMO Communications} \label{sec:future MimoNoma}

As discussed in the survey sections on the integration of MIMO related technologies with NOMA, a lot of work already exists in this area including in mMIMO-NOMA systems. However, there are still a few interesting research directions to pursue in these fields that we outline in this section, followed by a more detailed discussion of future work in NOMA integrated with the other advanced multi-antenna architectures operating in the higher end of the mmWave spectrum as well as THz frequencies in subsequent sections. 

For the conventional MIMO-NOMA systems, investigating rate optimization schemes in multi-carrier settings like OFDM requires further consideration as the user ordering will be different across the different blocks of OFDM sub-carriers, depending on the coherence bandwidth of the channel. Considering channel assignment, along with PA and US adds a significant level of complexity to the rate-optimization problem, e.g.,~\cite{he2019RL, 8653277}. Secondly, uplink MIMO-NOMA communications systems offer interesting research directions when the power budget of the UL transmitters is considered in the problem, especially the low-cost ones typical of the massive machine type connectivity scenarios in 5G~\cite{NR_5G}. As a result, the PA and BF strategy has to consider the end-user capabilities in the problem, particularly in multi-carrier settings like OFDM, where the peak-to-average-power-ratio (PAPR) constraints come into play~\cite{rajasekaran19PAPR}.

\subsection{Advanced Multi-antenna Architectures} 
\label{sec:future mMIMO and Advanced Multi-antenna Architectures}

Compared to the integration of MIMO with NOMA which is a well-studied area, the combination of NOMA with other advanced multi-antenna architectures is still in its infancy with a large room for contributions. In what follows, we outline some promising research directions for the integration of NOMA with each of these architectures. 

\subsubsection{Cell-free mMIMO (CF-mMIMO)}
\label{sec:future Cell-free mMIMO}
As discussed in Section~\ref{sec:body Cell-free mMIMO}, the integration of NOMA and CF-mMIMO is in its infancy~\cite{8368267, 8756881} and much work remains to be done in this area. The user pairing problem brings some unique flavors in such CF-mMIMO-NOMA settings that are different from other NOMA settings. In particular, due to the lack of definition of a cell, the users can be served by a large number of geographically distributed access points (AP's). To identify good NOMA pairs, i.e., correlated in the angle domain but sufficiently different in channel gains, can be quite challenging in such settings. On the contrary, compared to a regular OMA-based CF-mMIMO setting, NOMA brings tremendous advantages as it avoids the need to form unique beams for correlated users with the distributed AP's, instead allowing them to be served through PD-NOMA in one cluster. This tradeoff between the advantages of OMA and NOMA in CF-mMIMO, in terms of complexity and spectral efficiency gains, is an area of future research. 

\subsubsection{Reconfigurable Antenna Systems}
\label{sec:future Reconfigurable Antenna Systems}
The advantage with reconfigurable antennas is that it allows for changing physical attributes of the antennas to favor the desired transmit beamforming. The integration of NOMA to such a setting means that the reconfigurable properties of the antennas can be exploited for desired cluster formations. As described in Section~\ref{sec:body Reconfigurable Antenna Systems}, two flavors of these techniques were proposed in~\cite{8756960, 8935164}. However, the question of how best to exploit the reconfigurable properties of the antennas to best suit NOMA cluster formation still has lots of room for contribution in the literature. For example, when a large number of users are clustered in an area using mmWave systems with largely LoS paths, reconfigurable antennas can be used to form two clusters with different polarizations, to even out the distribution of users between the two clusters. 

\subsubsection{Large Intelligent Surfaces}
\label{sec:future Large Intelligent Surfaces}
The large intelligent surfaces (LIS), also described in Section~\ref{sec: bkd advMimoNoma} can be combined with NOMA to further enhance the spectral efficiency. In traditional LIS systems, despite the use of a large surface full of transmitting antennas, beams that are narrow enough to distinguish the different users are challenging to form~\cite{Hu2018activeLIS}. This is particularly true for massive connectivity scenarios. A NOMA-LIS system that groups clusters of users through the LIS and then serves the users within a cluster through NOMA is an area worthy of future work. The LIS by nature is an expensive solution, so processors capable of a large amount of signal processing can be expected. As a result, it is a promising solution in the uplink as it might be feasible to combine a large number of users in a cluster. The large number of antennas on the surface can be exploited for advanced inter-cluster interference mitigation purposes. 

\subsubsection{3-D MIMO}
\label{sec:future 3-D MIMO}
A system that integrates 3-D MIMO with NOMA can solve the typical challenges 3-D MIMO systems face on their own. To separate users on different floors of a building through vertical BF with a limited antenna array size, the challenge is to form narrow enough beams that keep the inter-beam interference under control~\cite{zhang20183dmimo}. The introduction of PD-NOMA to such a setting can help alleviate this problem, as users within a beam can be separated using NOMA, thereby allowing for much wider beams to be formed without affecting the overall sum-rate.

\subsection{mmWave and THz Communications}
\label{sec:future mmwave + THz Communications}

Most of the existing work in mmWave communications is studied assuming the one-path model where the LoS path or non-dominant LoS path dominates. The performance of the proposed user clustering and power allocation schemes in more generic mmWave models that might apply in scenarios where a LoS path is absent is a possible future work. As discussed in the literature survey on mmWave-NOMA, due to the high correlation in users' channels in the angle domain, ML techniques that identify clusters of highly correlated users is an emerging trend. There is still lots of room for more innovative solutions that exploit these features of mmWave channels for the benefit of NOMA cluster formation, through both ML and traditional optimization.

Further, since B5G systems are likely to use frequency bands above 100 GHz~\cite{8732419}, where the main challenge is the high path loss experienced during signal propagation. Through the use of extremely large antenna arrays, beamforming solutions help increase the coverage area. The integration of NOMA serves as a great complement to such systems, as it helps cover a large set of users within such a beam. In this way, a large number of users can be supported at very high data rates, a key requirement on B5G networks. Particularly, since analog or hybrid BF is often employed in very large antenna systems to reduce the cost~\cite{xiao2019user}, the BS is only able to produce one beam at a time with analog BF. The beam and user selection problems in such a setting are promising research topics in the context of rate optimal NOMA-enabled schemes operating beyond 100 GHz with very large antenna arrays. 

In fact, the study by Rappaport~\textit{et al.} in~\cite{8732419} highlights some peculiarities in bands above 100 GHz, where some bands are prone to high attenuation, while some other bands suffer surprisingly little loss compared to sub-6 GHz bands. In~\cite{8732419}, the authors advocate that such bands above 100 GHz that suffer less loss and so can provide good coverage are candidates for deployment in high-speed 6G networks. The integration of NOMA to such bands is an area of future research work. 

\subsection{CoMP-NOMA Communications Schemes} \label{sec:futureCompNoma}

Rate optimization is an attractive topic for future research on CoMP-NOMA systems. This is particularly true as both schemes target enhancing the spectral efficiency of future dense wireless networks. As highlighted in Section~\ref{Subsection: background: CoMP}, the promising performance gains of joint transmission (please see Table~\ref{Table: CoMP-NOMA}) are limited in practice by the demanding requirements on the back-haul capacity and the tight synchronization among the coordinated BSs/APs. This fact motivates consideration of the coordinated beamforming version of CoMP-NOMA schemes for future investigation. Further, the performance trade-offs of hybrid CoMP-NOMA and CoMP-OMA schemes need further study, especially in heterogeneous ultra-dense networks. In particular, as reported in~\cite{zhang2018heterogeneous, noma-het-1}, CoMP-OMA schemes tend to outperform CoMP-NOMA schemes for high-density micro-BSs networks due to the high interference conditions. Such optimized hybrid schemes are attractive for the evolution of the current 4G and 5G OFDMA-based networks towards B5G networks. 
Another related interesting direction to investigate is the achievable rates of the recently proposed partial NOMA scheme~\cite{pnoma-1}, which allows a partial overlap of the multiplexed users' signals to reduce the interference among them, with CoMP schemes in heterogeneous networks.   

An interesting direction of research for reliability-demanding applications in future networks is to integrate other advanced interference mitigation schemes such as interference alignment schemes. These interestingly would allow a linear scaling of the sum rate with the number of users and can be implemented using linear trans-receive beamforming schemes~\cite{interference-alignment-1, interference-alignment-2, interference-alignment-4}, in CoMP-NOMA systems to reduce the inter-cluster interference, as in~\cite{8725542} for a basic multi-cell environment with two users per cell, especially in case of non-orthogonal channel assignment to these NOMA clusters and/or the interference at the non-CoMP users to further enhance the overall network throughput.

\subsection{Cooperative NOMA Communications}

The three adopted types of cooperation in cooperative NOMA communications systems in the literature discussed in Section~\ref{Cooperative- existing work}, are envisioned to play a prominent role in future wireless networks. For relay-assisted and user-assisted cooperative NOMA schemes, an area worthy of investigation is the adoption of rate-optimal schemes in V2X systems under low and high mobility scenarios to improve the spectral efficiency as well as to accommodate more vehicles over limited spectrum resources. Furthermore, few number of existing works have studied the rate-optimal cooperative V2X-NOMA systems in downlink setups, for example, the study by Chen~\textit{et al.}~\cite{chen2017performance}, while rate-optimal cooperative uplink V2X-NOMA systems with single and multi-carrier settings is still an unexplored area. The adoption of V2V cooperative NOMA schemes in V2X systems is highly promising and yet is not well explored. 

Other interesting open research problems can be listed as follows: i) The majority of the works in cooperative NOMA systems have investigated either half-duplex or full-duplex communications is a separate fashion. Only two works were found~\cite{chen2019optimal,liu2018hybrid} that  have studied the potential of dynamic switching between the half-duplex and the full-duplex modes in relay-assisted and user-assisted cooperation types, respectively. Studying the applicability of such a hybrid mode for D2D cooperative schemes is essential. ii) Several researchers have investigated DF and AF relaying protocols in cooperative NOMA schemes; however, other relaying protocols such as CF and analog network coding (ANC) have been only studied under specific NOMA settings. Particularly, CF relaying protocol with the NOMA scheme has been only explored by So and Sung~\cite{so2016improving} under relay broadcast channel and by Li~\textit{et al.}~\cite{li2018optimized} under an overlay cooperative-cognitive radio network. Studying CF, ANC, and possibly other relaying protocols for different settings and comparing their performance with its counterpart system models that adopted DF and AF relaying protocols are valuable to guide the research community to the optimal relaying protocol for each setting. iii) Liu~\textit{et al.}~\cite{liu2015band} and Nwankwo~\textit{et al.}~\cite{8115161} discussed in detail several SI mitigation technologies for in-band FD relaying. Investigating the rate performance of in-band FD-NOMA enabled systems under different SI mitigation technologies is an important avenue for future research. iv) Cooperative relay sharing NOMA scheme is an interesting topic of research. In this system model, two source-destination user-pairs share a dedicated FD relay between them. This comprises two source-user pairs that utilize the uplink-NOMA scheme to transmit signals to the relay, then the relay utilizes the superposition coding in the downlink-NOMA scheme to deliver the signals to its intended destination user-pairs. Such a system can reduce the network deployment cost and also the end-to-end delay. The investigation of the rate-optimal scheme for such setup needs to be conducted. v) Alsaba~\textit{et al.}~\cite{alsaba2018full} have investigated a full-duplex user-assisted cooperative NOMA scheme with perfect and imperfect SI cancellation cases. These cases can be extended to relay-assisted cooperative NOMA systems for different architectures.

\subsection{Cognitive Radio NOMA Communications}  

As illustrated in Table~\ref{Table: CR-NOMA},  rate-optimal CR-NOMA schemes  outperform the corresponding CR-OMA schemes for single-cell scenarios. More practical system models in multi-cell settings, with single and multi- carrier deployments, under MIMO or mMIMO setups, need to be explored to further enhance the network performance and its trade-offs. Furthermore, as stated in Section~\ref{Rate-optimal CR-NOMA}, rate-optimal CR-NOMA schemes have been adopted only in heterogeneous networks as well as with relay-assisted and user-assisted cooperative communications in order to achieve better spectrum utilization and to add multiple design dimensions to the previously mentioned networks. In addition to that, more research on the adoption of rate-optimal CR-NOMA schemes in other emerging communications systems such as V2X communications, D2D cooperative communications, internet of things networks, etc., is required. Also, in~\cite{li2018optimized}, a hybrid NOMA-OMA transmission protocol in an overlay cooperative-cognitive network has been proposed, and applying such a transmission scheme for other CR paradigms and architectures can be an interesting future research direction as such a scheme can offer better system performance than NOMA or OMA schemes separately. Moreover, in~\cite{8753517}, two uplink information transmission CR-NOMA schemes under the conditions of two different CR paradigms, namely underlay and overlay, have been proposed. Extending such uplink schemes to different CR architectures can be an interesting topic of research. 

\subsection{VLC-NOMA}

As evident in Table~\ref{Table: NOMA-VLC}, only a few works have considered the optimization of the achievable data rates in VLC-NOMA systems and networks. Moreover, most of them have considered the single-cell scenario which motivates future work for the more practical multi-cell scenarios for both VLC and hybrid RF/VLC networks along with the complexity analysis as compared to RF networks. This should be carried out for the different trans-receive arrangements and ICI mitigation techniques in multi-cell VLC systems such as fractional frequency reuse (FFR), the use of red-green-blue (RGB) LEDs, angle-diversity receivers (by adjusting the orientation of the PDs), CoMP, and ZF pre-coding or combinations of them~\cite{9062301}, especially for outdoor applications where channel conditions may change fast and more interference sources are present as compared to those in indoor applications. Also, the use of VLC in V2V and in general V2X communications~\cite{vlc-v2v-1, vlc-v2v-2}, to replace or complement the existing dedicated short-range communication (DSRC) scheme, has attained high data rates. This would enable future high-definition and real-time applications required for safe, secure, and energy-efficient better informed or autonomous driving that have appeared in 5G networks and are expected to be fully realized in 6G networks. However, weather, road conditions, background solar radiation, and other light sources tend to limit the reliability of long-range V2V VLC which motivates the adoption of NOMA schemes in dense V2V VLC networks~\cite{noma-v2v}. This will also enhance the spectral efficiency and guarantee user fairness. The low  SNR at the receiver, and changing channel conditions,  require the design of robust VLC-NOMA systems. This will  ensure the quality of SIC decoding, along with rate-maximizing user clustering, beamforming, and power allocation schemes, as well as the LED's placement and the optical receiver arrangement (receiver type, FoV size and orientation, optical filtering, etc.) in dense and dynamic vehicular networks. Furthermore, cooperative VLC-NOMA and CR-VLC-NOMA can be utilized to extend the communication range, enhance the spectral efficiency, and increase the achievable rates in these networks. An example would be the use of multi-hop relaying in platooning applications in NOMA-enabled V2V VLC networks.

\subsection{UAV-NOMA Assisted Communications}

As can be seen  from Table~\ref{tab:uavNoma} in Section~\ref{sec:litsuvUAV}, the combination of UAV and NOMA is a very new field with a large number of unexplored research directions. Most existing rate optimization schemes are in the context of one UAV acting as a flying BS and serving multiple users. In~\cite{zhao2019joint}, this model was extended to consider the interference caused to users served by a ground BS. However, as described in Section~\ref{sec:litsuvUAV} and evidenced by the tutorial in~\cite{mozaffari2019tutorial}, the UAV-NOMA system has multiple  ways of using the UAV and sum-rate optimization problems in each of these frameworks are promising directions of future research. As an example, in~\cite{mei2019uplink}, the use case with the UAV as a flying user is explored. The most interesting of these scenarios arguably is the introduction of NOMA to a multi-tier terrestrial and aerial framework involving ground BSs for long term deployment, LAP flying UAV BSs for flexible short-term deployment, and some HAP UAV BSs for medium-term deployment as described in~\cite{mozaffari2019tutorial}. Tackling the sum-rate optimization problem with NOMA included in such a complex system model is an area of future research. 

Further, the rate optimization schemes from other sections surveyed in this paper can be applied to a UAV-NOMA system with the additional flexibility of the UAV altitude and placement. For example, MIMO-NOMA and mMIMO systems can be enhanced to consider the UAV placement problem in them. The ground BSs, UAVs, or the users can all be equipped with multiple antennas, leading to beam design, power allocation, UAV placement, and user scheduling optimization problems. Similarly, cooperative schemes that involve relays can be adapted to have the UAV act as a relay. This gives additional degrees of freedom because the location of the relay can be optimized to be closer to the BS or to the user as required. As a relay link, the UAV could even move between the different hops of the transmission. This leads to an interesting power allocation coefficient design problems. CoMP-NOMA systems with UAV is also an important unexplored area, since the BS and UAV can co-operate to jointly set the design variables involved in a UAV-NOMA system that aims to maximize the sum rate. The UAV-NOMA model can also be applied to other B5G technologies, for example, the authors in~\cite{farajzadeh2019UAVNOMA} have applied the UAV-NOMA model to self-sustaining backscatter networks that do not require external power supply~\cite{lu2018backscatter}.

\subsection{Other Enabling Technologies} \label{sec:Other_future}

\subsubsection{Backscatter Communications (BackCom)}

Rate optimization schemes in BackCom networks is a new and attractive research area for future wireless networks. Similar to traditional multi-cell cellular networks, system models with multiple RF sources or backscatter receivers need to be considered. BDs will need to choose which source carrier wave to modulate their information on. Further, as reported in~\cite{lyu2017optimal}, OMA outperforms NOMA for BDs operating in harvest-then-transmit (HTT) mode. However, NOMA was shown to enhance the sum rate when the BDs are only transmitting on top of the emitted RF waves from the source. Hence, hybrid models where some BDs are in HTT mode and some are only modulating on top of carrier waves is an avenue for future research.

\subsubsection{Intelligent Reflecting Surfaces (IRS)}

As discussed in  Section~\ref{Subsection: body-Intelligent Reflecting Surfaces (IRS)}, recent works in the literature~\cite{mu2019exploiting} and~\cite{yang2019intelligent} studied rate optimization problems in IRS-NOMA settings. Both these works were conducted in simple single-cell SISO and MISO NOMA systems with IRS passive reflectors. Applying advanced multi-antenna settings, particularly multiple antennas at the receiver in  MIMO or even mMIMO settings adds additional challenges to the BF weight design and is therefore a possible future research direction. In particular, the impact of multiple antennas at the receiver on the NOMA user ordering problem as well as the cluster formation problem are of interest in an IRS-NOMA system.  Further, with multi-cell setups in such IRS-NOMA enabled systems, the impact of the inter-cluster interference between NOMA clusters of different cells caused by the common passive reflectors offers interesting design challenges that are worth exploring.

\subsubsection{Mobile Edge Computing (MEC) and Edge Caching}

The increasing growth of limited battery IoT devices in existing and future wireless networks has open the door for the integration of NOMA with MEC networks. Particularly, the NOMA scheme can achieve efficient communication by supporting massive connections and enhance the spectral efficiency of such networks. In~\cite{8761991}, a resource allocation
problem in a downlink NOMA-enabled MEC network that maximizes the weighted sum rate of NOMA users has been proposed. Later an extension to~\cite{8761991} that jointly optimizes the resource allocation and power allocation has been analyzed in~\cite{8891423}. The joint user clustering, resource allocation, and power allocation problem for downlink and uplink NOMA-enabled MEC network is an attractive topic for future research.

As mentioned in Section~\ref{Subsection: body-MEC}, NOMA assisted caching realizes the update of local content servers during on-peak periods through \textit{push-then-deliver} and \textit{push-and-deliver} strategies~\cite{8368286}. In~\cite{8368286}, the main performance criterion was the cache hit probability. Optimizing NOMA power allocation coefficients in order to increase the system content delivery rate and consequently enhance users' quality-of-experience is an interesting area of future research.

\subsubsection{Integrated Terrestrial-Satellite Networks}

The use of NOMA in satellite-based communications networks has strong appeal for B5G networks as it enables ubiquitous coverage with high spectral efficiency. In this regard, rate optimization problems for different flavors of users and system models can be considered for research. In the existing works, the earth BS co-operates with the satellite to support ground users, or alternatively, a multicast link is established between the satellite and earth BS. As illustrated in~\cite{zhu2017nonSat}, if the same frequency band is used for the satellite-BS and BS-users links, it leads to interference management challenges that are different from a multi-cell ground network. This framework can be extended with the UAV-based framework, where users can be supported via a 3-tier system involving the satellite, UAVs, and the ground BSs. NOMA can be integrated with all the three tiers of the system, but interference management becomes an even bigger challenge. Different frequency bands can be used for some links, e.g., the mmWave band for the satellite or UAV communication paths with good LoS links.

\subsubsection{Underwater Communications}

An attractive area for future research in underwater communications systems can be to examine the sum-rate performance of underwater acoustic NOMA system with the joint optimization of power allocation, resource allocation, and user clustering. Furthermore, the adoption of rate-optimal NOMA scheme in underwater optical wireless communications with multi-hop cooperative communications is another interesting area of future research. In~\cite{9071997}, the authors envisioned a software-defined hybrid underwater optical-acoustic network architecture. This architecture targets achieving high-speed low-latency and ubiquitous network performance. Exploring the potential of adopting NOMA scheme with such opto-acoustic network architecture can be an interesting topic of research.

\subsection {Rate Optimization of NOMA-enabled Systems with Machine Learning}
\label{sec: future ML_NOMA}

For NOMA-enabled next-generation wireless systems, a common theme to all categories of the rate optimization problems surveyed in this paper is that the number of design variables becomes prohibitively large to configure as the complexity of the system model grows. The multiple design variables are hard to jointly optimize due to the combinatorial complexity. Hence, there is a large potential to apply machine learning (ML) to solve the types of optimization problems for NOMA-enabled systems surveyed in this paper. In~\cite{hussain2019machine}, applications of ML and DL are discussed to resource optimization problems in IoT and other cellular networks, including a brief description of the applications to NOMA systems. Similarly, the survey by Vaezi~\textit{et al.} in~\cite{8792153} presents some discussion on NOMA integrated with ML and deep learning (DL).

To address the problem of combinatorial complexity, the most common approach in the NOMA literature is to divide the problem into a set of sub-problems. ML techniques can be used to aid one or more of these sub-problems as appropriate. For example, in~\cite{cui2018unsupervised} and~\cite{8856221}, the user selection sub-problem is tackled through an unsupervised clustering algorithm and the power allocation problem is addressed through conventional optimization. In~\cite{he2019RL}, where a multi-carrier setting is studied, the channel assignment is tackled through a DRL algorithm while the power allocation is again addressed through conventional optimization. When further variables are added to the problem, ML techniques can be applied to the power optimization too. For example, the authors in~\cite{xiao2018reinforcement} use an RL algorithm for power allocation when an intentional jammer is present. The takeaway message is that as the complexity of the model grows introducing a prohibitively large number of design variables, as is typical for NOMA-enabled systems, ML techniques can be used to solve a subset of these problems and can be used in tandem with traditional optimization techniques. For example, in~\cite{9075277}, an artificial neural network is used in conjunction with a traditional optimization approach to solve a joint problem of power allocation and UAV\textquotesingle s placement to
maximize the sum rate of all users. As complexity is introduced to the NOMA enabled systems through multi-cell, multi-carrier, cooperative settings, etc., the number of design variables can even grow too large for ML algorithms. In such scenarios, an interesting research direction can be to investigate a deep learning neural network such as the one proposed in~\cite{7996587} to identify the parameters that have the largest impact on the sum-rate performance. The selected optimization variables can then be set using conventional optimization approaches or through other ML techniques.

While applying ML to NOMA-enabled systems comes with many attractive advantages, it has its challenges also. One of the big concerns with ML algorithms is the computational power required to run some of these ML algorithms. However, emerging trends such as quantum machine learning (QML) for 6G systems are being studied to aid this. The authors in~\cite{quantum2019} specifically discuss how QML can significantly speed up multi-objective optimization problems that involve tweaking a large number of parameters and their constraints, a typical setting for all NOMA-enabled systems. Another challenge with applying ML algorithms is the large amount of data required. However, communications systems collect and discard a large amount of data today, e.g., CSI, user locations, etc., using them only for instantaneous scheduling decisions. With big data processing developing rapidly, these can be fed to ML algorithms in NOMA-enabled systems. 

Another challenge with applying ML at the physical layer is that the channel changes so fast that an ML algorithm does not have enough time to collect meaningful data to learn from. In particular, this makes applying supervised algorithms that learn from past data challenging at the physical layer. However, in mMIMO systems, supervised learning algorithms have been studied for channel feedback and estimation, MIMO detection, and other related problems~\cite{wen2015MLforChEst, bjornson2019massive}. However, in what follows, we focus on how the other ML techniques, namely unsupervised learning, reinforcement learning, and deep learning, outlined in Section~\ref{sec:Bkgd_ML}, can be applied to rate optimization problems in the NOMA-enabled system models surveyed in this paper. These discussions offer several future research avenues for applying ML in the context of NOMA enabled systems.

Unsupervised ML algorithms do not rely on past training data. In particular, clustering algorithms are a natural fit for NOMA-enabled systems due to user selection sub-problem. As we discussed earlier, due to the combinatorial complexity of the joint optimization of a large number of design variables, the typical approach in NOMA literature is to divide the problem into several sub-problems. The user clustering or user pairing is a typical first sub-problem that researchers tackle. Clustering algorithms such as K-means clustering can be used to tackle this sub-problem. As described in Section~\ref{sec:body mmwave and THz Communications}, the works in~\cite{cui2018unsupervised} and~\cite{8856221} study a mmWave-NOMA system and exploit the high correlation amongst users' channels and the fact that mmWave propagation is dominated by the LoS path to effectively employ K-means clustering. System models that are dominated by a LoS path favor the use of K-means clustering as the problem breaks down to finding spatially correlated users. This type of channel model appears in several NOMA-enabled systems. For example, in UAV or satellite communications, the link between the UAV/satellite and ground users is LoS-dominated and offers an opportunity for user clustering based on unsupervised ML algorithms like K-means clustering. For example, in~\cite{liu2019uav}, K-means clustering is used to find an initial clustering of spatially correlated users after which a Q-learning algorithm (RL algorithm) is used for the 3-D placement of the UAV BS. It is, however, more challenging to apply such a K-means clustering in a rich multipath environment such as in lower frequency bands because good user pairs are not necessarily correlated users in space in such a setting. An area worth investigating is if techniques that infer the user location from the reported CSI such as the channel charting proposed in~\cite{studer2018ChannelCharting} can be fed to a K-means clustering algorithm to form good user clusters for NOMA-enabled systems operating in a rich multipath environment.

Deep Learning (DL) is the more powerful form of machine learning as it involves multiple layers and can extract a set of features in the data, before performing tasks like classification~\cite{zhang2019thirty}. However, due to the fast-changing nature of the physical channel, it becomes difficult to implement a DL approach that first extracts the relevant features from the channel and then applies it to NOMA enabled systems. However, the power of deep learning can still be extracted in several ways in NOMA-enabled systems. In~\cite{8500158}, a deep recurrent neural network is constructed to provide optimal resource allocation results for the NOMA heterogeneous IoT with fast convergence and low computational complexity. We discussed earlier in this section how a neural network such as the one in~\cite{7996587} can be used to extract the most important parameters when the number of design variables and objectives grows very large. However, the most promising use of DL for rate optimization problems in NOMA-enabled systems is when used in conjunction with reinforcement learning, such that the agent employs a multi-layered neural network to make decisions when interacting with the environment. This is termed as deep reinforcement learning (DRL) to highlight the joint use of DL techniques with RL.

DRL algorithms have been studied for several resource allocation problems in next-gen wireless communications systems~\cite{luong2019ApplicationsML}. Such ideas can easily be extended to the NOMA-enabled systems surveyed in this paper. We described the two works of~\cite{he2019RL} and~\cite{xiao2018reinforcement} earlier, where DRL agents are used for channel assignment and power allocation respectively in NOMA systems. For the NOMA-enabled B5G technologies surveyed in this paper, the DRL agent can be either the BS, UAV, users, relay nodes, etc. that need to make autonomous decisions based on their interaction with the other nodes in the system. For example, in~\cite{huang2019reinforcement}, a RL agent is employed for UAV positioning. Similar to the idea of ML clustering being used to solve the user selection sub-problem, the DRL can be used to solve certain sub-problems for the overall rate optimization objective. Potential avenues worth investigating include using a RL agent for the sub-problems of relay selection in NOMA relay networks, spectrum selection in CR-NOMA networks, or UAV placement in UAV-NOMA networks. In this way, DRL can be applied to the NOMA-enabled versions of each of these systems to complement NOMA-specific optimization algorithms. Another interesting problem where a RL algorithm can be used is for systems employing a hybrid MU-MIMO and NOMA approach such as~\cite{senel2019_mMimovsNOMA_bjornson}. Here, a RL agent can be used to switch between the two spectrum sharing schemes, depending on the favorable conditions it detects based on its past experience of interacting with the system.

Another important class of ML algorithms, called online learning, can help address the issue of system flexibility in NOMA-enabled systems, i.e., to be able to adapt to new users or slight changes in the system without too much overhead. For example, the authors in~\cite{cui2018unsupervised} designed an online ML clustering algorithm that can handle new users entering the system up to a certain threshold. Applying online ML algorithms to all the different NOMA-enabled rate optimization problems surveyed in this paper is an important direction of future work, as it helps make the systems practically implementable.

\section{Conclusion}
In this article, a comprehensive review of the literature is conducted on the integration of PD-NOMA with the main candidate communications schemes and technologies for high-data-rate future wireless networks including MISO, MIMO, mMIMO, advanced multi-antenna architectures, mmWave and THz communications, CoMP, cooperative communications, cognitive radio, VLC, UAV communications and other enabling technologies, namely, BackCom, IRS, MEC and edge caching, integrated terrestrial-satellite networks, and underwater communications. Particularly, the survey has investigated the system models and the various utilized optimization methods for each NOMA-enabled technology and the combinations of these technologies, and revealed the increased achievable rates as well as the associated trade-offs. Furthermore, the envisioned role of machine learning in future NOMA-enabled networks and a set of possible directions of future research are presented. It should be emphasized here that although the list of the investigated references, up to the best of the authors' knowledge, is quite comprehensive, it is yet not exhaustive and the reader might find some other related references to rate maximization in the combination of PD-NOMA and the various enabling schemes and technologies considered in this paper. 

\section*{\textbf{Acknowledgments}}
The authors would like to thank King Fahd University of Petroleum \& Minerals (KFUPM) for supporting this work through grant number SB191049. Also, the authors would like to thank Ericsson Canada Inc. and the Discovery Grant of the Natural Sciences and Engineering Research Council of Canada for support.

\bibliographystyle{IEEEtran}
\bibliography{main}

\vskip -2\baselineskip plus -1fil

\begin{IEEEbiography}[{\includegraphics[width=1in,height=1.25in,clip,keepaspectratio]{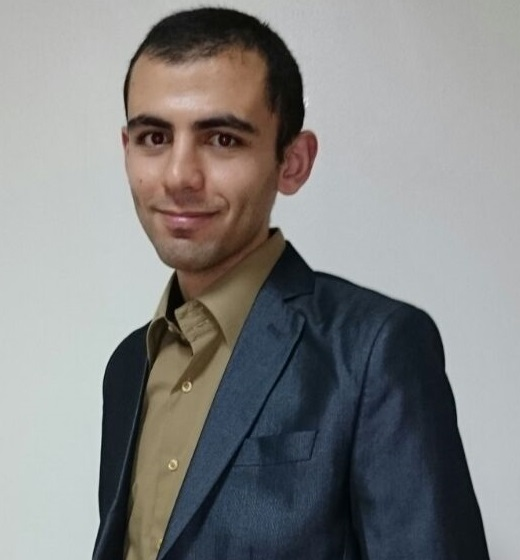}}]{\textbf{Omar Maraqa}}
has received his B.S. degree in Electrical Engineering from Palestine Polytechnic University, Palestine, in 2011, and his M.S. degree in Computer Engineering from King Fahd University of Petroleum \& Minerals (KFUPM), Dhahran, Saudi Arabia, in 2016. He is currently pursuing a Ph.D. degree in Electrical Engineering at KFUPM, Dhahran, Saudi Arabia. His research interests include performance analysis and optimization of wireless communications systems.
\end{IEEEbiography}


\begin{IEEEbiography}[{\includegraphics[width=1in,height=1.25in,clip,keepaspectratio]{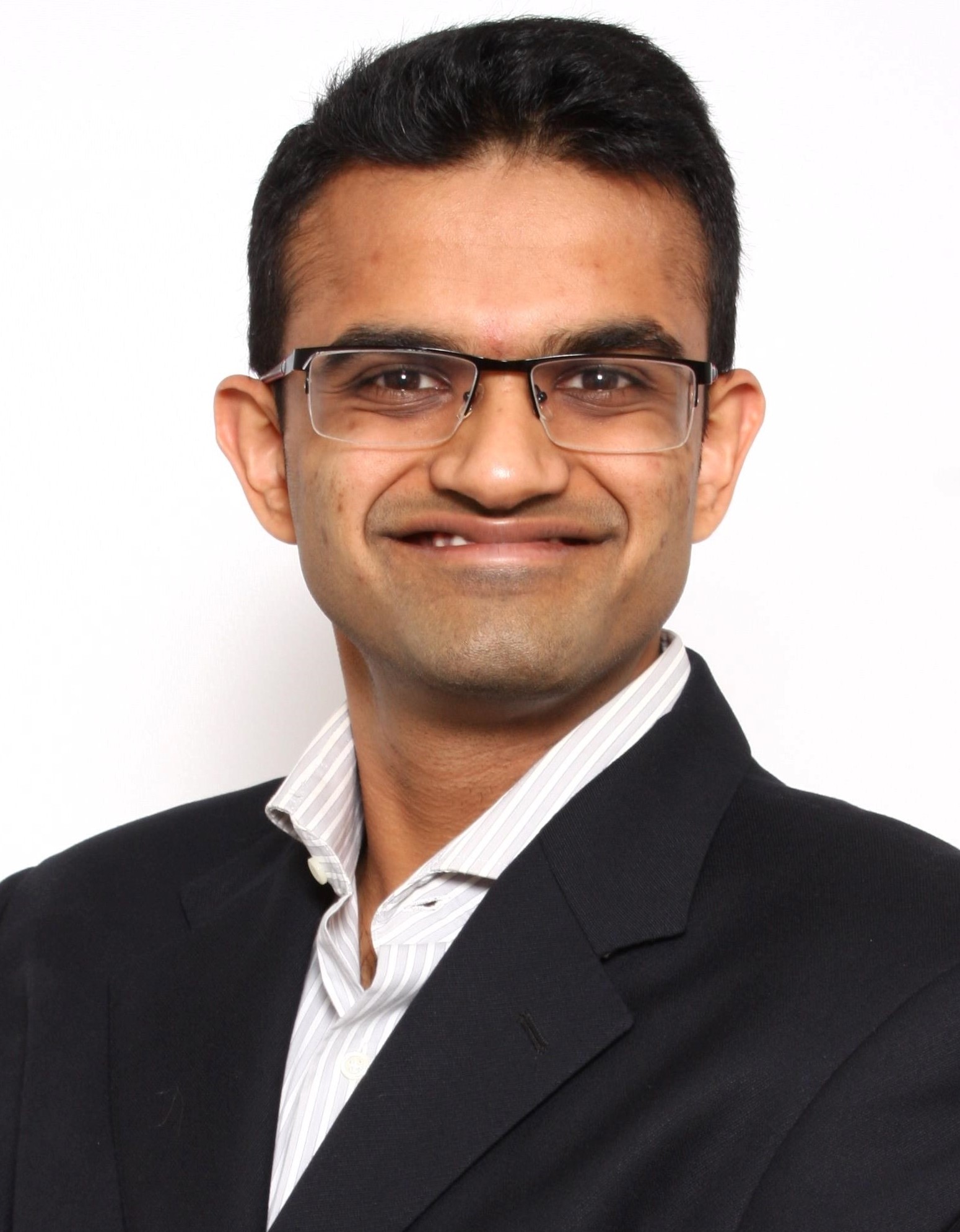}}]{Aditya S. Rajasekaran} (M'18) received the B.Eng (with High Distinction) and M.Eng degree in Systems and Computer Engineering from Carleton University, Ottawa, ON, Canada, in 2014 and 2017, respectively. He is currently pursuing his Ph.D. degree, also in Systems and Computer Engineering at Carleton University. His research interests include  wireless technology solutions aimed towards 5G and beyond cellular networks, including non-orthogonal multiple access solutions.

He is also with Ericsson Canada, where he has been working as a software developer since 2014. He is currently involved in the physical layer development work for Ericsson's 5G New Radio (NR) solutions. 
\end{IEEEbiography}


\begin{IEEEbiography}[{\includegraphics[width=1in,height=1.25in,clip,keepaspectratio]{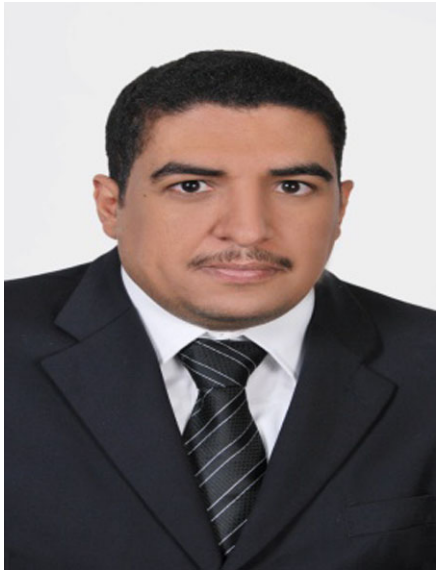}}]{\textbf{Saad Al-Ahmadi}}
has received his M.Sc. in Electrical Engineering from King Fahd University of Petroleum \& Minerals (KFUPM), Dhahran, Saudi Arabia, in 2002 and his Ph.D. in Electrical and Computer Engineering from Ottawa-Carleton Institute for ECE (OCIECE), Ottawa, Canada, in 2010. He is currently with the Department of Electrical Engineering at KFUPM as an Associate Professor. His current research interests include channel characterization, design, and performance analysis of wireless communications systems and networks.
\end{IEEEbiography}


\begin{IEEEbiography}[{\includegraphics[width=1in,height=1.25in,clip,keepaspectratio]{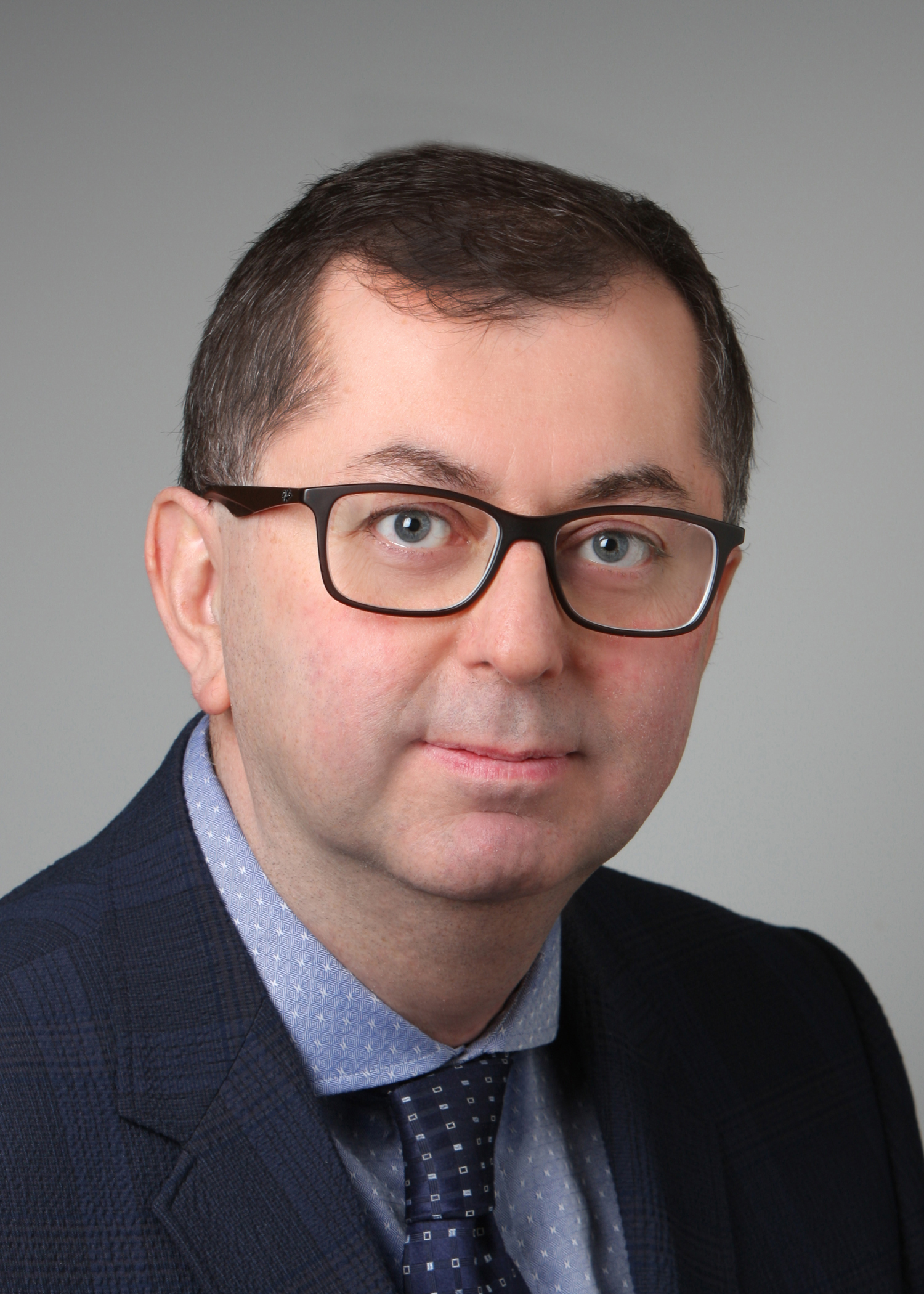}}]{Halim Yanikomeroglu} (F'17) is a Professor at Carleton University, Canada. His research covers many aspects of communications technologies with emphasis on wireless networks. He is a Fellow of IEEE, a Fellow of the Engineering Institute of Canada (EIC), a Fellow of the Canadian Academy of Engineering, a Distinguished Speaker for the IEEE Communications Society and the IEEE Vehicular Technology Society. He has been one of the most frequent tutorial presenters in the leading international IEEE conferences. He has had extensive collaboration with industry which resulted in 35 granted patents (plus more than a dozen applied). During 2012-2016, he led one of the largest academic-industrial collaborative research projects on pre-standards 5G wireless, sponsored by the Ontario Government and the industry. He served as the General Chair and Technical Program Chair of several major international IEEE conferences. He is currently serving as the Chair of the Steering Board of the IEEEs flagship wireless conference, WCNC (Wireless Communications and Networking Conference). He supervised 24 PhD students (all completed with theses).
\end{IEEEbiography}


\begin{IEEEbiography}[{\includegraphics[width=1in,height=1.25in,clip,keepaspectratio]{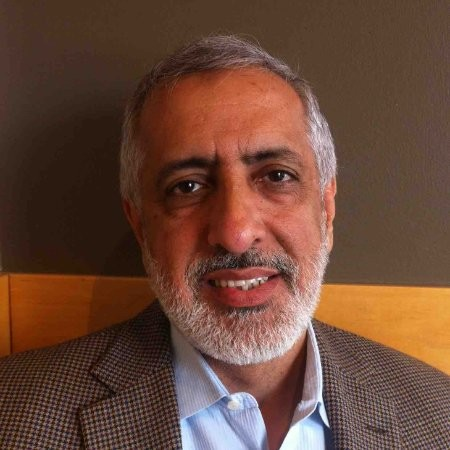}}]{\textbf{Sadiq M. Sait}}
(SM'02) was born in Bengaluru. He received the bachelor's degree in electronics engineering from Bangalore University in 1981, and the master's and Ph.D. degrees in electrical engineering from the King Fahd University of Petroleum \& Minerals (KFUPM) in 1983 and 1987, respectively. He is currently a Professor of Computer Engineering and the Director of the Center for Communications and IT Research, Research Institute, KFUPM. He has authored over 200 research papers, contributed chapters to technical books, and lectured in over 25 countries. He is also the Principle Author of two books. He received the Best Electronic Engineer Award from the Indian Institute of Electrical Engineers, Bengaluru, in 1981.
\end{IEEEbiography}

\end{document}